\documentclass[longauth]{aaEC}  
\usepackage{threeparttable}
\usepackage{txfonts}

\usepackage{graphicx}
\usepackage{natbib}
\usepackage{scalerel}
\usepackage{amsmath}	% Advanced maths commands
\usepackage{dsfont}
\usepackage{newtxtext,newtxmath}
\usepackage{tikz}
\usetikzlibrary{angles, quotes}
\usetikzlibrary{shapes.arrows,calc}
\usepackage[normalem]{ulem}
\usepackage[switch, modulo]{lineno}



\usepackage{graphicx}   % Including figure files

\usepackage[colorlinks,citecolor=blue, urlcolor=blue, linkcolor=blue]{hyperref}
\usepackage{euclid}

\newcommand{\Msun}{ h^{-1}{\rm M_{ \odot}}}

\newcommand{\ihMpc}{ h\,{\rm Mpc}^{-1}}

\newcommand{\MapMapMap}{{\expval{\Map^3}}}

\newcommand{\diracd}{\delta_\mathrm{D}}

\newcommand{\dd}{\mathrm{d}}

\newcommand{\expval}[1]{\langle #1 \rangle}

\newcommand{\code}{\tt{}}
\newcommand{\mycomment}[1]{}

\renewcommand{\arraystretch}{1.2}

\begin{document}

% Title of the paper, and the short title which is used in the headers.
% Keep the title short and informative.
\title{\Euclid\/: An emulator for baryonic effects on the matter bispectrum\thanks{This paper is published on behalf of the Euclid Consortium.}}

% \author{P. A. Burger
%           \inst{1,2}, 
%           G. Aric{\`o}\inst{3,4}\fnmsep\thanks{E-mail: \href{mailto:arico@bo.infn.it}{arico@bo.infn.it}},

%%%% Version Tuesday 23rd of September 2025 01:29:03 PM UT												
%%%% Please do not edit the author list -- contact ECEB Bureau for changes
\newcommand{\orcid}[1]{} %% if already defined in aa.cls: comment, or use renewcommand			   
\author{P.~A.~Burger\orcid{0000-0001-8637-6305}\inst{\ref{aff1},\ref{aff2}}
\and G.~Aric\`o\orcid{0000-0002-2802-2928}\thanks{\email{arico@bo.infn.it}}\inst{\ref{aff3},\ref{aff4}}
\and L.~Linke\orcid{0000-0002-2622-8113}\inst{\ref{aff5}}
\and R.~E.~Angulo\orcid{0000-0003-2953-3970}\inst{\ref{aff6},\ref{aff7}}
\and J.~C.~Broxterman\inst{\ref{aff8},\ref{aff9}}
\and J.~Schaye\orcid{0000-0002-0668-5560}\inst{\ref{aff9}}
\and M.~Schaller\orcid{0000-0002-2395-4902}\inst{\ref{aff9},\ref{aff8}}
\and M.~Zennaro\orcid{0000-0002-4458-1754}\inst{\ref{aff10}}
\and A.~Halder\orcid{0000-0002-0352-9351}\inst{\ref{aff11},\ref{aff12},\ref{aff13},\ref{aff14}}
\and L.~Porth\orcid{0000-0003-1176-6346}\inst{\ref{aff15}}
\and S.~Heydenreich\inst{\ref{aff16},\ref{aff15}}
\and M.~J.~Hudson\orcid{0000-0002-1437-3786}\inst{\ref{aff1},\ref{aff2},\ref{aff17}}
\and A.~Amara\inst{\ref{aff18}}
\and S.~Andreon\orcid{0000-0002-2041-8784}\inst{\ref{aff19}}
\and C.~Baccigalupi\orcid{0000-0002-8211-1630}\inst{\ref{aff20},\ref{aff21},\ref{aff22},\ref{aff23}}
\and M.~Baldi\orcid{0000-0003-4145-1943}\inst{\ref{aff24},\ref{aff25},\ref{aff3}}
\and A.~Balestra\orcid{0000-0002-6967-261X}\inst{\ref{aff26}}
\and S.~Bardelli\orcid{0000-0002-8900-0298}\inst{\ref{aff25}}
\and A.~Biviano\orcid{0000-0002-0857-0732}\inst{\ref{aff21},\ref{aff20}}
\and E.~Branchini\orcid{0000-0002-0808-6908}\inst{\ref{aff27},\ref{aff28},\ref{aff19}}
\and M.~Brescia\orcid{0000-0001-9506-5680}\inst{\ref{aff29},\ref{aff30}}
\and S.~Camera\orcid{0000-0003-3399-3574}\inst{\ref{aff31},\ref{aff32},\ref{aff33}}
\and V.~Capobianco\orcid{0000-0002-3309-7692}\inst{\ref{aff33}}
\and C.~Carbone\orcid{0000-0003-0125-3563}\inst{\ref{aff34}}
\and V.~F.~Cardone\inst{\ref{aff35},\ref{aff36}}
\and J.~Carretero\orcid{0000-0002-3130-0204}\inst{\ref{aff37},\ref{aff38}}
\and S.~Casas\orcid{0000-0002-4751-5138}\inst{\ref{aff39}}
\and M.~Castellano\orcid{0000-0001-9875-8263}\inst{\ref{aff35}}
\and G.~Castignani\orcid{0000-0001-6831-0687}\inst{\ref{aff25}}
\and S.~Cavuoti\orcid{0000-0002-3787-4196}\inst{\ref{aff30},\ref{aff40}}
\and K.~C.~Chambers\orcid{0000-0001-6965-7789}\inst{\ref{aff41}}
\and A.~Cimatti\inst{\ref{aff42}}
\and C.~Colodro-Conde\inst{\ref{aff43}}
\and G.~Congedo\orcid{0000-0003-2508-0046}\inst{\ref{aff44}}
\and L.~Conversi\orcid{0000-0002-6710-8476}\inst{\ref{aff45},\ref{aff46}}
\and Y.~Copin\orcid{0000-0002-5317-7518}\inst{\ref{aff47}}
\and F.~Courbin\orcid{0000-0003-0758-6510}\inst{\ref{aff48},\ref{aff49}}
\and H.~M.~Courtois\orcid{0000-0003-0509-1776}\inst{\ref{aff50}}
\and A.~Da~Silva\orcid{0000-0002-6385-1609}\inst{\ref{aff51},\ref{aff52}}
\and H.~Degaudenzi\orcid{0000-0002-5887-6799}\inst{\ref{aff53}}
\and S.~de~la~Torre\inst{\ref{aff54}}
\and G.~De~Lucia\orcid{0000-0002-6220-9104}\inst{\ref{aff21}}
\and F.~Dubath\orcid{0000-0002-6533-2810}\inst{\ref{aff53}}
\and C.~A.~J.~Duncan\orcid{0009-0003-3573-0791}\inst{\ref{aff44},\ref{aff55}}
\and X.~Dupac\inst{\ref{aff46}}
\and S.~Dusini\orcid{0000-0002-1128-0664}\inst{\ref{aff56}}
\and S.~Escoffier\orcid{0000-0002-2847-7498}\inst{\ref{aff57}}
\and M.~Farina\orcid{0000-0002-3089-7846}\inst{\ref{aff58}}
\and R.~Farinelli\inst{\ref{aff25}}
\and S.~Ferriol\inst{\ref{aff47}}
\and F.~Finelli\orcid{0000-0002-6694-3269}\inst{\ref{aff25},\ref{aff59}}
\and P.~Fosalba\orcid{0000-0002-1510-5214}\inst{\ref{aff60},\ref{aff61}}
\and N.~Fourmanoit\orcid{0009-0005-6816-6925}\inst{\ref{aff57}}
\and M.~Frailis\orcid{0000-0002-7400-2135}\inst{\ref{aff21}}
\and E.~Franceschi\orcid{0000-0002-0585-6591}\inst{\ref{aff25}}
\and M.~Fumana\orcid{0000-0001-6787-5950}\inst{\ref{aff34}}
\and S.~Galeotta\orcid{0000-0002-3748-5115}\inst{\ref{aff21}}
\and B.~Gillis\orcid{0000-0002-4478-1270}\inst{\ref{aff44}}
\and C.~Giocoli\orcid{0000-0002-9590-7961}\inst{\ref{aff25},\ref{aff3}}
\and J.~Gracia-Carpio\inst{\ref{aff14}}
\and A.~Grazian\orcid{0000-0002-5688-0663}\inst{\ref{aff26}}
\and F.~Grupp\inst{\ref{aff14},\ref{aff62}}
\and S.~V.~H.~Haugan\orcid{0000-0001-9648-7260}\inst{\ref{aff63}}
\and H.~Hoekstra\orcid{0000-0002-0641-3231}\inst{\ref{aff9}}
\and W.~Holmes\inst{\ref{aff64}}
\and I.~M.~Hook\orcid{0000-0002-2960-978X}\inst{\ref{aff65}}
\and F.~Hormuth\inst{\ref{aff66}}
\and A.~Hornstrup\orcid{0000-0002-3363-0936}\inst{\ref{aff67},\ref{aff68}}
\and K.~Jahnke\orcid{0000-0003-3804-2137}\inst{\ref{aff69}}
\and M.~Jhabvala\inst{\ref{aff70}}
\and B.~Joachimi\orcid{0000-0001-7494-1303}\inst{\ref{aff71}}
\and E.~Keih\"anen\orcid{0000-0003-1804-7715}\inst{\ref{aff72}}
\and S.~Kermiche\orcid{0000-0002-0302-5735}\inst{\ref{aff57}}
\and M.~Kilbinger\orcid{0000-0001-9513-7138}\inst{\ref{aff73}}
\and B.~Kubik\orcid{0009-0006-5823-4880}\inst{\ref{aff47}}
\and M.~Kunz\orcid{0000-0002-3052-7394}\inst{\ref{aff74}}
\and H.~Kurki-Suonio\orcid{0000-0002-4618-3063}\inst{\ref{aff75},\ref{aff76}}
\and A.~M.~C.~Le~Brun\orcid{0000-0002-0936-4594}\inst{\ref{aff77}}
\and S.~Ligori\orcid{0000-0003-4172-4606}\inst{\ref{aff33}}
\and P.~B.~Lilje\orcid{0000-0003-4324-7794}\inst{\ref{aff63}}
\and V.~Lindholm\orcid{0000-0003-2317-5471}\inst{\ref{aff75},\ref{aff76}}
\and I.~Lloro\orcid{0000-0001-5966-1434}\inst{\ref{aff78}}
\and G.~Mainetti\orcid{0000-0003-2384-2377}\inst{\ref{aff79}}
\and D.~Maino\inst{\ref{aff80},\ref{aff34},\ref{aff81}}
\and E.~Maiorano\orcid{0000-0003-2593-4355}\inst{\ref{aff25}}
\and O.~Mansutti\orcid{0000-0001-5758-4658}\inst{\ref{aff21}}
\and O.~Marggraf\orcid{0000-0001-7242-3852}\inst{\ref{aff15}}
\and M.~Martinelli\orcid{0000-0002-6943-7732}\inst{\ref{aff35},\ref{aff36}}
\and N.~Martinet\orcid{0000-0003-2786-7790}\inst{\ref{aff54}}
\and F.~Marulli\orcid{0000-0002-8850-0303}\inst{\ref{aff82},\ref{aff25},\ref{aff3}}
\and R.~Massey\orcid{0000-0002-6085-3780}\inst{\ref{aff83}}
\and E.~Medinaceli\orcid{0000-0002-4040-7783}\inst{\ref{aff25}}
\and S.~Mei\orcid{0000-0002-2849-559X}\inst{\ref{aff84},\ref{aff85}}
\and M.~Melchior\inst{\ref{aff86}}
\and M.~Meneghetti\orcid{0000-0003-1225-7084}\inst{\ref{aff25},\ref{aff3}}
\and E.~Merlin\orcid{0000-0001-6870-8900}\inst{\ref{aff35}}
\and G.~Meylan\inst{\ref{aff87}}
\and A.~Mora\orcid{0000-0002-1922-8529}\inst{\ref{aff88}}
\and M.~Moresco\orcid{0000-0002-7616-7136}\inst{\ref{aff82},\ref{aff25}}
\and L.~Moscardini\orcid{0000-0002-3473-6716}\inst{\ref{aff82},\ref{aff25},\ref{aff3}}
\and C.~Neissner\orcid{0000-0001-8524-4968}\inst{\ref{aff89},\ref{aff38}}
\and S.-M.~Niemi\orcid{0009-0005-0247-0086}\inst{\ref{aff90}}
\and C.~Padilla\orcid{0000-0001-7951-0166}\inst{\ref{aff89}}
\and S.~Paltani\orcid{0000-0002-8108-9179}\inst{\ref{aff53}}
\and F.~Pasian\orcid{0000-0002-4869-3227}\inst{\ref{aff21}}
\and K.~Pedersen\inst{\ref{aff91}}
\and W.~J.~Percival\orcid{0000-0002-0644-5727}\inst{\ref{aff2},\ref{aff1},\ref{aff17}}
\and V.~Pettorino\inst{\ref{aff90}}
\and S.~Pires\orcid{0000-0002-0249-2104}\inst{\ref{aff73}}
\and G.~Polenta\orcid{0000-0003-4067-9196}\inst{\ref{aff92}}
\and M.~Poncet\inst{\ref{aff93}}
\and L.~A.~Popa\inst{\ref{aff94}}
\and F.~Raison\orcid{0000-0002-7819-6918}\inst{\ref{aff14}}
\and A.~Renzi\orcid{0000-0001-9856-1970}\inst{\ref{aff95},\ref{aff56}}
\and J.~Rhodes\orcid{0000-0002-4485-8549}\inst{\ref{aff64}}
\and G.~Riccio\inst{\ref{aff30}}
\and E.~Romelli\orcid{0000-0003-3069-9222}\inst{\ref{aff21}}
\and M.~Roncarelli\orcid{0000-0001-9587-7822}\inst{\ref{aff25}}
\and R.~Saglia\orcid{0000-0003-0378-7032}\inst{\ref{aff62},\ref{aff14}}
\and Z.~Sakr\orcid{0000-0002-4823-3757}\inst{\ref{aff96},\ref{aff97},\ref{aff98}}
\and A.~G.~S\'anchez\orcid{0000-0003-1198-831X}\inst{\ref{aff14}}
\and D.~Sapone\orcid{0000-0001-7089-4503}\inst{\ref{aff99}}
\and B.~Sartoris\orcid{0000-0003-1337-5269}\inst{\ref{aff62},\ref{aff21}}
\and P.~Schneider\orcid{0000-0001-8561-2679}\inst{\ref{aff15}}
\and T.~Schrabback\orcid{0000-0002-6987-7834}\inst{\ref{aff5}}
\and A.~Secroun\orcid{0000-0003-0505-3710}\inst{\ref{aff57}}
\and E.~Sefusatti\orcid{0000-0003-0473-1567}\inst{\ref{aff21},\ref{aff20},\ref{aff22}}
\and G.~Seidel\orcid{0000-0003-2907-353X}\inst{\ref{aff69}}
\and S.~Serrano\orcid{0000-0002-0211-2861}\inst{\ref{aff60},\ref{aff100},\ref{aff61}}
\and P.~Simon\inst{\ref{aff15}}
\and C.~Sirignano\orcid{0000-0002-0995-7146}\inst{\ref{aff95},\ref{aff56}}
\and G.~Sirri\orcid{0000-0003-2626-2853}\inst{\ref{aff3}}
\and A.~Spurio~Mancini\orcid{0000-0001-5698-0990}\inst{\ref{aff101}}
\and L.~Stanco\orcid{0000-0002-9706-5104}\inst{\ref{aff56}}
\and J.~Steinwagner\orcid{0000-0001-7443-1047}\inst{\ref{aff14}}
\and P.~Tallada-Cresp\'{i}\orcid{0000-0002-1336-8328}\inst{\ref{aff37},\ref{aff38}}
\and A.~N.~Taylor\inst{\ref{aff44}}
\and I.~Tereno\orcid{0000-0002-4537-6218}\inst{\ref{aff51},\ref{aff102}}
\and S.~Toft\orcid{0000-0003-3631-7176}\inst{\ref{aff103},\ref{aff104}}
\and R.~Toledo-Moreo\orcid{0000-0002-2997-4859}\inst{\ref{aff105}}
\and F.~Torradeflot\orcid{0000-0003-1160-1517}\inst{\ref{aff38},\ref{aff37}}
\and I.~Tutusaus\orcid{0000-0002-3199-0399}\inst{\ref{aff97}}
\and L.~Valenziano\orcid{0000-0002-1170-0104}\inst{\ref{aff25},\ref{aff59}}
\and J.~Valiviita\orcid{0000-0001-6225-3693}\inst{\ref{aff75},\ref{aff76}}
\and T.~Vassallo\orcid{0000-0001-6512-6358}\inst{\ref{aff62},\ref{aff21}}
\and G.~Verdoes~Kleijn\orcid{0000-0001-5803-2580}\inst{\ref{aff106}}
\and A.~Veropalumbo\orcid{0000-0003-2387-1194}\inst{\ref{aff19},\ref{aff28},\ref{aff27}}
\and Y.~Wang\orcid{0000-0002-4749-2984}\inst{\ref{aff107}}
\and J.~Weller\orcid{0000-0002-8282-2010}\inst{\ref{aff62},\ref{aff14}}
\and G.~Zamorani\orcid{0000-0002-2318-301X}\inst{\ref{aff25}}
\and E.~Zucca\orcid{0000-0002-5845-8132}\inst{\ref{aff25}}
\and C.~Burigana\orcid{0000-0002-3005-5796}\inst{\ref{aff108},\ref{aff59}}
\and L.~Gabarra\orcid{0000-0002-8486-8856}\inst{\ref{aff10}}
\and A.~Pezzotta\orcid{0000-0003-0726-2268}\inst{\ref{aff19}}
\and V.~Scottez\orcid{0009-0008-3864-940X}\inst{\ref{aff109},\ref{aff110}}
\and M.~Viel\orcid{0000-0002-2642-5707}\inst{\ref{aff20},\ref{aff21},\ref{aff23},\ref{aff22},\ref{aff111}}}
										   
%%%% please do not edit the affiliation list -- contact ECEB Bureau for changes
\institute{Department of Physics and Astronomy, University of Waterloo, Waterloo, Ontario N2L 3G1, Canada\label{aff1}
\and
Waterloo Centre for Astrophysics, University of Waterloo, Waterloo, Ontario N2L 3G1, Canada\label{aff2}
\and
INFN-Sezione di Bologna, Viale Berti Pichat 6/2, 40127 Bologna, Italy\label{aff3}
\and
Department of Astrophysics, University of Zurich, Winterthurerstrasse 190, 8057 Zurich, Switzerland\label{aff4}
\and
Universit\"at Innsbruck, Institut f\"ur Astro- und Teilchenphysik, Technikerstr. 25/8, 6020 Innsbruck, Austria\label{aff5}
\and
Donostia International Physics Center (DIPC), Paseo Manuel de Lardizabal, 4, 20018, Donostia-San Sebasti\'an, Guipuzkoa, Spain\label{aff6}
\and
IKERBASQUE, Basque Foundation for Science, 48013, Bilbao, Spain\label{aff7}
\and
Institute Lorentz, Leiden University, Niels Bohrweg 2, 2333 CA Leiden, The Netherlands\label{aff8}
\and
Leiden Observatory, Leiden University, Einsteinweg 55, 2333 CC Leiden, The Netherlands\label{aff9}
\and
Department of Physics, Oxford University, Keble Road, Oxford OX1 3RH, UK\label{aff10}
\and
Institute of Astronomy, University of Cambridge, Madingley Road, Cambridge CB3 0HA, UK\label{aff11}
\and
Kavli Institute for Cosmology Cambridge, Madingley Road, Cambridge, CB3 0HA, UK\label{aff12}
\and
University Observatory, LMU Faculty of Physics, Scheinerstrasse 1, 81679 Munich, Germany\label{aff13}
\and
Max Planck Institute for Extraterrestrial Physics, Giessenbachstr. 1, 85748 Garching, Germany\label{aff14}
\and
Universit\"at Bonn, Argelander-Institut f\"ur Astronomie, Auf dem H\"ugel 71, 53121 Bonn, Germany\label{aff15}
\and
Department of Astronomy and Astrophysics, University of California, Santa Cruz, 1156 High Street, Santa Cruz, CA 95064, USA\label{aff16}
\and
Perimeter Institute for Theoretical Physics, Waterloo, Ontario N2L 2Y5, Canada\label{aff17}
\and
School of Mathematics and Physics, University of Surrey, Guildford, Surrey, GU2 7XH, UK\label{aff18}
\and
INAF-Osservatorio Astronomico di Brera, Via Brera 28, 20122 Milano, Italy\label{aff19}
\and
IFPU, Institute for Fundamental Physics of the Universe, via Beirut 2, 34151 Trieste, Italy\label{aff20}
\and
INAF-Osservatorio Astronomico di Trieste, Via G. B. Tiepolo 11, 34143 Trieste, Italy\label{aff21}
\and
INFN, Sezione di Trieste, Via Valerio 2, 34127 Trieste TS, Italy\label{aff22}
\and
SISSA, International School for Advanced Studies, Via Bonomea 265, 34136 Trieste TS, Italy\label{aff23}
\and
Dipartimento di Fisica e Astronomia, Universit\`a di Bologna, Via Gobetti 93/2, 40129 Bologna, Italy\label{aff24}
\and
INAF-Osservatorio di Astrofisica e Scienza dello Spazio di Bologna, Via Piero Gobetti 93/3, 40129 Bologna, Italy\label{aff25}
\and
INAF-Osservatorio Astronomico di Padova, Via dell'Osservatorio 5, 35122 Padova, Italy\label{aff26}
\and
Dipartimento di Fisica, Universit\`a di Genova, Via Dodecaneso 33, 16146, Genova, Italy\label{aff27}
\and
INFN-Sezione di Genova, Via Dodecaneso 33, 16146, Genova, Italy\label{aff28}
\and
Department of Physics "E. Pancini", University Federico II, Via Cinthia 6, 80126, Napoli, Italy\label{aff29}
\and
INAF-Osservatorio Astronomico di Capodimonte, Via Moiariello 16, 80131 Napoli, Italy\label{aff30}
\and
Dipartimento di Fisica, Universit\`a degli Studi di Torino, Via P. Giuria 1, 10125 Torino, Italy\label{aff31}
\and
INFN-Sezione di Torino, Via P. Giuria 1, 10125 Torino, Italy\label{aff32}
\and
INAF-Osservatorio Astrofisico di Torino, Via Osservatorio 20, 10025 Pino Torinese (TO), Italy\label{aff33}
\and
INAF-IASF Milano, Via Alfonso Corti 12, 20133 Milano, Italy\label{aff34}
\and
INAF-Osservatorio Astronomico di Roma, Via Frascati 33, 00078 Monteporzio Catone, Italy\label{aff35}
\and
INFN-Sezione di Roma, Piazzale Aldo Moro, 2 - c/o Dipartimento di Fisica, Edificio G. Marconi, 00185 Roma, Italy\label{aff36}
\and
Centro de Investigaciones Energ\'eticas, Medioambientales y Tecnol\'ogicas (CIEMAT), Avenida Complutense 40, 28040 Madrid, Spain\label{aff37}
\and
Port d'Informaci\'{o} Cient\'{i}fica, Campus UAB, C. Albareda s/n, 08193 Bellaterra (Barcelona), Spain\label{aff38}
\and
Institute for Theoretical Particle Physics and Cosmology (TTK), RWTH Aachen University, 52056 Aachen, Germany\label{aff39}
\and
INFN section of Naples, Via Cinthia 6, 80126, Napoli, Italy\label{aff40}
\and
Institute for Astronomy, University of Hawaii, 2680 Woodlawn Drive, Honolulu, HI 96822, USA\label{aff41}
\and
Dipartimento di Fisica e Astronomia "Augusto Righi" - Alma Mater Studiorum Universit\`a di Bologna, Viale Berti Pichat 6/2, 40127 Bologna, Italy\label{aff42}
\and
Instituto de Astrof\'{\i}sica de Canarias, V\'{\i}a L\'actea, 38205 La Laguna, Tenerife, Spain\label{aff43}
\and
Institute for Astronomy, University of Edinburgh, Royal Observatory, Blackford Hill, Edinburgh EH9 3HJ, UK\label{aff44}
\and
European Space Agency/ESRIN, Largo Galileo Galilei 1, 00044 Frascati, Roma, Italy\label{aff45}
\and
ESAC/ESA, Camino Bajo del Castillo, s/n., Urb. Villafranca del Castillo, 28692 Villanueva de la Ca\~nada, Madrid, Spain\label{aff46}
\and
Universit\'e Claude Bernard Lyon 1, CNRS/IN2P3, IP2I Lyon, UMR 5822, Villeurbanne, F-69100, France\label{aff47}
\and
Institut de Ci\`{e}ncies del Cosmos (ICCUB), Universitat de Barcelona (IEEC-UB), Mart\'{i} i Franqu\`{e}s 1, 08028 Barcelona, Spain\label{aff48}
\and
Instituci\'o Catalana de Recerca i Estudis Avan\c{c}ats (ICREA), Passeig de Llu\'{\i}s Companys 23, 08010 Barcelona, Spain\label{aff49}
\and
UCB Lyon 1, CNRS/IN2P3, IUF, IP2I Lyon, 4 rue Enrico Fermi, 69622 Villeurbanne, France\label{aff50}
\and
Departamento de F\'isica, Faculdade de Ci\^encias, Universidade de Lisboa, Edif\'icio C8, Campo Grande, PT1749-016 Lisboa, Portugal\label{aff51}
\and
Instituto de Astrof\'isica e Ci\^encias do Espa\c{c}o, Faculdade de Ci\^encias, Universidade de Lisboa, Campo Grande, 1749-016 Lisboa, Portugal\label{aff52}
\and
Department of Astronomy, University of Geneva, ch. d'Ecogia 16, 1290 Versoix, Switzerland\label{aff53}
\and
Aix-Marseille Universit\'e, CNRS, CNES, LAM, Marseille, France\label{aff54}
\and
Jodrell Bank Centre for Astrophysics, Department of Physics and Astronomy, University of Manchester, Oxford Road, Manchester M13 9PL, UK\label{aff55}
\and
INFN-Padova, Via Marzolo 8, 35131 Padova, Italy\label{aff56}
\and
Aix-Marseille Universit\'e, CNRS/IN2P3, CPPM, Marseille, France\label{aff57}
\and
INAF-Istituto di Astrofisica e Planetologia Spaziali, via del Fosso del Cavaliere, 100, 00100 Roma, Italy\label{aff58}
\and
INFN-Bologna, Via Irnerio 46, 40126 Bologna, Italy\label{aff59}
\and
Institut d'Estudis Espacials de Catalunya (IEEC),  Edifici RDIT, Campus UPC, 08860 Castelldefels, Barcelona, Spain\label{aff60}
\and
Institute of Space Sciences (ICE, CSIC), Campus UAB, Carrer de Can Magrans, s/n, 08193 Barcelona, Spain\label{aff61}
\and
Universit\"ats-Sternwarte M\"unchen, Fakult\"at f\"ur Physik, Ludwig-Maximilians-Universit\"at M\"unchen, Scheinerstrasse 1, 81679 M\"unchen, Germany\label{aff62}
\and
Institute of Theoretical Astrophysics, University of Oslo, P.O. Box 1029 Blindern, 0315 Oslo, Norway\label{aff63}
\and
Jet Propulsion Laboratory, California Institute of Technology, 4800 Oak Grove Drive, Pasadena, CA, 91109, USA\label{aff64}
\and
Department of Physics, Lancaster University, Lancaster, LA1 4YB, UK\label{aff65}
\and
Felix Hormuth Engineering, Goethestr. 17, 69181 Leimen, Germany\label{aff66}
\and
Technical University of Denmark, Elektrovej 327, 2800 Kgs. Lyngby, Denmark\label{aff67}
\and
Cosmic Dawn Center (DAWN), Denmark\label{aff68}
\and
Max-Planck-Institut f\"ur Astronomie, K\"onigstuhl 17, 69117 Heidelberg, Germany\label{aff69}
\and
NASA Goddard Space Flight Center, Greenbelt, MD 20771, USA\label{aff70}
\and
Department of Physics and Astronomy, University College London, Gower Street, London WC1E 6BT, UK\label{aff71}
\and
Department of Physics and Helsinki Institute of Physics, Gustaf H\"allstr\"omin katu 2, University of Helsinki, 00014 Helsinki, Finland\label{aff72}
\and
Universit\'e Paris-Saclay, Universit\'e Paris Cit\'e, CEA, CNRS, AIM, 91191, Gif-sur-Yvette, France\label{aff73}
\and
Universit\'e de Gen\`eve, D\'epartement de Physique Th\'eorique and Centre for Astroparticle Physics, 24 quai Ernest-Ansermet, CH-1211 Gen\`eve 4, Switzerland\label{aff74}
\and
Department of Physics, P.O. Box 64, University of Helsinki, 00014 Helsinki, Finland\label{aff75}
\and
Helsinki Institute of Physics, Gustaf H{\"a}llstr{\"o}min katu 2, University of Helsinki, 00014 Helsinki, Finland\label{aff76}
\and
Laboratoire d'etude de l'Univers et des phenomenes eXtremes, Observatoire de Paris, Universit\'e PSL, Sorbonne Universit\'e, CNRS, 92190 Meudon, France\label{aff77}
\and
SKAO, Jodrell Bank, Lower Withington, Macclesfield SK11 9FT, United Kingdom\label{aff78}
\and
Centre de Calcul de l'IN2P3/CNRS, 21 avenue Pierre de Coubertin 69627 Villeurbanne Cedex, France\label{aff79}
\and
Dipartimento di Fisica "Aldo Pontremoli", Universit\`a degli Studi di Milano, Via Celoria 16, 20133 Milano, Italy\label{aff80}
\and
INFN-Sezione di Milano, Via Celoria 16, 20133 Milano, Italy\label{aff81}
\and
Dipartimento di Fisica e Astronomia "Augusto Righi" - Alma Mater Studiorum Universit\`a di Bologna, via Piero Gobetti 93/2, 40129 Bologna, Italy\label{aff82}
\and
Department of Physics, Institute for Computational Cosmology, Durham University, South Road, Durham, DH1 3LE, UK\label{aff83}
\and
Universit\'e Paris Cit\'e, CNRS, Astroparticule et Cosmologie, 75013 Paris, France\label{aff84}
\and
CNRS-UCB International Research Laboratory, Centre Pierre Bin\'etruy, IRL2007, CPB-IN2P3, Berkeley, USA\label{aff85}
\and
University of Applied Sciences and Arts of Northwestern Switzerland, School of Engineering, 5210 Windisch, Switzerland\label{aff86}
\and
Institute of Physics, Laboratory of Astrophysics, Ecole Polytechnique F\'ed\'erale de Lausanne (EPFL), Observatoire de Sauverny, 1290 Versoix, Switzerland\label{aff87}
\and
Telespazio UK S.L. for European Space Agency (ESA), Camino bajo del Castillo, s/n, Urbanizacion Villafranca del Castillo, Villanueva de la Ca\~nada, 28692 Madrid, Spain\label{aff88}
\and
Institut de F\'{i}sica d'Altes Energies (IFAE), The Barcelona Institute of Science and Technology, Campus UAB, 08193 Bellaterra (Barcelona), Spain\label{aff89}
\and
European Space Agency/ESTEC, Keplerlaan 1, 2201 AZ Noordwijk, The Netherlands\label{aff90}
\and
DARK, Niels Bohr Institute, University of Copenhagen, Jagtvej 155, 2200 Copenhagen, Denmark\label{aff91}
\and
Space Science Data Center, Italian Space Agency, via del Politecnico snc, 00133 Roma, Italy\label{aff92}
\and
Centre National d'Etudes Spatiales -- Centre spatial de Toulouse, 18 avenue Edouard Belin, 31401 Toulouse Cedex 9, France\label{aff93}
\and
Institute of Space Science, Str. Atomistilor, nr. 409 M\u{a}gurele, Ilfov, 077125, Romania\label{aff94}
\and
Dipartimento di Fisica e Astronomia "G. Galilei", Universit\`a di Padova, Via Marzolo 8, 35131 Padova, Italy\label{aff95}
\and
Institut f\"ur Theoretische Physik, University of Heidelberg, Philosophenweg 16, 69120 Heidelberg, Germany\label{aff96}
\and
Institut de Recherche en Astrophysique et Plan\'etologie (IRAP), Universit\'e de Toulouse, CNRS, UPS, CNES, 14 Av. Edouard Belin, 31400 Toulouse, France\label{aff97}
\and
Universit\'e St Joseph; Faculty of Sciences, Beirut, Lebanon\label{aff98}
\and
Departamento de F\'isica, FCFM, Universidad de Chile, Blanco Encalada 2008, Santiago, Chile\label{aff99}
\and
Satlantis, University Science Park, Sede Bld 48940, Leioa-Bilbao, Spain\label{aff100}
\and
Department of Physics, Royal Holloway, University of London, Surrey TW20 0EX, UK\label{aff101}
\and
Instituto de Astrof\'isica e Ci\^encias do Espa\c{c}o, Faculdade de Ci\^encias, Universidade de Lisboa, Tapada da Ajuda, 1349-018 Lisboa, Portugal\label{aff102}
\and
Cosmic Dawn Center (DAWN)\label{aff103}
\and
Niels Bohr Institute, University of Copenhagen, Jagtvej 128, 2200 Copenhagen, Denmark\label{aff104}
\and
Universidad Polit\'ecnica de Cartagena, Departamento de Electr\'onica y Tecnolog\'ia de Computadoras,  Plaza del Hospital 1, 30202 Cartagena, Spain\label{aff105}
\and
Kapteyn Astronomical Institute, University of Groningen, PO Box 800, 9700 AV Groningen, The Netherlands\label{aff106}
\and
Infrared Processing and Analysis Center, California Institute of Technology, Pasadena, CA 91125, USA\label{aff107}
\and
INAF, Istituto di Radioastronomia, Via Piero Gobetti 101, 40129 Bologna, Italy\label{aff108}
\and
Institut d'Astrophysique de Paris, 98bis Boulevard Arago, 75014, Paris, France\label{aff109}
\and
ICL, Junia, Universit\'e Catholique de Lille, LITL, 59000 Lille, France\label{aff110}
\and
ICSC - Centro Nazionale di Ricerca in High Performance Computing, Big Data e Quantum Computing, Via Magnanelli 2, Bologna, Italy\label{aff111}}    

\date{Received June 23rd, 2025; accepted October 20th, 2025}

% Abstract of the paper
\abstract
{
The \Euclid mission and other next-generation large-scale structure surveys will enable high-precision measurements of the cosmic matter distribution. Understanding the impact of baryonic processes such as star formation and active galactic nuclei (AGN) feedback on matter clustering is crucial to ensure precise and unbiased cosmological inference. Most theoretical models of baryonic effects to date focus on two-point statistics, neglecting higher-order contributions. This work develops a fast and accurate emulator for baryonic effects on the matter bispectrum, a key non-Gaussian statistic in the nonlinear regime. We employ high-resolution $N$-body simulations from the BACCO suite and apply a combination of cutting-edge techniques such as cosmology scaling and baryonification to efficiently span a large cosmological and astrophysical parameter space. A deep neural network is trained to emulate baryonic effects on the matter bispectrum measured in simulations, capturing modifications across various scales and redshifts relevant to \Euclid. We validate the emulator accuracy and robustness using an analysis of \Euclid mock data, employing predictions from the state-of-the-art FLAMINGO hydrodynamical simulations. The emulator reproduces baryonic suppression in the bispectrum to better than 2$\%$ for the $68\%$ percentile across most triangle configurations for $k \in [0.01, 20]\,\ihMpc$  and ensures consistency between cosmological posteriors inferred from second- and third-order weak lensing statistics. These results demonstrate that our emulator meets the high-precision requirements of the \Euclid mission for at least the first data release and provides reliable forecasts of the cosmological information contained in the small-scale matter bispectrum. This underscores the potential of emulation techniques to bridge the gap between complex baryonic physics and observational data, maximising the scientific output of \Euclid.
}

%%%%%%%%%%%%%%%%%%%%%%%%%%%%%%%%%%%%%%%%%%%%%%%%%

\titlerunning{}
\authorrunning{P.A. Burger et al.} 

\maketitle

%%%%%%%%%%%%%%%%% BODY OF PAPER %%%%%%%%%%%%%%%%%%

\section{Introduction}
\label{sec:intro}

The European Space Agency mission \Euclid \citep{EuclidSkyOverview} is set to revolutionise our understanding of the large-scale structure (LSS) of the Universe by providing high-precision measurements of cosmic shear, galaxy clustering, and the distribution of matter on cosmological scales \citep[e.g.][]{heymans2021,DES2021,2023PhRvD.108l3517M}. These data will enable tight constraints on key cosmological parameters, such as the equation of state of dark energy, the sum of neutrino masses, and the amplitude of primordial density fluctuations \citep{EuclidSkyOverview}.

The classical cosmic shear analysis involves measuring the correlations between galaxy ellipticities as a function of their separation, quantified by the two-point correlation function of the shear \cite[$\gamma$-2PCF, e.g.][]{Asgari:2020,2022PhRvD.105b3514A,HSC2023}. This statistic benefits from a well-established theoretical framework, 
 primarily capturing the multiscale variance of the lensing field. However, gravitational collapse introduces non-Gaussian features into the shear field, which are not fully captured by $\gamma$-2PCF or other second-order statistics.

As a result, two-point statistics cannot extract all cosmological information, leading to parameter degeneracies, such as the one between the matter density parameter $\Omega_\mathrm{m}$ and normalisation of the power spectrum $\sigma_8$, which limit constraints to combinations such as $S_8 = \sigma_8 \sqrt{\Omega_\mathrm{m}/0.3}$. Various higher-order statistics (HOS) have been proposed in the literature to access additional information \citep[see e.g.][]{Ajani-EP29}.

A compelling HOS for probing non-Gaussian features of the LSS is the matter bispectrum \citep[e.g.][]{2003A&A...397..405B,2003MNRAS.344..857T}, which captures three-point correlations in the density field and provides complementary information to the widely studied two-point statistics, such as the power spectrum \citep{Halder:2021,Halder:2022,Heydenreich2023,Burger2024a,2025arXiv250303964G}.

One of the biggest challenges of HOS is the precise modelling of intermediate to small scales, as HOS extract most additional information on nonlinear scales \citep{2020MNRAS.498.2887F}. However, these small scales are heavily affected by baryonic feedback processes, such as gas cooling, star formation, and feedback from active galactic nuclei (AGN), which modify the distribution of matter \citep{vanDaalen2011, 2018MNRAS.480.3962C}. These processes significantly affect the matter distribution, especially in the nonlinear regime, and must be accounted for in cosmological analyses to avoid biased constraints on fundamental physics \citep{2011MNRAS.417.2020S, 2019JCAP...03..020S}. 

\cite{2013MNRAS.434..148S} used a halo-model approach to predict the effect of baryons on second-order shear statistics and a perturbation theory approach to predict the effect of baryons on the three-point shear statistics. By applying their method to the OverWhelmingly Large Simulations \citep{2010MNRAS.402.1536S}, they found that, similar to second-order statistics, the effect of baryonic processes on third-order statistics is not negligible. Moreover, the two statistics have different sensitivities and dependences on the baryonic parameters. Therefore, the combination of second- and third-order statistics has the potential to self-calibrate baryonic feedback parameters, thus improving constraining power on cosmological parameters and reducing projection effects that arise when marginalising high-dimensional posterior distributions over subsets of parameters, which lead to shifts in one- or two-dimensional marginal posteriors that do not reflect the true structure of the full joint distribution.

Hydrodynamical simulations \citep[e.g.][]{2014MNRAS.441.1270L,2015A&C....13...12N,2015MNRAS.446..521S,2018MNRAS.476.2999M,Schaye2023,2024MNRAS.528.2308P} offer the most detailed insight into baryonic processes and their effects on the power spectrum \citep{Schaller2024_emu}. However, their computational expense makes exploring the high-dimensional parameter space necessary for future cosmological surveys impractical. A promising method for incorporating baryonic effects involves modifying Gravity-Only (GrO) simulations to mimic the influence of baryons using physically motivated models. These approaches, often referred to as Baryon Correction Models (BCM) or baryonification techniques \citep{2015JCAP...12..049S,2019JCAP...03..020S,2020MNRAS.495.4800A}, are computationally efficient and enable the generation of extensive training sets. Such datasets are instrumental in building fast emulators that predict baryonic suppression as a function of cosmological parameters and a few physically motivated baryonic parameters \citep{2021MNRAS.506.4070A,2021JCAP...12..046G}. Using these tools, the baryonic suppression effects observed in various hydrodynamical simulations can be replicated with percent-level accuracy down to scales of $k = 5\mathrm{-}10 \,\ihMpc$. Although much progress has been made in emulating baryonic effects on the matter power spectrum, the bispectrum or higher-order statistics have remained comparatively underexplored \citep[see e.g.][]{2021MNRAS.503.3596A,Anbajagane2024, Zhou2025}. Given the additional cosmological information contained in the bispectrum and its sensitivity to non-Gaussianities, we develop here, as an essential extension, an emulator that can accurately capture baryonic effects on the bispectrum and is fast enough to be used in Bayesian pipelines. 

We do this by expanding the work of \cite{2021MNRAS.506.4070A} on the matter power spectrum and present the first emulator predicting the impact of baryonic processes on the matter bispectrum. 
We model the relevant baryonic processes employing high-resolution simulations from the BACCO suite \citep{2021MNRAS.507.5869A}, post-processed using the baryonification technique developed in \cite{2020MNRAS.495.4800A} and the cosmology scaling of \cite{2010MNRAS.405..143A}. We train deep neural networks to predict the power spectrum and bispectrum of our simulations. Although an emulator of baryonic effects on the power spectrum was already presented in \cite{2021MNRAS.506.4070A}, we expand their work to have consistent modelling for the power and bispectrum with a broader parameter space and a scale that extends to larger $k$-values. We validate these emulators against the FLAMINGO set of high-resolution hydrodynamical simulations \citep{Schaye2023,Kugel2023}, each incorporating different physical prescriptions for feedback mechanisms, star formation, and gas dynamics. Training on various scales and redshifts ensures that our emulators cover the relevant parameter space for the \Euclid mission.

Specifically, we show that our BCM emulator can accurately reproduce the FLAMINGO baryonic effects on the second- and third-order moments of the aperture mass lensing statistic \citep{Jarvis:2004,Schneider:2005} across different configurations of box sizes, particle masses, and strengths of AGN feedback. Furthermore, we combine second- and third-order weak lensing statistics to forecast cosmological constraints under the impact of baryonic effects for a non-tomographic analysis of the \Euclid first data release (DR1), quantifying the importance of accounting for baryonic effects in the power spectrum and bispectrum measurements. Our results demonstrate that ignoring baryonic effects could lead to significant biases in interpreting \Euclid data or require scale cuts, leading to a severe reduction in constraining power, highlighting the necessity of incorporating these effects into cosmological analyses. 

This paper is organised as follows. In Sect.~\ref{sec:sim}, we describe the suites of simulations used for training, validating, and building the emulator. We also describe the GrO simulations used for estimating the \Euclid-DR1 covariance matrix for the weak lensing statistics. Section \ref{sec:bispec} presents the estimation of the bispectrum and its validation. In Sect.~\ref{sec:train}, we present the training of the BCM emulator of the power spectrum and bispectrum. In Sect.~\ref{sec:theory}, we present the methodology for converting the spectra to weak lensing statistics, which we then use in Sect.~\ref{sec:forecasts} to perform the non-tomographic \Euclid-DR1 forecast. Finally, we discuss the implications of our findings and potential future extensions in Sect.~\ref{sec:conclusions}.

\section{Simulations}
\label{sec:sim}

\subsection{BACCO simulations}
\label{sec:bacco_sim}
This section describes the BACCO $N$-body simulations used in this study to build a BCM and quantify the baryonic effects on the matter power and bispectrum. The BACCO simulation suite is specifically designed to provide highly accurate predictions of nonlinear cosmic structure formation as a function of cosmology. The gravitational evolution is computed using \texttt{L-Gadget3} \citep{Springel2005,2012MNRAS.426.2046A,2021MNRAS.507.5869A}, a variant of the Gadget code, with a Plummer-equivalent softening length of $\epsilon = 5\,h^{-1} \mathrm{kpc}$. The numerical parameters of the suite, including force and mass resolution, were chosen to achieve convergence at the 1$\%$ level in the nonlinear power spectrum at $z=0$ and $k \sim 10 \,\ihMpc$. This setup has been validated against other $N$-body codes, performing exceptionally well in the \Euclid Code Comparison project \citep{2016JCAP...04..047S}, where it agreed within 2$\%$ with most codes up to $k \sim 10 \,\ihMpc$ \citep{2021MNRAS.507.5869A}. Each simulation explores a distinct cosmology, with parameters carefully selected to maximise the coverage of the parameter space when combined with rescaling algorithms \citep{Contreras2020,2021MNRAS.507.5869A}.

In this work, we employ a single gravity-only simulation, a higher-resolution version of TheOne simulation presented in \cite{2021MNRAS.506.4070A}. This simulation follows the evolution of $2288^3$ particles of mass $m_\mathrm{p} \approx 9.55 \times 10^8 \Msun$ within a periodic box of $L = 512 \, h^{-1} \mathrm{Mpc}$ on a side. We report its cosmology in Table \ref{tab:D3A}. The initial conditions were generated at $z=49$ using second-order Lagrangian perturbation theory (2LPT). To minimise cosmic variance, the amplitudes of Fourier modes were fixed to the ensemble average of the linear power spectrum, and we average the summary statistics of two realisations having opposite phases, following the fix and pair approach described by \cite{2016MNRAS.462L...1A}. The particle catalogue was downsampled by a factor of $4^3$ for computational and memory efficiency. We refer to \cite{2021MNRAS.503.3596A} for detailed convergence tests of resolution and cosmic variance, which show that baryonic effects on the power spectrum and bispectrum produced with the BCM on top of this simulation are converged at the percent level.

\begin{table}
    \renewcommand{\arraystretch}{1.2}
    % \centering
    \begin{threeparttable}
    \caption{Cosmological parameters of the simulations used in this work.}
    \begin{tabular}{ l | l| l| l| l| l }
         & $h$ & $\Omega_\mathrm{m}$ & $\Omega_\mathrm{b}$ & $\sigma_8$ & $n_\mathrm{s}$ \\
        \hline
        TheOne & 0.680 & 0.307 & 0.0480 & 0.900 & 0.960 \\  
        D3A & 0.681 & 0.306 & 0.0486 & 0.807 & 0.967 \\
        LS8 & 0.682 & 0.305 & 0.0473 & 0.760 & 0.965 \\
        \Planck & 0.673 & 0.316 & 0.0494 & 0.812 & 0.966 \\
        WMAP9 & 0.700 & 0.279 & 0.0460 & 0.820 & 0.970 
    \end{tabular}
        \tablefoot{The table reports the cosmologies of the simulations used in this work. “TheOne” is from the BACCO simulation suite. “D3A” corresponds to the FLAMINGO simulation run with the DES Y3 “$3\times2$pt+ All Ext.” $\Lambda$CDM cosmology \citep{DES2021}. “LS8” refers to a FLAMINGO run with low $\sigma_8$, and “\Planck” uses the \citet{planck2020} cosmology. The Takahashi simulations adopt a cosmology approximately consistent with WMAP9 \citep{Hinshaw2013}. The parameters shown are: the dimensionless Hubble parameter $h$, the matter and baryon density parameters $\Omega_\mathrm{m}$ and $\Omega_\mathrm{b}$, the amplitude of linear fluctuations $\sigma_8$, and the spectral index $n_\mathrm{s}$.}
    \label{tab:D3A}
    \end{threeparttable}
\end{table}

\subsubsection{Cosmology rescaling}
We explore the cosmological space with the cosmology rescaling algorithm originally proposed in \cite{2010MNRAS.405..143A}, and further refined by \cite{Zennaro2019} and \cite{Contreras2020}. This method computes and applies optimal space and time scale factors, which minimise the linear mass variance between two different cosmologies. Furthermore, after applying these scale factors, one can also correct the large-scale flow motions and the small-scale internal structure of haloes. This algorithm can accurately reproduce the matter and halo clustering in a wide cosmology parameter space, including massive neutrinos and dynamical dark energy, with just a few simulations and in only a few seconds. In this work, we only vary the cosmological parameters $\Omega_\mathrm{m}$, $\Omega_\mathrm{b}$, $\sigma_8$, that have been shown in several works to be the dominant cosmological dependence of baryonic effects, at the level of the matter power spectrum \citep{2019JCAP...03..020S,2021MNRAS.506.4070A}. We tested in Appendix \ref{app:BCMdependence} with the BACCO simulation suites that this statement is also valid for the matter bispectrum. 

\subsubsection{Baryonic correction model}
\label{sec:BCM}
To incorporate the effects of baryonic physics into the mass distribution in GrO simulations, we use the `baryonification' framework \citep{2015JCAP...12..049S,2019JCAP...03..020S,2020MNRAS.495.4800A}. This method applies physically motivated prescriptions for processes such as star formation, gas cooling, and AGN feedback to perturb the positions of particles in gravity-only $N$-body simulations, thereby modifying the mass distribution.

One key advantage of the baryonification approach is its flexibility in exploring possible modifications to the matter power and bispectrum, accounting for the uncertainties inherent in baryonic physics. Model parameters can be constrained using observational data and tested with hydrodynamical simulations. In this work, we adopt the implementation detailed in \cite{2021MNRAS.503.3596A}, which has been successfully tested along with the cosmology rescaling algorithms in simultaneously reproducing the power spectra and bispectra of different hydrodynamical simulations. 
The model considers several baryonic components. 
The first component is the bound gas in haloes, whose density is described by a double power law with the transition radius and slopes being free parameters. In this work, we adopt only one free parameter for the bound gas shape, $\theta_\mathrm{inn}$, which represents the inner characteristic scale where the gas slope varies. The outer characteristic scale, $\theta_\mathrm{out}$, was fixed to one, which we verify in Appendix \ref{app:BCMdependence}. Both characteristic scales are reported in units of the halo radius, $r_{200}$, which in turn is defined as the radius where the halo's density exceeds the Universe's critical density by a factor of 200. The bound gas mass fraction, $f_\mathrm{BG}$, is parametrised as
\begin{equation}
        f_\mathrm{BG}(M_{200}) = \frac{f_\mathrm{b}-f_\mathrm{CG}-f_\mathrm{SG}}{1 - (M_\mathrm{c}/M_{200})^\beta},
\end{equation}
where $f_\mathrm{b} = \Omega_\mathrm{b}/\Omega_\mathrm{m}$ is the cosmic baryon fraction, $M_{200}$ is the mass inside radius $r_{200}$, $M_\mathrm{c}$ and $\beta$ are free parameters, and $f_\mathrm{CG,SG}$ are the central and satellite galaxy stellar mass fractions, respectively. An abundance-matching parametrisation gives the central and satellite galaxy fractions \citep{2013ApJ...770...57B}, which have the same parametric form in halo mass and redshifts and their parameters are assumed to be linearly dependent. 

In this work, we follow \cite{2021MNRAS.506.4070A} and free only one parameter for the stellar component, $M_{1,z0,\mathrm{cen}}$. This parameter represents the characteristic halo mass for which the central galaxy to halo mass fraction is set to $\epsilon_{0,\mathrm{cen}}=0.023$ at $z=0$.

A fraction of the gas is assumed to be ejected by feedback, with a constant density and an exponential cutoff at a given scale, parametrised by a free parameter $\eta$. 
The sum of stellar and gas mass fractions, including the ejected gas, is by construction equivalent to the cosmic baryon fraction $f_\mathrm{b}$. We report the priors on the varied BCM parameters in Table \ref{tab:BCM_prior}. Moreover, we fix to a fiducial value the BCM parameters $M_{\mathrm{inn}}=2.3\times 10^{13} \, \Msun$, which are verified in Appendix \ref{app:BCMdependence}. Finally, we vary the scale factor $a=1/(1+z)$ to account for the redshift evolution of baryonic effects. 
We note that in the following, we write $M_1$ instead of $M_{1,z0,\mathrm{cen}}$.

\begin{table}
    \renewcommand{\arraystretch}{1.3}
    \centering
    \caption{Priors of the varied BCM parameters.}
    \begin{tabular}{l|c}
    Parameter & Prior Range \\
        \hline
        $\log_{10} (M_\mathrm{c} / h^{-1} \mathrm{M}_\odot)$ & [10, 16] \\
        $\log_{10} (M_1 / h^{-1} \mathrm{M}_\odot)$          & [9, 13] \\
        $\log_{10} (\beta)$                                  & [$-1.0$, 0.7] \\
        $\log_{10} (\eta)$                                   & [$-0.7$, 0.2] \\
        $\log_{10} (\theta_\mathrm{inn})$                    & [$-2.0$, 0.0] \\
        $\Omega_\mathrm{m}$                             & [0.23, 0.40] \\
        $\Omega_\mathrm{b}$                             & [0.04, 0.06] \\
        $\sigma_8$                                      & [0.73, 0.90] \\
        $a$                                             & [0.24, 1.01] \\
    \end{tabular}
    \tablefoot{Units are shown inside the logarithmic expressions where applicable; all other parameters are dimensionless. $a$ refers to the scale factor.}
    \label{tab:BCM_prior}
\end{table}

\subsection{FLAMINGO simulations}
\label{sec:Flamingo_sim}
To test our baryon model, we use the FLAMINGO simulations,\footnote{\url{https://flamingo.strw.leidenuniv.nl/}} in particular, we use the convergence maps developed and described in \cite{Jeger2024}. A detailed description of the FLAMINGO simulations, their performance, and calibration strategy can be found in \cite{Schaye2023}, \cite{Kugel2023}, and \cite{McCarthy2023}.

The simulations were conducted using the SWIFT hydrodynamic code \citep{Schaller2024} with the SPHENIX smoothed particle hydrodynamics (SPH) implementation \citep{Borrow2022}. Neutrinos were modelled as massive particles using the method developed by \cite{Elbers2021}, which minimises particle shot noise. The simulations incorporate radiative cooling and heating on an element-by-element basis \citep{Ploeckinger2020}, star formation following the prescription of \cite{Schaye2008}, and time-dependent stellar mass loss as outlined by \cite{Wiersma2009}. Supernova and stellar feedback are implemented kinetically, with neighbouring particles receiving kicks that conserve energy, linear momentum, and angular momentum, as described in \cite{Chaikin2023}. Gas accretion onto supermassive black holes and subsequent thermal AGN feedback are modelled based on \cite{Booth2009}, while kinetic jet feedback follows the AGN jet implementation of \cite{Husko2022}, where gas particles are accelerated to a fixed jet velocity aligned with the black hole’s spin.

In this work, we use the box with 2.8 comoving Gpc side length, $5040^3$ dark matter (DM) and baryon particles, and $2800^3$ massive neutrino particles, where the baryonic particle mass is $1.07 \times 10^9\,\mathrm{M}_\odot$ and the DM particle mass is $5.65\times 10^9 \,\mathrm{M}_\odot$. We use all eight available light cones with different observers. The cosmology of these runs was taken from the Dark Energy Survey year three “$3\times2$pt+ All Ext.” $\Lambda$CDM cosmology \citep{DES2021}, which we denote as ‘D3A’. Complementary to this run, we use all nine baryon descriptions available for the D3A cosmology given in Table \ref{tab:D3A}. These additional runs have the same resolution, but with a box size of 1 Gpc, $1800^3$ baryonic particles, and $1000^3$ massive neutrino particles. We also use the higher and lower mass resolution for the fiducial baryon feedback description (L1$\_$m8 and L1$\_$m10). In the following, we label the gravity-only simulations as `GrO' and those including baryonic feedback as `Hydro'.

\begin{table}
    \renewcommand{\arraystretch}{1.2}
    \centering
    \caption{Baryon descriptions of the FLAMINGO simulations.}
    \begin{tabular}{ l | l | l | l | l }
        Identifier & $\Delta M^*$ & $\Delta f_\mathrm{gas}$ & AGN mode & $N_\mathrm{b}$ \\
        \hline
        L2p8$\_$m9 & \phantom{$-$}0 & \phantom{$-$}0 & thermal & $5040^3$ \\
        L1$\_$m9 & \phantom{$-$}0 & \phantom{$-$}0 & thermal & $1800^3$ \\
        L1$\_$m10 & \phantom{$-$}0 & \phantom{$-$}0 & thermal & $900^3$\\
        L1$\_$m8 & \phantom{$-$}0 & \phantom{$-$}0 & thermal & $3600^3$\\
        fgas$+2\sigma$ & \phantom{$-$}0 & $+$2 & thermal & $1800^3$\\
        fgas$-2\sigma$ & \phantom{$-$}0 & $-$2 & thermal & $1800^3$\\
        fgas$-4\sigma$ & \phantom{$-$}0 & $-$4 & thermal & $1800^3$\\
        fgas$-8\sigma$ & \phantom{$-$}0 & $-$8 & thermal & $1800^3$\\
        M$^*$--$\sigma$ & $-$1 & \phantom{$-$}0 & thermal & $1800^3$\\ 
        M$^*$--$\sigma\_$fgas--4$\sigma$ & $-$1 & $-$4 & thermal & $1800^3$\\ 
        Jet & \phantom{$-$}0 & \phantom{$-$}0 & jet & $1800^3$\\ 
        Jet$\_$fgas--4$\sigma$ & \phantom{$-$}0 & $-$4 & jet & $1800^3$\\ 
    \end{tabular}
    \tablefoot{The cosmology of these simulations is the D3A cosmology given in Table \ref{tab:D3A}.}
    \label{tab:baryon_variation_table}
\end{table}

At the D3A cosmology, eight runs were calibrated to different galaxy stellar mass functions ($M_\star$) and/or gas fractions in clusters ($f_\mathrm{gas}$), or they were varied in their overall AGN subgrid feedback prescriptions. The values of the subgrid parameters were adjusted so that the stellar mass function and/or the gas fractions in clusters deviated by specific standard deviations ($\Delta M_\star$ and $\Delta f_\mathrm{gas}$) from the fiducial model, as shown in the second and third columns of Table \ref{tab:baryon_variation_table}.

Four of these eight runs focused exclusively on varying gas fractions in clusters. These are identified as fgas$\pm$ $n\sigma$, where $n$ represents the number of standard deviations by which the gas fraction is modified. One additional run was calibrated to match a stellar mass function shifted to a lower mass by $1\sigma$ (M$^*$--$\sigma)$. Another run, labelled M$^*$--$\sigma\_$fgas--4$\sigma$, incorporated variations in both observables. The fiducial AGN feedback model uses a thermal implementation, but two alternative models employ a kinetic jet feedback mechanism. These are denoted as Jet and Jet$\_$fgas--4$\sigma$, with the latter targeting reduced gas fractions in clusters. The jet models provide a way to evaluate the sensitivity of observables to variations in subgrid models calibrated to the same data. Lastly, we also make use of a simulation at a lower $\sigma_8$ than the D3A, which is reported in Table \ref{tab:D3A}, denoted as LS8, and uses the baryon description of the fiducial case L1$\_$m9.

\subsubsection{Weak lensing convergence maps}

The weak lensing convergence maps $\kappa(\pmb{\vartheta},\chi)$ were created using ray-tracing for each comoving slice, where $\pmb{\vartheta}$ refers to the sky coordinates and $\chi$ to the comoving distance \citep{Jeger2024}. A key quantity of gravitational lensing is the magnification or Jacobian matrix defined as
\begin{equation}
    A_{ij}(\pmb{\vartheta},\chi) =  
\frac{\partial \beta_i}{\partial \vartheta_j}
\;=\;
\delta_{\mathrm{K},ij}-\frac{\partial^{2}\phi}{\partial\theta_i\,\partial\theta_j}
\;\approx
\begin{pmatrix}
  1-\phi_{,11} & -\phi_{,12} \\
  -\phi_{,12} & 1-\phi_{,22}
\end{pmatrix} \, ,
\end{equation}
where $\delta_{\mathrm{K}}$ is the Kronecker delta symbol, $_{,ij} = \partial/\partial\theta_i\,\partial/\partial\theta_j$, $\pmb{\beta}$ denotes the original position of the ray-traced light ray, and can be computed as an integral along the comoving distance over the gradient of the deflection potential $\phi$ \citep[see][]{Bartelmann:2001}. Given the fact that $\nabla^2 \phi = 2 \kappa$, we can also express the magnification as 
\begin{equation}
    A(\pmb{\vartheta},\chi) =  
\begin{pmatrix}
  1-\kappa - \gamma_1 & -\gamma_2 \\
  -\gamma_2 & 1-\kappa + \gamma_1
\end{pmatrix} \, ,
\end{equation}
where 
\begin{equation}
    \gamma_1(\pmb{\vartheta},\chi) = \frac{1}{2}[\phi_{,11} - \phi_{,22}](\pmb{\vartheta},\chi) \quad \& \quad  \gamma_2(\pmb{\vartheta},\chi) = \phi_{,12}(\pmb{\vartheta},\chi) \, .
\end{equation}

Based on the magnification matrix derived from density maps in {\code HEALPix} format \citep{Gorskietal2005} with an $N_\mathrm{side}= 8192$, which corresponds to a pixel size of $\ang{;0.43;}$, the shear signal $\kappa(\pmb{\vartheta},\chi)$ and $\gamma(\pmb{\vartheta},\chi)$ are computed at each pixel location. Next, these convergence maps are combined using the Limber integration \citep{Limber:1954}
\begin{equation}
    \kappa(\pmb{\vartheta}) = \int_0^{\chi_\mathrm{max}} \dd \chi\,  n[z(\chi)]\, \frac{\dd z}{\dd \chi} \, \kappa(\pmb{\vartheta},\chi), 
\end{equation}
where $n(z)$ is the redshift probability distribution of source galaxies given in \cite{Blanchard-EP7} as
\begin{equation}
    n(z) \propto \left(\frac{z}{z_0}\right) \exp\left[-\left(\frac{z}{z_0} \right)^{3/2}\right]
    \label{eq:Euclid_nz}
\end{equation}
with $z_0 = 0.9/\sqrt{2}$. This redshift distribution is cut at $z_{\rm max}= 3$, so that $\chi_\mathrm{max}=\chi(z_{\rm max})$ and it is normalised.

\subsection{Takahashi simulations}
\label{sec:T17_description}

As in \cite{Burger2023}, we use the \cite{Takahashi2017} simulations to estimate a \Euclid-like covariance matrix for our lensing statistics. 
The simulations by \citet{Takahashi2017} track the nonlinear evolution of $2048^3$ particles within a series of nested cosmological volumes. These volumes start with a side length of $L=450\,h^{-1}\mathrm{Mpc}$ at low redshifts and grow progressively larger at higher redshifts, resulting in 108 full-sky realizations. These simulations were generated using the Gadget-3 $N$-body code \citep{Springel2005} and are publicly available.\footnote{\url{http://cosmo.phys.hirosaki-u.ac.jp/takahasi/allsky_raytracing/}} The cosmological parameters are set to values consistent with WMAP 9 \citep{Hinshaw2013} and reported in Table \ref{tab:D3A}. Lensing information from these simulations is provided using the Born approximation as 108 full-sky realisations of the convergence and shear, each subdivided into 38 redshift slices in ascending order. To construct a non-tomographic \Euclid-DR1-like covariance matrix, we built the weighted average of the first 30 $\kappa$ maps with ascending redshift. The weight for each redshift slice is measured by integrating the $n(z)$ given in Eq.~\eqref{eq:Euclid_nz} over the corresponding width of the redshift slice.  
To have a footprint of roughly 2000\,deg$^2$, we divided the 30 full-sky maps into 19 patches, each with a size of roughly 2146\,deg$^2$. Shape noise is added to the convergence maps by drawing random numbers from a Gaussian distribution with a vanishing mean and a standard deviation, as follows:
\begin{equation}
\label{eq:shapenoise}
    \sigma=\frac{\sigma_\epsilon}{\sqrt{n_\mathrm{eff}\, A_\mathrm{pix}}}\;, 
\end{equation}
with the pixel area as $A_\mathrm{pix}=0.18\,\mathrm{arcmin}^2$ $(N_\mathrm{side}=8192)$, $n_\mathrm{eff} = 30\,\mathrm{arcmin}^{-2}$ as the effective galaxy number density, and shape noise contribution $\sigma_\epsilon=0.3$ as stated in \citet{Blanchard-EP7}.

\section{Estimation of the bispectrum}
\label{sec:bispec}
We measure the bispectra and power spectra employing the estimator by \citet{Scoccimarroe:2000}, which defines the bispectrum at $k$-bins $\vec{k} = \{k_1, k_2, k_3\}$ for a simulation of volume $V$ as
\begin{equation}
    B(k_1, k_2, k_3) = \frac{\prod_{i=1}^3 \int_{\Delta k_i} \dd^3 {q_i}\,\hat{\delta}(\vec{q}_i)\,\diracd(\vec{q}_1+\vec{q}_2+\vec{q}_3)}{\prod_{i=1}^3 \int_{\Delta k_i} \dd^3 {q_i}\,\diracd(\vec{q}_1+\vec{q}_2+\vec{q}_3)}\;,
\end{equation}
where $\hat{\delta}$ is the Fourier transform of $\delta$, $\diracd$ is the Dirac delta distribution, and $\Delta k_i$ are the bins of wavenumber $k_i$. The power spectrum is defined by
\begin{equation}
    P(k) = \frac{ \int_{\Delta k}\dd^3{q}\,|\hat{\delta}(\vec{q})|^2}{ \int_{\Delta k}\dd^3{q}}\;.
\end{equation}
 We further use 
\begin{equation}
    \diracd(\vec{q}_1+\vec{q}_2+\vec{q}_3) = \int_{L_\mathrm{box}} \frac{\dd^3{x}}{(2\pi)^3} \mathrm{e}^{-\mathrm{i}\vec{x}\cdot(\vec{q}_1+\vec{q}_2+\vec{q}_3) }\;,
\end{equation}
so the $q$-integrals become separable, and we obtain
\begin{equation}
     B(k_1, k_2, k_3) = \frac{\int_{L_\mathrm{box}} \frac{\dd^3{x}}{(2\pi)^3} \prod_{i=1}^3 I_{k_i}^{\delta}(\vec{x}) }{\int_{L_\mathrm{box}} \frac{\dd^3{x}}{(2\pi)^3} \prod_{i=1}^3 I^{\vec{1}}_{k_i}(\vec{x}) }\;,
\end{equation}
where $L_\mathrm{box}$ is the length of the simulation box.
\begin{align}
    I_{k}^\delta(\vec{x}) = \int_{\Delta k} \dd^3 q\, \hat{\delta}(\vec{q})\, \mathrm{e}^{-\mathrm{i}\vec{q}\cdot\vec{x}}\quad \mathrm{and}  \quad  I_{k}^{\vec{1}}(\vec{x}) = \int_{\Delta k} \dd^3 q\, \mathrm{e}^{-\mathrm{i}\vec{q}\cdot\vec{x}}\;.
\end{align}
The $I_{k}^\delta(\vec{x})$ can be quickly obtained using Fast Fourier Transforms (FFT).
To make feasible the estimation of thousands of spectra using the large grids required for this project, we produced a new implementation of these FFTs that runs on GPUs using the JAX framework,  which provides a speed-up of a factor up to 10 with respect to CPUs codes. We publicly released the code BiG \citep{BiG}, and show in Appendix~\ref{app:accuracyBiG} how its estimates for the bispectrum agree with those of CPUs codes such as \texttt{bskit} \citep{2020MNRAS.498.2887F} to better than $0.01\%$.

This estimator averages over the number of triangle configurations inside a $(k_1, k_2, k_3)$ bin. However, as we are interested in a bispectrum prediction for a specific configuration, we need to define which `effective' triangle configuration any measured bispectrum value corresponds to. Following \citet{Oddo:2020}, we assign each bin an effective triangle configuration $(k_{1,\mathrm{eff}}, k_{2,\mathrm{eff}}, k_{3,\mathrm{eff}})$ as 
\begin{equation}
        k_{i,\mathrm{eff}}(k_1, k_2, k_3) = \frac{\prod_{j=1}^3 \int_{\Delta k_j}\dd^3{q_j}\, {q}_i\,\diracd(\vec{q}_1+\vec{q}_2+\vec{q}_3)}{\prod_{i=1}^3 \int_{\Delta k_i}\dd^3{q_i}\,\diracd(\vec{q}_1+\vec{q}_2+\vec{q}_3)}\;.
\end{equation}
We use these effective $k$s when creating the emulator in the following sections and, for simplicity, drop the `eff' subscript.

For a given density field with $N^3$ grid points and box length $L_\mathrm{box}$, the theoretically accessible $k$ modes are $k \in [2k_\mathrm{ny}, N k_\mathrm{ny}]$, where $k_\mathrm{ny} = \pi/L_\mathrm{box}$ is the Nyquist frequency. In our case, we use $N=858$, which leads to $k_\mathrm{max}=5.3\,\ihMpc$. 

To access smaller scales (larger $k$) without increasing the number of grid points, we use the folding technique presented in \cite{Jenkins1998} and \cite{Colombi:2009} and validated for the bispectrum by \cite{2021MNRAS.503.3596A}. The particle distribution is `folded' by wrapping particles into a sub-box of side length $L/f$. Given the same grid and the $k_\mathrm{ny}$ defined in the full box, the new Nyquist frequency is $k_\mathrm{ny}' = f \, k_\mathrm{ny}$. In this work, we deploy three foldings $f=[1,2,4]$, such that the lowest $k_\mathrm{min}$ is $0.0123 \, \ihMpc$ and the largest $k_\mathrm{max}$ is $21.058 \, \ihMpc$. While combining the three foldings for the power spectrum is straightforward, it is more complicated for the bispectrum. The reason is the missing squeezed triangles, where one of the $k_i$ values can only be measured in one folding, while the other $k_i$ values are only measurable in the other. In the following, we will allow the emulator to extrapolate to these missing squeezed triangles and validate this on the lensing analysis below, since we currently cannot measure bispectra where all $k$-configurations fit in a single grid. We also notice that we use the symmetries of the bispectrum $B(k_1,k_2,k_3) = B(k_2,k_3,k_1) = B(k_3,k_2,k_1) = B(k_3,k_1,k_2) = B(k_1,k_3,k_2) = B(k_2,k_1,k_3)$ and that the three sides need to build a triangle, $k_3 \leq k_1 + k_2$, which reduces the required measurements. We average the measurements from two opposite-phase realizations to suppress cosmic variance for both power spectra and bispectra. We also subtract the shot noise contribution, which is given as $\bar{n}^{-1}=L_\mathrm{box}^3/N_\mathrm{par}$ for the power spectrum and $B_\mathrm{sn}(k_1,k_2,k_3) = \bar{n}^{-2} + \bar{n}^{-1} [P(k_1)+P(k_2)+P(k_3)]$ for the bispectrum \citep{Peebles1980}. Lastly, we note that we do not deconvolve the window function introduced by the mass assignment scheme, but we checked that the effect is smaller than $0.02\%$ when taking ratios of power spectra and bispectra as in our case.

\section{Training and validation of the emulator}
\label{sec:train}
After measuring the power and bispectra for the GrO and baryonified cases for 1200 nodes distributed in a Latin Hypercube defined by the prior mentioned in Sect.~\ref{sec:sim}, we are ready to build the BCM of the matter power spectrum, $S(k) = P^\mathrm{Hydro}/P^\mathrm{GrO}$, and bispectra, $R(\vec{k}) = B^\mathrm{Hydro}/B^\mathrm{GrO}$. For our emulation, we use the \texttt{CosmoPower} neural network (NN) library described in \cite{COSMOPOWER2022}. In principle, we could use only the six baryon parameters as input and then interpolate between different $k$-values. However, we found that it is more convenient, accurate and faster to use also the $k$-value as input parameters, as this allows us to predict the power and bispectra for any $k$-values even if they were not measured in the first place. Moreover, in this way, we can clean our training set more efficiently, removing the single $B(k_1,k_2,k_3)$ configurations where $B(k_1,k_2,k_3) < 2\,B_\mathrm{sn}(k_1,k_2,k_3)$ without having to exclude the whole sampling node. With this procedure, we discard only about $2.5\%$ of the total measurements. We note that we emulate the three-dimensional power spectrum and bispectrum, as opposed to the projected two-dimensional lensing statistics presented in the next sections, to have the flexibility to employ the emulators with different survey setups and projections, without the need to perform the training of more emulators. To further speed up the cosmological analyses, specific emulators for a given survey configuration can be easily developed starting from the ones presented here. 

We randomly split our samples of 1200 nodes into 200 points used for testing and 1000 for training. After some iterations, we found that an architecture with four hidden layers and 32 neurons, the Adam optimizer, a mean squared difference loss function, a learning rate that starts at 0.01 and decreases by a factor of 10 if the loss does not improve for 100 epochs of training, and a sigmoid activation, gives the best results. The reason for using as few layers and neurons as possible is to prevent overfitting and attempt to improve the extrapolation to $k$-values not present in the training. These are, for instance, the squeezed triangles that do not lay in any of the three foldings used and $k$-values smaller than $k_\mathrm{min}=0.0123\, \ihMpc$. We note that, while for the power spectrum, the BCM to large scales, $k<0.0123\, \ihMpc$, can straightforwardly be set to one, this is not obvious for the bispectrum given its sensitivity to different scales (see Fig.~\ref{fig:bispec_dependence_squeezed}). Therefore, a more complex extrapolation is needed. 
We implicitly test the accuracy of our NN extrapolation by performing mock analyses using the expected characteristics of \Euclid DR1 in Sect.~\ref{sec:forecasts}, leaving a more thorough explicit test for future works. 

\begin{figure}[ht]
\includegraphics[width=\columnwidth]{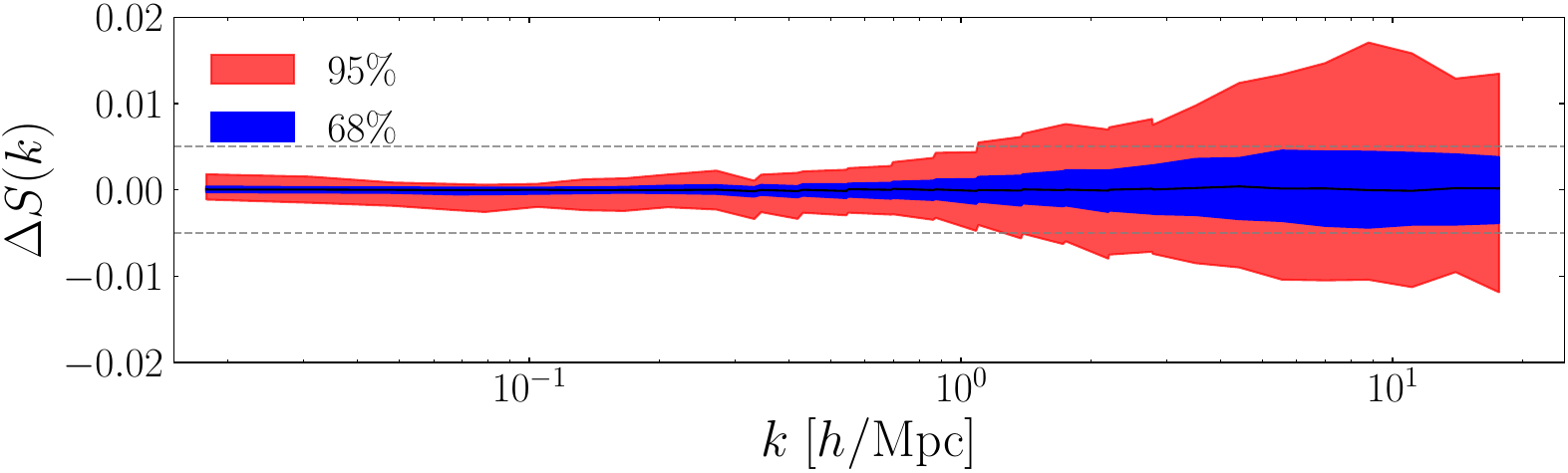}\\
\caption{Accuracy of the emulator for baryonic effects on the matter power spectrum. The $y$-axis is defined as $\Delta S(k) = S^\mathrm{emu}/ S^\mathrm{true}-1$, where $S(k)=P_{\rm BCM}(k)/P_{\rm GrO}(k)$. The horizontal dashed lines show the 0.5$\%$ region. }
\label{fig:emulator_acc}
\end{figure}

\begin{figure*}[ht]
\includegraphics[width=\textwidth]{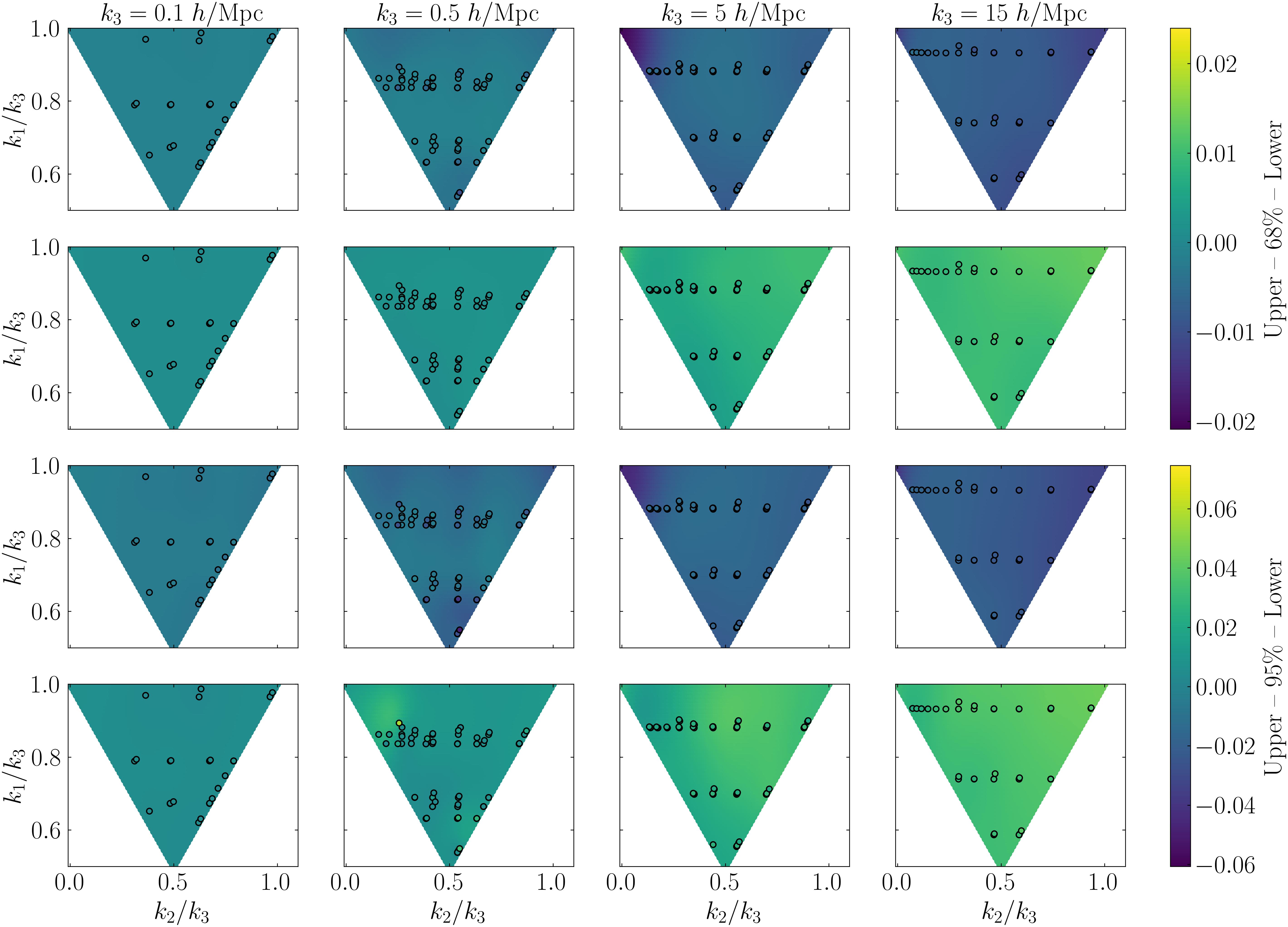}\\
\caption{Accuracy of the emulator for baryonic effects on the matter bispectrum $R(\vec{k})$. The upper two rows show the lower and upper $68\%$ percentiles of $\Delta R(\vec{k})$, and the two lower rows show the lower and upper $95\%$ percentiles of $\Delta R(\vec{k})$. Each column is for a different $k_3$ value, which is by construction larger than $k_2$ and $k_1$ but smaller than $k_1+k_2$. The black circles show the actual location where we have measured $B(k_1,k_2,k_3)$. The background is coloured using a linear interpolation/extrapolation of the percentiles between the measured $k$-values.}
\label{fig:emulator_acc_grid}
\end{figure*}

We first show the accuracy of our BCM power spectrum emulator in Fig.~\ref{fig:emulator_acc}. We find an accuracy on $S(k)$, which we define as  $\Delta S(k) = S^\mathrm{emu}/ S^\mathrm{true}-1$, smaller than $0.5\%$ for the $68\%$ percentile, and $3\%$ for the $95\%$ percentile. In other words, this means that $68\%$ of the predictions are better than $0.5\%$ for the power spectrum. We show the accuracy of the bispectrum emulator, defined as $\Delta R(\vec{k}) = R^\mathrm{emu}/ R^\mathrm{true}-1$, in Fig.~\ref{fig:emulator_acc_grid}, as a scatter plot where the corresponding $68\%$ and $95\%$ percentiles of $\Delta R(\vec{k})$ are shown in the upper two and lower two rows, respectively. We used linear interpolation between the measured $k$-values for a better illustration. The linear extrapolation into regions without measurements, such as the upper left corner for $k_3 = 5 \, \ihMpc$, should be taken cautiously. Furthermore, we show in Figs.~\ref{fig:emulator_acc_deltaR} and \ref{fig:emulator_acc_deltaR_kgrid} two different illustrations of the accuracy of the bispectrum emulator, where it is seen that the accuracy is smaller than $2\%$ for the $68\%$ percentile and similar across the different foldings and not particularly problematic at the transition from one folding to the next. Since we took 200 nodes for testing, which are also distributed in a Latin hypercube, the accuracies shown here marginalize over all parameters. However, we found only a minimal dependency of the accuracy towards larger $M_\mathrm{c}$. All the other parameters showed no dependency at all. Overall, we conclude that the emulator is less accurate for very squeezed triangles, especially where the measurement is at the edge of the $k$-range.

Remarkably, the overall accuracy of the emulator $\Delta R(\vec{k})$ is better than $2\%$ for the $68\%$ percentile. In the $95\%$ percentile, the accuracy of isolated triangle configurations seems to be worse than $5\%$. We have checked that this error arises from residual noise in the measurements. Given that we made sure that our emulator does not overfit this noise, we expect its accuracy to be better than the quoted values in those regions. 
We stress that these values refer to the accuracy of the emulation process. The underlying physical model was found to reproduce hydrodynamical simulations at a level of $1\%$ to $3\%$ in \cite{2021MNRAS.503.3596A} down to scales of $5 \, \ihMpc$. In the next sections, we explore whether this accuracy is enough to provide an unbiased cosmological inference in a \Euclid DR1 lensing analysis.   
For completeness, we show in Figs.~\ref{fig:bispec_dependence} and \ref{fig:bispec_dependence_squeezed} how $S(k)$ and $R(\vec{k})$ depend on BCM and cosmological parameters.

\section{From spectra to weak lensing statistics}
\label{sec:theory}
This section reviews the second- and third-order weak lensing statistics used to perform a forecast of a non-tomographic analysis of \Euclid DR1.
For a detailed review of gravitational lensing statistics, we refer to \citet{Bartelmann:2001}, \citet{Hoekstra:2008}, \citet{Munshi:2008}, \citet{Bartelmann:2010}, and \cite{Kilbinger2015}. In this work, we follow the description in \cite{Burger2023} and repeat only the essential details. 

\subsection{Limber projections of the power spectrum and bispectrum}
\label{subsec:power_and_bispectra_theory}

The projected power- and bispectrum at wavenumber, $\ell$, can be computed using the extended Limber approximation \citep{Limber:1954,Kaiser:1997,Bernardeau:1997,Schneider:1998,LoVerde2008}, up to the maximum distance defined by source redshift distribution (Eq.~\ref{eq:Euclid_nz}),
\begin{equation}
P_{\kappa\kappa} (\ell;\Theta) = \int_0^{\chi_\mathrm{max}} \dd \chi\;\frac{g^2(\chi)}{a^2(\chi)}\, P\left[\frac{\ell+1/2}{\chi},z(\chi);\Theta\right] \;, \label{eq:pkappa_defn}\\
\end{equation}
\begin{align}
{}&{} B_{\kappa\kappa\kappa}(\ell_1,\ell_2,\ell_3;\Theta) =  \int_0^{\chi_\mathrm{max}}\dd \chi\; \frac{g^3(\chi)}{a^3(\chi)\,\chi} \nonumber \\ {}&{}\times B\left[\frac{\ell_1+1/2}{\chi},\frac{\ell_2+1/2}{\chi},\frac{\ell_3+1/2}{\chi},z(\chi);\Theta\right] \label{eq:bkappa_defn}\, ,
\end{align}
where $\Theta = (\Theta_\mathrm{cos},\Theta_\mathrm{BCM})$ is the set of cosmological and BCM parameters. The 3-dimensional $P(k,z;\Theta)$ is defined as 
\begin{equation}
    P(k,z;\Theta) = P^\mathrm{GrO}(k,z;\Theta_\mathrm{cos}) \, S^\mathrm{BCM}(k,z;\Theta_\mathrm{BCM})
\end{equation}
and $P^\mathrm{GrO}(k,z;\Theta_\mathrm{cos})$ is the matter power spectrum computed using \texttt{Halofit} \citep{2012ApJ...761..152T}
that depends only on cosmological parameters $\Theta_\mathrm{cos}$, and $S^\mathrm{BCM} = P^\mathrm{Hydro}/P^\mathrm{GrO}$ is the BCM described in Sects.~\ref{sec:bispec} and \ref{sec:train} that depends on the BCM parameters $\Theta_\mathrm{BCM}$. In principle, we can use any implementation of $P^\mathrm{Gro}$, but we wanted to make use of $k>10 \, \ihMpc$ and $a<0.4$, which the current baccoemu does not provide. In analogy, the 3-dimensional bispectrum $B(\vec{k},z,\Theta)$ is computed as
\begin{equation}
    B(\vec{k},z;\Theta) = B^\mathrm{GrO}(\vec{k},z;\Theta_\mathrm{cos}) \, R^\mathrm{BCM}(\vec{k},z;\Theta_\mathrm{BCM})\, ,
\end{equation}
where $B^\mathrm{GrO}(\vec{k},z;\Theta_\mathrm{cos})$ is the nonlinear matter bispectrum computed using \texttt{BiHalofit} \citep{Takahashi:2020}, and $R^\mathrm{BCM} = B^\mathrm{Hydro}/B^\mathrm{GrO}$ is the BCM for the bispectrum.

Lastly, we are only considering closed triangles, $\ell_3 = |\pmb{\ell}_1+\pmb{\ell}_2|$, and $g(\chi)$ denotes the lensing efficiency and is defined as
\begin{equation}
g(\chi) = \frac{3\Omega_\mathrm{m}H_0^2}{2c^2}\int_\chi^{\chi_\mathrm{max}} \dd \chi'\; n(\chi') \,\frac{\chi'-\chi}{\chi'} 
\label{eq:lensing_efficiency_defn} \, ,
\end{equation}
with $n(\chi) = n(z) \,\dd z / \dd \chi$ and $n(z)$ being the redshift probability distribution given in Eq.~\eqref{eq:Euclid_nz}.

\subsection{Aperture mass statistics}
\label{subsec:background_map3}
The aperture mass $\Map$ at position $\pmb{\vartheta}$ with filter radius $\theta_\mathrm{ap}$ is defined through the convergence $\kappa$, as follows:
\begin{equation}
\label{eq:definition_aperture_masskappa}
\Map(\pmb{\vartheta};\theta_\mathrm{ap})=\int_{\mathbb{R}^2}\dd^2\vartheta'\; U_{\theta_\mathrm{ap}}(|\pmb{\vartheta'}|)\, \kappa(\pmb{\vartheta}+\pmb{\vartheta'}) \; ,
\end{equation}
where $U_{\theta_\mathrm{ap}}(\vartheta')$ is a compensated filter such that ${\int_{\mathbb{R}} \dd\vartheta'\,\vartheta'\, U_{\theta_\mathrm{ap}}(\vartheta') = 0}$. 

We define $U_{\theta_\mathrm{ap}}(\vartheta)=\theta_\mathrm{ap}^{-2}u(\vartheta/\theta_\mathrm{ap})$, denote by $\hat{u}(\alpha)$ the Fourier transform of $u$, and use the filter function introduced in \citet{Crittenden:2002},
\begin{align}
u(x)= \frac{1}{2\pi}\left(1-\frac{x^2}{2}\right)\exp\left(-\frac{x^2}{2}\right),\quad \hat{u}(\alpha) = \frac{\alpha^2}{2}\exp\left(-\frac{\alpha^2}{2}\right)\;.
\end{align}

\subsection{Modelling aperture mass moments}
\label{subsec:modelling_map}

While the expectation value of the aperture mass $\expval{\Map}(\theta_\mathrm{ap})$ vanishes by construction, the second-order (variance) of the aperture mass is nonzero and can be calculated as
\begin{equation}
\expval{\Map^2}(\theta_\mathrm{ap}) = \int_{\mathbb{R}_+}\frac{\dd \ell\;\ell}{2\pi}\,P_{\kappa\kappa}(\ell)\,\hat{u}^2(\theta_\mathrm{ap}\ell)\; .
\end{equation}
Equivalently, the third-order moment of the aperture statistics $\MapMapMap$, can be computed from the convergence bispectrum via \citep{Jarvis:2004,Schneider:2005}
\begin{align}
&\expval{\Map^3}(\theta_{\mathrm{ap},1},\theta_{\mathrm{ap},2},\theta_{\mathrm{ap},3}) =  \int_{\mathbb{R}_+^2}\frac{\dd^2 \ell_1}{(2\pi)^2}\int_{\mathbb{R}_+^2}\frac{\dd^2 \ell_2}{(2\pi)^2}\ \nonumber \\ & \quad  \times B_{\kappa\kappa\kappa}(\ell_1,\ell_2,\ell_3)\,  \hat{u}(\theta_{\mathrm{ap},1}\ell_1) \, \hat{u}(\theta_{\mathrm{ap},2}\ell_2)\,\hat{u}(\theta_{\mathrm{ap},3}\ell_3)\; ,
\label{eq:Ma3_def}
\end{align}
where $\ell_3 = |\pmb{\ell}_1+\pmb{\ell}_2|$. 
In the two lower panels of Fig.~\ref{fig:Map23_dependence_GrO}, we show how $\expval{\Map^2}$ and $\expval{\Map^3}$ react to changes in the baryonic parameters, while all other parameters are fixed to the fiducial parameters given in Tables \ref{tab:D3A} and \ref{tab:baryon_fid}. We divide the $\expval{\Map^n}$ predictions by GrO predictions, $\expval{\Map^n}^\mathrm{GrO}$, to highlight the baryonic effects. 
As expected, $M_\mathrm{c}$ has the largest impact on both $\expval{\Map^2}$ and $\expval{\Map^3}$. This baryonic parameter sets the characteristic halo mass for which half of the gas mass is ejected by feedback, thus regulating the quantity of gas retained in haloes. As expected, the galaxy parameter $M_1$ and the inner gas slope scale $\theta_{\rm inn}$ have a larger effect on small apertures, whereas the feedback scale $\eta$ has a larger impact on larger apertures. 
We note that, comparatively, $\beta$ has a significantly larger effect on $\expval{\Map^3}$ than on $\expval{\Map^2}$. 
Looking at the cosmological parameters, we find that the dependence on $\sigma_8$ is negligible for $\expval{\Map^3}$, but we observe some effect in the second-order statistics. We note that the dependence on $\Omega_\mathrm{m}$ and $\Omega_\mathrm{b}$ can be better captured by their ratio, $\Omega_\mathrm{b}/\Omega_\mathrm{m}$, which comes from the fact that the impact of the baryon feedback, at fixed strength, depends to zeroth order on the relative amounts of baryons to the dark matter.

\begin{table}
    \renewcommand{\arraystretch}{1.3}
    \centering
    \caption{Fiducial baryon parameters. }
    \begin{tabular}{l|c}
        Parameter & Fiducial Value \\
        \hline
        $\log_{10} (M_\mathrm{c} / h^{-1} \mathrm{M}_\odot)$ & 14.0 \\
        $\log_{10} (M_1 / h^{-1} \mathrm{M}_\odot)$          & 11.0 \\
        $\log_{10} (\beta)$                                  & $-0.35$ \\
        $\log_{10} (\eta)$                                   & $-0.35$ \\
        $\log_{10} (\theta_\mathrm{inn})$                    & $-1.0$ \\
    \end{tabular}
    \tablefoot{Units are shown inside the logarithmic expressions where applicable; all parameters are dimensionless except for the mass terms.}
    \label{tab:baryon_fid}
\end{table}

\subsection{Modelling convergence correlation functions}
\label{subsec:modelling_xikappa}
Since the $\expval{\Map^2}$ statistics smooth much of the small scales, we decided to add the information from the two-point correlation function $\xi_\kappa$ itself, which is defined by
\begin{equation}
\xi_\kappa(\theta) = \int_{\mathbb{R}_+}\frac{\dd \ell\;\ell}{2\pi}\,P_{\kappa\kappa}(\ell)\,J_0(\theta \ell)\; ,
\end{equation}
where $J_0$ is the zeroth Bessel function of the first kind. In the lower panel of Fig.~\ref{fig:Map23_dependence_GrO}, we show how $\xi_\kappa$ reacts to changes in the baryonic parameters. As for the $\expval{\Map^n}$ moments, we divide the $\xi_\kappa$ predictions by GrO predictions, $\xi_\kappa^\mathrm{GrO}$, to highlight the role of baryons. Similarly as for $\expval{\Map^2}$, we observe that the most important baryonic parameters are $M_\mathrm{c}$ and $\eta$.

\subsection{Flamingofied data vector}
\label{sec:flamingofying}
\begin{figure}[ht]
\centering
\includegraphics[width=0.49\textwidth]{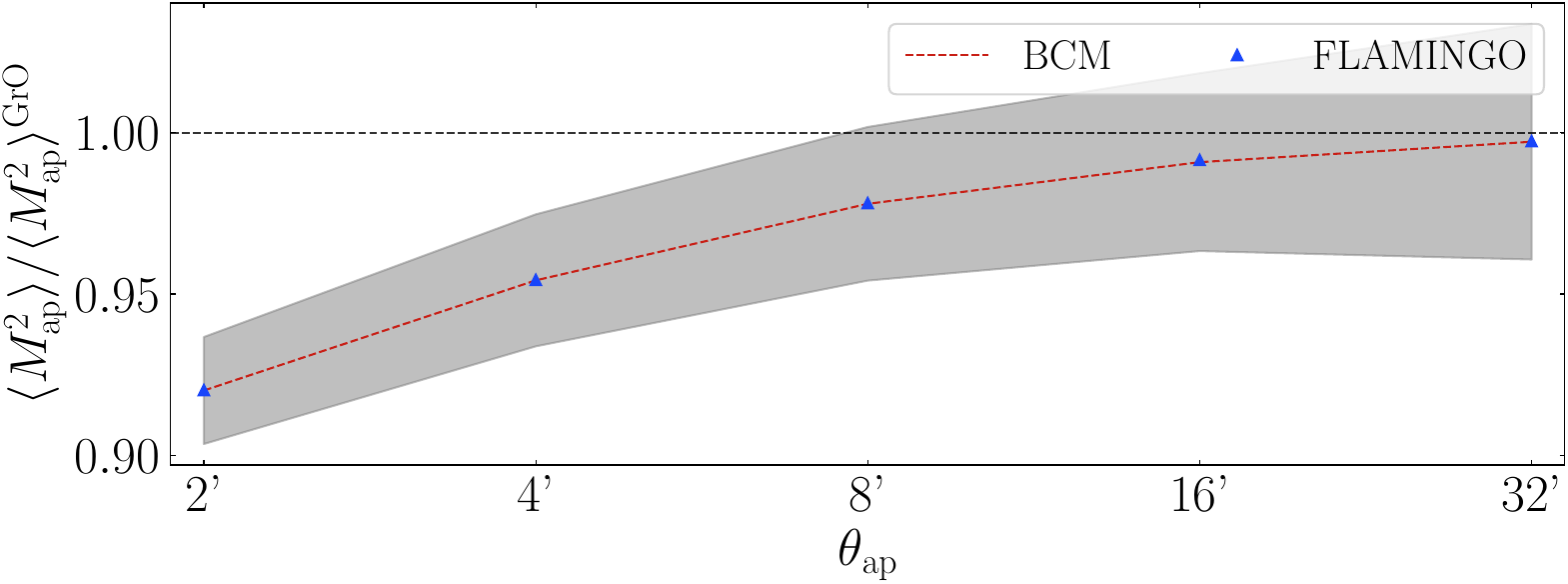}\\
\includegraphics[width=\columnwidth]{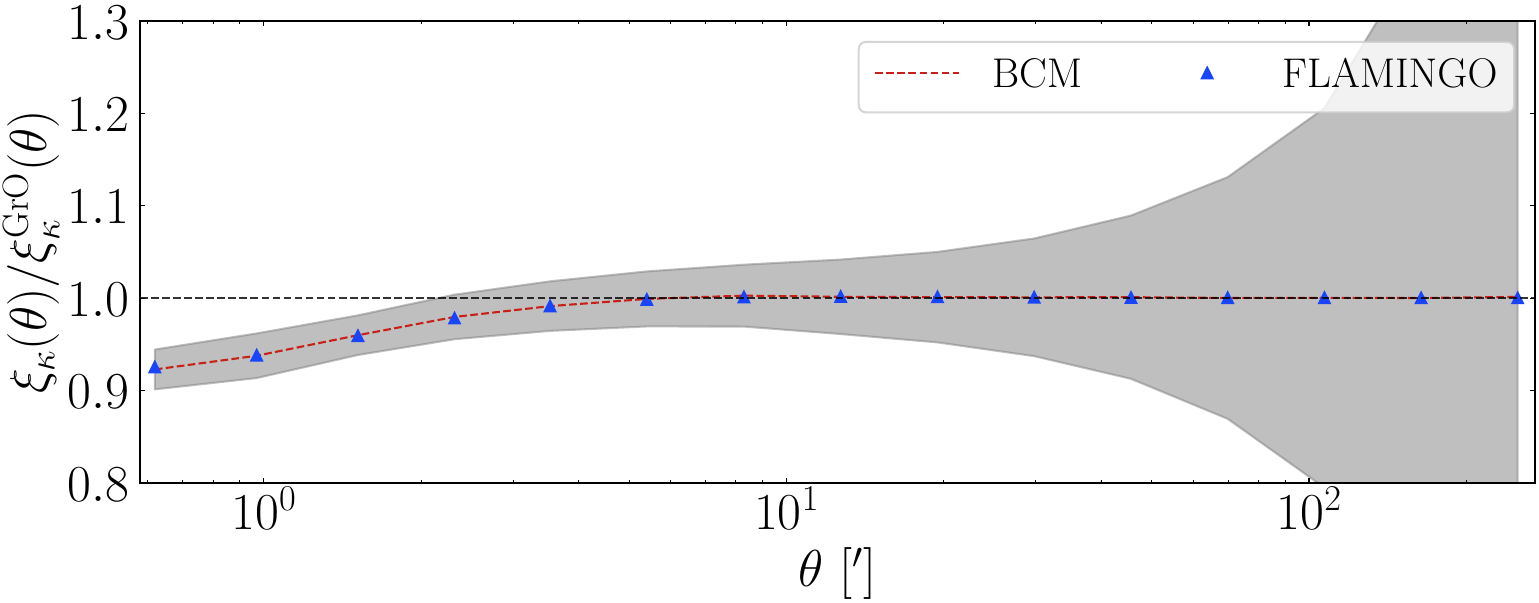}
\caption{Baryonic effects on $\expval{\Map^2}$ (upper panel) and on $\xi_\kappa$ (lower panel), illustrated by ratios of measurements with baryonic physics to GrO measurements. We show with blue triangles the fiducial FLAMINGO measurements, and the grey shaded area represents the \Euclid DR1 uncertainty. We overplot with a red dashed line the best-fitting BCM to the flamingofied data vector, found by employing our emulators with fixed cosmological parameters.}
\label{fig:xikappa_Map2_Flamingo}
\end{figure}

\begin{figure*}[ht]
\centering
\includegraphics[width=\textwidth]{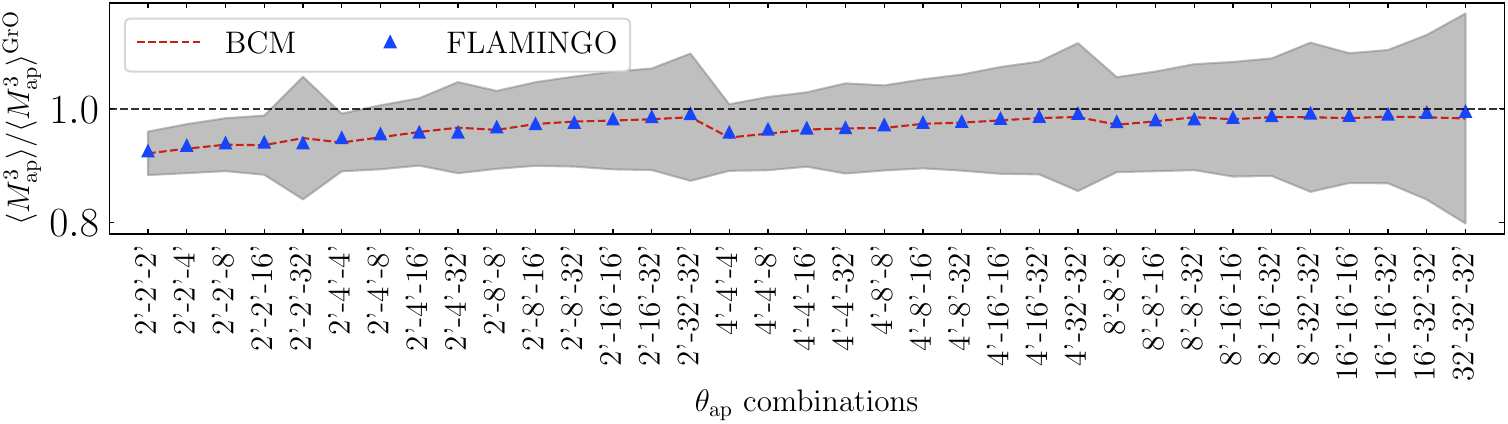}
\caption{Same as Fig.~\ref{fig:xikappa_Map2_Flamingo} but for the $\expval{\Map^3}$ statistics. The $x$-axis shows different combinations of $\theta_\mathrm{ap,1}$-$\theta_\mathrm{ap,2}$-$\theta_\mathrm{ap,3}$.}
\label{fig:Map3_Flamingo}
\end{figure*}

In this section, we describe the process of `flamingofying' a GrO reference model. A flamingofied reference model of $\expval{\Map^n}$ is built by scaling the prediction of a GrO model at the cosmology of the corresponding FLAMINGO simulation with the ratio of a hydrodynamic-GrO pair of FLAMINGO simulations
\begin{equation}
    \expval{\Map^n}^\mathrm{flamingofied} = \expval{\Map^n}^\mathrm{GrO} \frac{\expval{\Map^n}^\mathrm{FLAMINGO,Hydro}}{\expval{\Map^n}^\mathrm{FLAMINGO,GrO}} \, .
\end{equation}
Analogously, we flamingofy the $\xi_\kappa$ in the following way
\begin{equation}
    \xi_\kappa^\mathrm{flamingofied} = \xi_\kappa^\mathrm{GrO} \frac{\xi_\kappa^\mathrm{FLAMINGO,Hydro}}{\xi_\kappa^\mathrm{FLAMINGO,GrO}} \, .
\end{equation}
In Figs.~\ref{fig:xikappa_Map2_Flamingo} and \ref{fig:Map3_Flamingo}, we show in blue the ratio Hydro/GrO of the FLAMINGO measurements for $\expval{\Map^n}$ and $\xi_\kappa$ for the L2p8$\_$m9 scenario. Comparing these ratios to the estimated uncertainty in grey indicates which scales are affected by baryonic effects. Furthermore, we show the red dashed line as the best-fitting model to fit the flamingofied reference model if we fix the cosmology to D3A, where we performed a $\chi^2$ minimisation for each statistic separately to find the best-fitting parameters. Our BCM model can perfectly fit the flamingofied reference model. This is an encouraging result for the cosmological analysis in the next section. 

We also show the effects of baryons from the FLAMINGO description L2p8$\_$m9 in each panel. The $\expval{\Map^n}$ in the FLAMINGO and Takahashi simulations are measured by applying Eq.~\eqref{eq:definition_aperture_masskappa}, using the \texttt{Healpy} smoothing option. We have decided to use aperture filter radii $\theta_\mathrm{ap} = \{\ang{;2;},\ang{;4;},\ang{;8;},\ang{;16;},\ang{;32;}\}$, where the lower value $\theta_\mathrm{ap} = \ang{;2;}$ is constrained by the chosen pixel resolution $\ang{;0.43;}$ of the FLAMINGO/T17 convergence maps. The $\xi_\kappa$ in the FLAMINGO and Takahashi simulations are measured with \texttt{TreeCorr} \citep{Jarvis:2004}, from $\ang{;0.5;} < \theta < \ang{;300;}$ in 15 logarithmic bins, where the lower edge is determined again by the resolution of the chosen convergence maps.

\section{\Euclid forecasts}
\label{sec:forecasts}

In this section, we perform a non-tomographic \Euclid DR1 forecast, where we vary the matter density $\Omega_\mathrm{m}$, the baryon density $\Omega_\mathrm{b}$, the normalisation of the power spectrum $\sigma_8$, and the five baryonic parameters described in Sect~\ref{sec:BCM}. The dependence of these parameters on $\expval{\Map^n}$ and the $\xi_\kappa$ are already discussed in Sect.~\ref{sec:theory} and are shown in Fig.~\ref{fig:Map23_dependence_GrO}. We have not yet mentioned that the computations of the integrals of these quantities are not efficient enough to perform several Markov Chain Monte Carlo (MCMC) analyses. Therefore, we decided to train another set of emulators on top of the three-dimensional power and bispectrum emulators, which predict the $\expval{\Map^n}$ and $\xi_\kappa$. We computed the $\expval{\Map^n}$ and the $\xi_\kappa$ for 2200 points distributed in a Latin hypercube. We estimate the error of the emulation, by building a set of emulators iteratively leaving out from the training 200 of the 2200 points at a time. Since the power and bispectrum are noisy quantities themselves, their noise also propagates into the summary statistics. Furthermore, the integration to $\expval{\Map^3}$ sometimes fails for very squeezed configurations, because we had to limit the number of $\ell$ values where we compute $B_{\kappa \kappa \kappa}$ for the sake of computational speed. 
Therefore, the emulator error accounts for emulator noise in the summary statistics but also for noise in the power and bispectrum estimations and their subsequent integration. We note that this procedure, however,  does not account for possible systematic errors in the emulation or integration. 
The error of this emulator is less than a tenth of the uncertainty estimated with the \cite{Takahashi2017} simulations. Although this might be neglected, we decided to build an emulator covariance matrix for correctly estimating the posterior, using the 2200 $\Delta m = m^\mathrm{emu}-m^\mathrm{true}$ predictions and add it to the \cite{Takahashi2017} covariance matrix as
\begin{equation}
    \label{eq:emu_cov}
    \tens C = \tens C^\mathrm{T17} + \tens C^\mathrm{emu} \,.
\end{equation}

Since the estimated covariance matrix $\tens C$ is a random variable itself, we follow \cite{Percival2021} to compute the Likelihood. Given a data vector $\vec{d}$ and covariance matrix $\tens C$ of rank $n_\mathrm{d}$ measured from $n_\mathrm{r}$ realizations, the posterior distribution of a model vector $\vec{m}(\Theta)$ that depends on $n_{\Theta}=12$ parameters is
\begin{equation}
\vec{P}\left(\vec{m}(\Theta)|\vec{d},\tens C\right) \propto |\tens C|^{-\frac{1}{2}} \left( 1 + \frac{\chi^2}{n_{\rm r}-1}\right)^{-m/2}\, ,
    \label{eq:t_distribution}
\end{equation}
where
\begin{equation}
\chi^2 =  \left[\vec{m}(\Theta)-\vec{d}\right]^{\rm T} \tens C^{-1} \left[\vec{m}(\Theta)-\vec{d}\right] \, .
\label{eq:chi2}
\end{equation}
The power-law index $m$ is 
\begin{equation}
    m = n_\Theta+2+\frac{n_\mathrm{r}-1+B(n_\mathrm{d}-n_\Theta)}{1+B(n_\mathrm{d}-n_\Theta)} \, ,
    \label{eq:m_power}
\end{equation}
and $B$ is defined as
\begin{equation}
    B = \frac{n_\mathrm{r}-n_\mathrm{d}-2}{(n_\mathrm{r}-n_\mathrm{d}-1)(n_\mathrm{r}-n_\mathrm{d}-4)} \, .
    \label{eq:B}
\end{equation}
and $n_\mathrm{r}=570$ is the number of realizations, $n_{\Theta}=8$ the number of free parameters, and $n_{\rm d}$ the rank of the covariance matrix. Although $n_\mathrm{r}$ for $C^\mathrm{emu}$ is not clearly defined, we used the same correction for $C$ as we used for $C^\mathrm{T17}$. Lastly, we used the nested sampler \texttt{nautilus} \citep{Lange2023} to perform the inference, employing 2000 live points. We validated the convergence of the posteriors by comparing them with a Hamiltonian Monte Carlo sampler and a standard Metropolis–Hastings algorithm.

\begin{figure}[ht]
\includegraphics[width=\columnwidth]{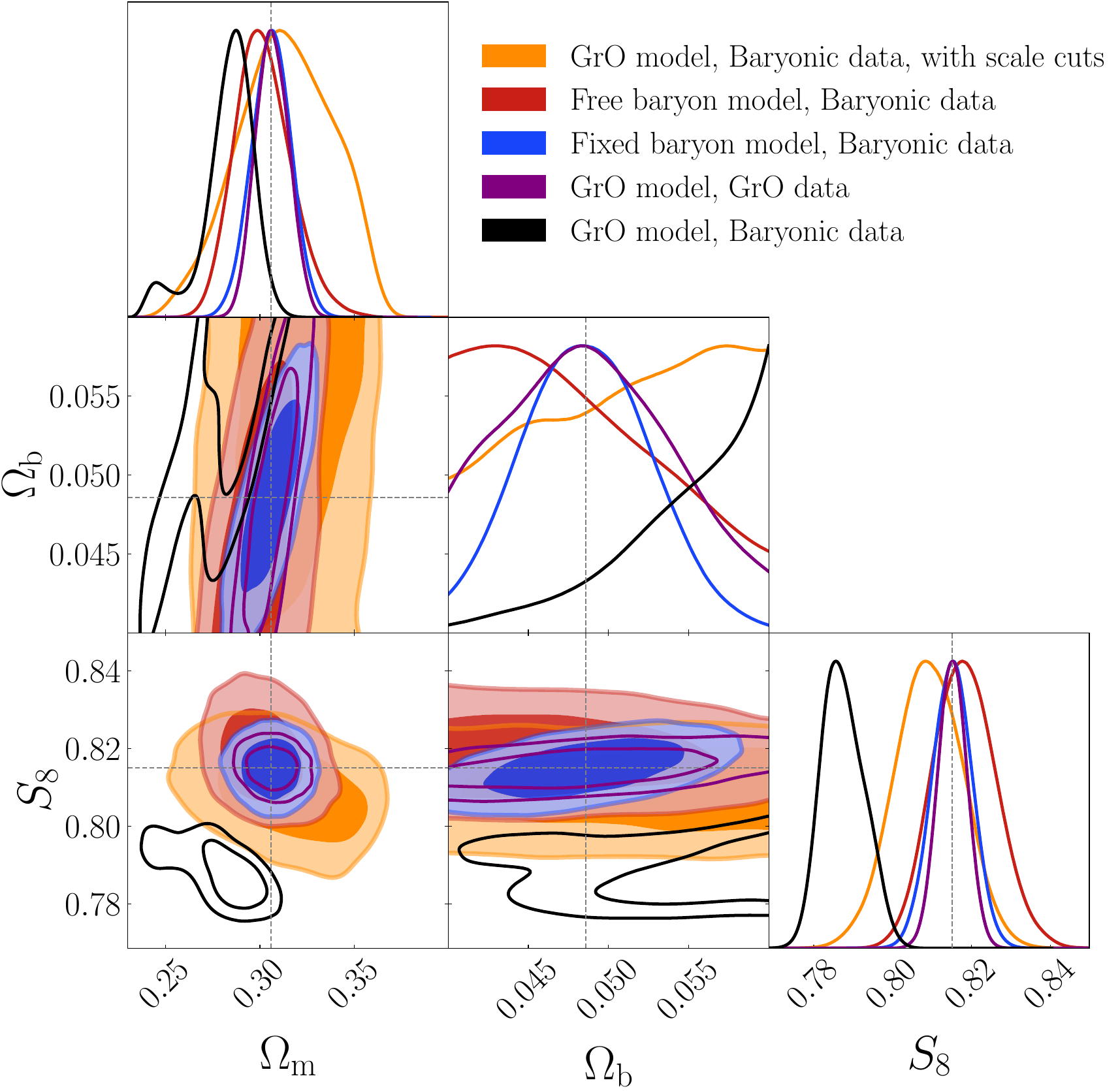}
\caption{Posteriors of the cosmological parameters $S_8$, $\Omega_\mathrm{m}$, and $\Omega_\mathrm{b}$, analysing a synthetic \Euclid DR1 joint data vector of second- and third-order weak lensing statistics. We consider several cases. First, a gravity-only (GrO) data vector was analysed with a GrO model (purple contours). We obtain the black contours when baryonic effects are included in the data vector but not in the model. In orange, we illustrate the constraints we get when all elements of the data vector are excluded where the baryonic data vector deviates by more than $0.4\sigma$ from the GrO data vector. Lastly, we show the results if the model includes baryonic feedback in red and blue, where we fixed the baryon parameters to the truth in the blue case and marginalised over all five parameters for the red. The grey dashed line indicates the underlying D3A cosmology.}
\label{fig:motivation}
\end{figure}

To start our investigation of the impact of baryonic feedback, we show in Fig.~\ref{fig:motivation} posteriors on the cosmological parameters $S_8$, $\Omega_\mathrm{m}$, and $\Omega_\mathrm{b}$, which we obtain by analysing a synthetic \Euclid DR1 data vector built from a combination $\xi_\kappa$, $\langle \Map^2 \rangle$, and $\langle \Map^3 \rangle$. The investigation here is built solely using \texttt{Halofit}, \texttt{BiHalofit}, and the BCM, which means we always know the true input of cosmological and BCM parameters. In the first case (black), we created a data vector using the BCM with the parameters given in Table~\ref{tab:baryon_fid} and analysed it with the GrO model, meaning we turned the BCM off in the modelling. This results in a $2.5\sigma$ shift in $S_8$. Substantial scale cuts on small scales are required to avoid such a bias, which we show as the orange contours. In terms of scales these are $\theta > \ang{;8;}$ for $\xi_\kappa$, $\theta_\mathrm{ap} = \ang{;32;}$ for $\expval{\Map^2}$ and $\theta_\mathrm{ap,1}  \in \{ \ang{;16;},\ang{;32;}\}$ with $\theta_\mathrm{ap,2,3} =\ang{;32;}$ for $\expval{\Map^3}$. However, this drastically reduces the precision on $S_8$ by $8\%$ and by almost $80\%$ on $\Omega_\mathrm{m}$ compared to modelling all five BCM parameters shown as the red contours. We obtain even tighter constraints by fixing these five BCM parameters to their true values, as shown in the blue contours. Having perfect knowledge of baryonic effects results in cosmological constraints almost identical to the case where we used a GrO data vector and analysed it with a GrO model, displayed with purple contours. We will show later that we probably do not need to vary all five BCM parameters for the expected \Euclid DR1 uncertainty.

\subsection{Validating the emulator on the FLAMINGO simulations}
\begin{figure}[ht]
\includegraphics[width=\columnwidth]{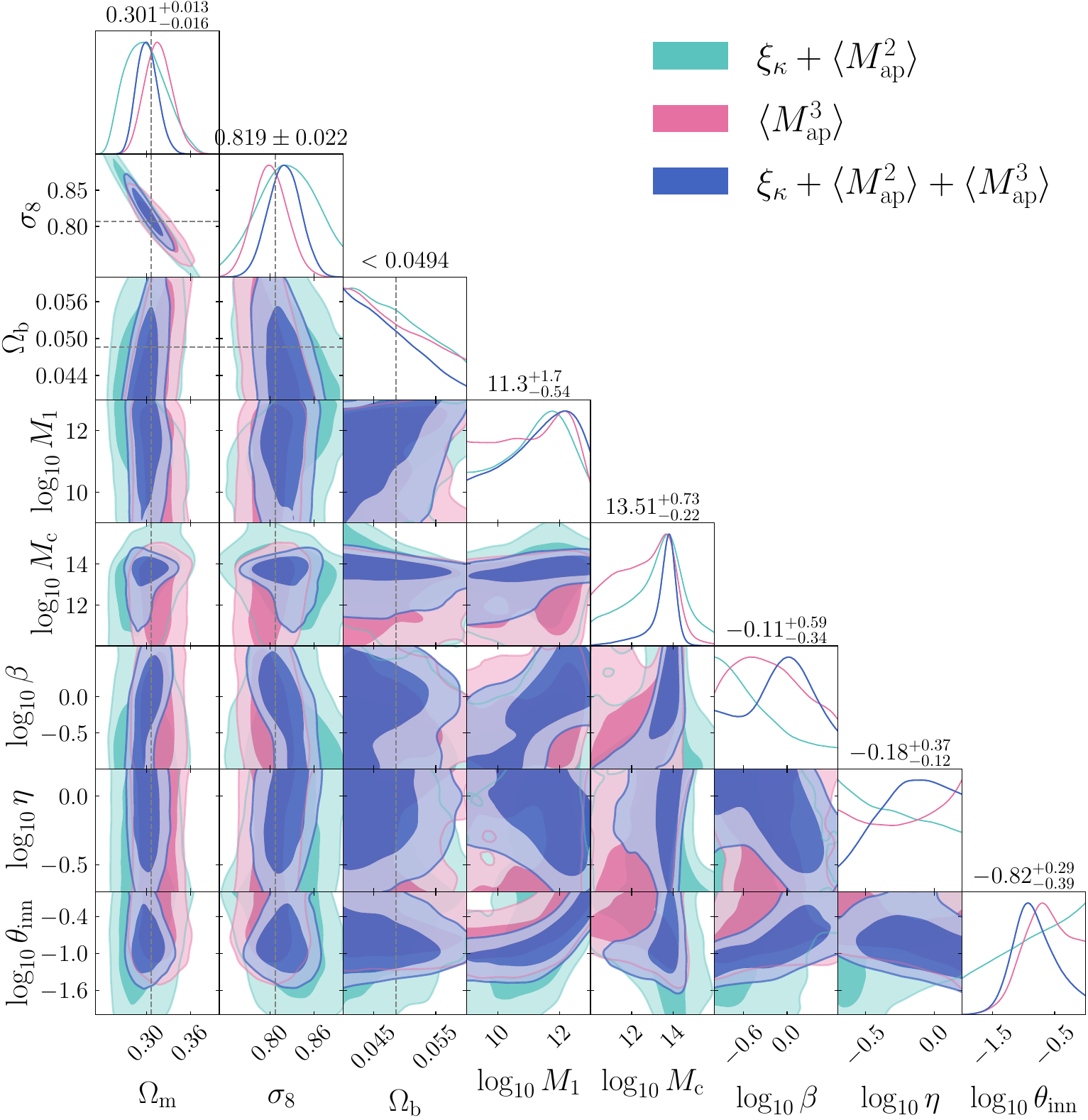}
\caption{\Euclid DR1 forecast, where we used a flamingofied reference data vector (see Sect.~\ref{sec:flamingofying}) estimated from the fiducial baryon description of the FLAMINGO simulations with eight light cones of the 2.8 Gpc box. The stated constraints correspond to the joint analysis of second- and third-order statistics. The grey dashed line indicates the underlying D3A cosmology.}
\label{fig:MCMC_flamingo}
\end{figure}

First, we use a flamingofied data vector estimated from eight light cones of the 2.8 Gpc box that uses the fiducial baryon description (see Sect.~\ref{sec:flamingofying}). This test aims to see if the combination of the three different statistics improves the constraining power and check if the BCM is flexible enough to fit the fiducial FLAMINGO simulation. The constraints on both cosmological and baryonic parameters are shown in Fig.~\ref{fig:MCMC_flamingo}. This is one of the main results of this paper and shows that our baryon model can describe the modifications due to baryons while resulting in unbiased cosmological results for all three statistics and their joint analysis. We note that $\expval{\Map^2}$ yields similar constraints as $\xi_\kappa,$ since both result from integrating the power spectrum, so we decided to show only the combination. Next, we notice that $\expval{\Map^3}$ constrains the cosmological parameters slightly better than the second-order statistics, and furthermore helps breaking parameter degeneracies. Compared to \cite{Burger2024a}, where second-order statistics were relatively more constraining, we expect that this comes from the fact that we include smaller scales with $\theta_\mathrm{ap} = \ang{;2;}$, where higher-order statistics become more important. 
We do not show the `true' FLAMINGO values for the BCM parameters. These can be measured in hydrodynamic simulations in principle; however, their measured values are redshift-dependent. Thus, interpreting our constraints is not straightforward since they are weighted in redshift according to our lensing kernel.
We leave for future work a thorough test of the BCM parameter constraints, as a correct inference of baryonic physics is essential, for example, in cross-correlations and multi-probes analyses.

\begin{figure}[ht]
\includegraphics[width=\columnwidth]{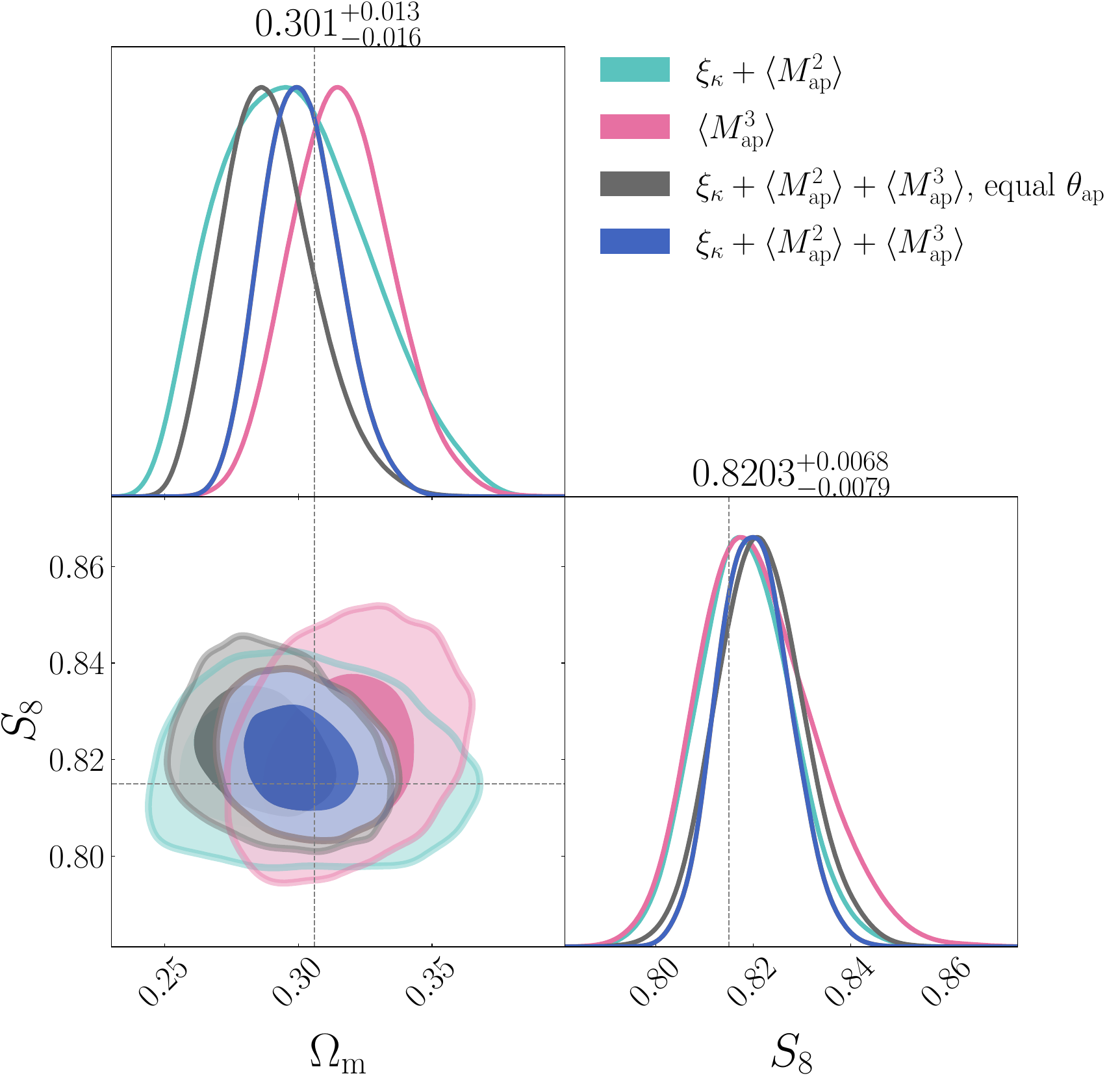}
\caption{\Euclid DR1 forecast, where we used a flamingofied reference data vector (see Sect.~\ref{sec:flamingofying}) estimated from the fiducial baryon description of the FLAMINGO simulations with eight light cones of the 2.8 Gpc box. The stated constraints correspond to the joint analysis of second- and third-order statistics. The posteriors are the same as those used in Fig.~\ref{fig:MCMC_flamingo}, meaning we marginalised over all BCM parameters. The stated constraints correspond to the joint analysis of second- and third-order statistics (blue contours). The grey dashed line indicates the underlying D3A cosmology.}
\label{fig:MCMC_Flamingo_OmS8}
\end{figure}

In Fig.~\ref{fig:MCMC_Flamingo_OmS8}, we compare the constraining power of second-order and third-order statistics in the $\Omega_\mathrm{m}$-$S_8 = \sigma_8 \sqrt{\Omega_\mathrm{m}/0.3}$ plane. We highlight the different degeneracy axes of the second- and third-order statistics, which are broken when both are combined, in agreement with \cite{2013MNRAS.434..148S}. Therefore, combining second- and third-order statistics increases the constraining power by $22\%$ for $S_8$ and even $50\%$ for $\Omega_\mathrm{m}$ compared to using second-order statistics alone. Lastly, we also show the case where we discard all the non-equal scale filter radii $\theta_\mathrm{ap}$. Interestingly, these non-equal scale filter radii $\theta_\mathrm{ap}$ play an essential role in the precision and accuracy of cosmological parameters, since discarding them reduces the constraining power by around $16\%$ for $S_8$ and $\Omega_\mathrm{m}$ compared to using all filter combinations. \cite{Burger2023} found that these non-equal scale filter radii are less significant. We interpret this as non-equal scale filter radii being important to constrain the BCM parameters, which in turn helps to constrain cosmological parameters. Another explanation for the discrepancies with \cite{Burger2023} could also lie in the fact that we include the $\theta_\mathrm{ap}=\ang{;2;}$ filter and that we used a different tomographic setup.

\begin{figure*}[ht]
\includegraphics[width=\textwidth]{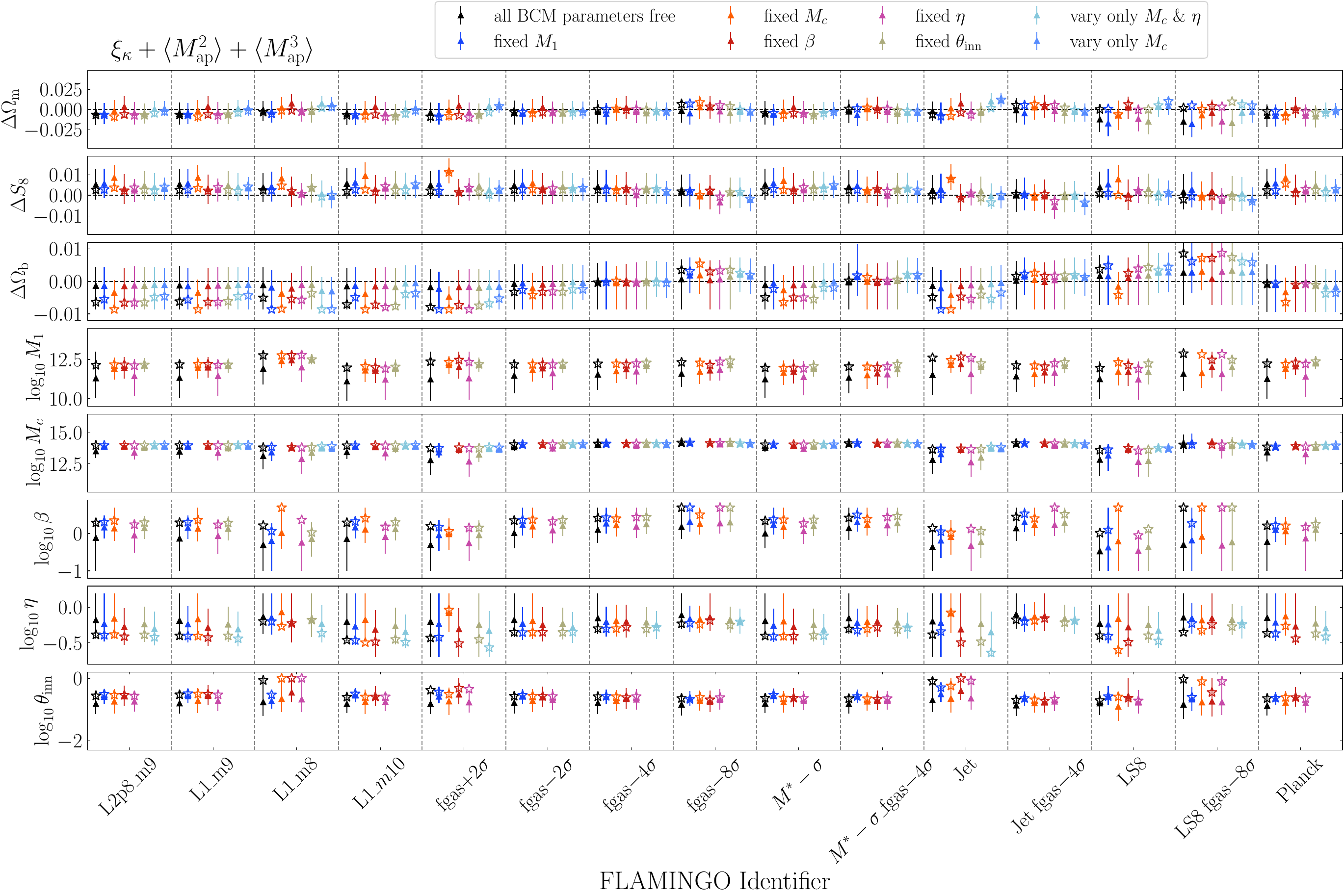}
\caption{Estimated weighted mean (triangles) and $68\%$ credible intervals from the MCMC chains for all FLAMINGO models described in Sect.~\ref{sec:Flamingo_sim}. Since the LS8 and \Planck cosmology is different we plot $\Delta \Omega_\mathrm{m} = \Omega^\mathrm{best}_\mathrm{m} - \Omega^\mathrm{true}_\mathrm{m}$ and in analogy $\Delta S_8$ and $\Delta \Omega_\mathrm{b}$. The different colours show cases where parameters are fixed. The stars indicate the best-fitting parameters resulting from a minimisation process that returns the lowest $\chi^2$. The figure is for $\xi_\kappa + \langle \Map^2 \rangle + \langle \Map^3 \rangle$. The corresponding figure for $\xi_\kappa + \langle \Map^2 \rangle$ is shown in Fig.~\ref{fig:Flamingo_variation_cosmology_2nd}.
}
\label{fig:Flamingo_variation_cosmology}
\end{figure*}

Testing only one astrophysical scenario is insufficient to ensure a robust cosmological inference. Therefore, we use all baryon feedback descriptions described in Sect.~\ref{sec:Flamingo_sim}. We build flamingofied data vectors as described in Sect.~\ref{sec:flamingofying}. We run an MCMC for each baryon feedback description once using only second-order statistics $\xi_\kappa + \langle \Map^2 \rangle$ and once combining second- and third-order $\xi_\kappa + \langle \Map^2 \rangle + \langle \Map^3 \rangle$, where we vary all baryon parameters. We present the result as the black symbols in Fig.~\ref{fig:Flamingo_variation_cosmology}, where the error bars represent the $1\sigma$ uncertainties estimated from the MCMC chains. The corresponding best-fitting parameters are shown as the stars, and we notice that the best-fitting parameters are much closer to the truth than the mean, which might indicate non-Gaussian parameter posteriors or significant projection effects. Remarkably, we get unbiased cosmological results for all FLAMINGO flavours and configurations for both the second-order alone and the second- and third-orders combined. This is the most important result of this paper, as it shows that our BCM model is flexible enough to get unbiased results for all FLAMINGO feedback variations. This also shows that the interpolation and extrapolation of our emulators for the BCM power spectrum and bispectrum are sufficient for the current setup. We expect the emulator accuracy requirements to be similar in a tomographic setup, such that our emulators are ready for a \Euclid-like DR1 analysis. However, a tomographic analysis might require an explicit redshift dependence on the baryonic parameters. We highlight that the three-dimensional power spectrum and bispectrum emulators developed in this work would remain valid and could be used with parametrised redshift dependencies without creating new training sets or emulators.

\subsection{Assessing the importance of the BCM parameters}

In this subsection, we exploit a wide range of feedback models of the FLAMINGO suite to assess the importance of each BCM parameter in a cosmological analysis. We aim to find the minimal BCM model required to deliver unbiased cosmological constraints in a Euclid DR1 lensing analysis. Fewer free parameters would help mitigate projection effects and allow for faster convergence of the posteriors. This analysis cannot be straightforwardly generalised to different hydrodynamical simulations and summary statistics. Nevertheless, this section provides us with valuable information about the most important BCM parameters that can capture the large variety of baryonic variations in FLAMINGO, even if care must be taken when fixing one or more baryonic parameters in data analyses.

To perform our test, we fix one or more BCM parameters to the weighted average of the best-fitting parameters over all FLAMINGO flavours and configurations, where the weights are estimated from the $1\sigma$ credible intervals of MCMC chains. We report those values in Table \ref{tab:baryon_best}. Next, we run an MCMC for each scenario, and show the best-fitting values, mean and standard deviation of the marginalised posteriors in Fig.~\ref{fig:Flamingo_variation_cosmology}. 
Focusing first on $\xi_\kappa + \langle \Map^2 \rangle + \langle \Map^3 \rangle$, we notice that the best-fitting values, especially for $S_8$, are closer to the truth than the weighted mean of the MCMC, indicating strong projection effects. We find that fixing $M_\mathrm{c}$ is particularly bad in some cases, where an incorrect estimation of $M_\mathrm{c}$ results in a biased $S_8$ inference. Next, we observe that fixing $\eta$ leads to small biases for the strongest feedback descriptions (jet fgas--$4\sigma$ and fgas--$8\sigma$). Fixing either $M_1$ or $\beta$ does result in unbiased results for all scenarios, which is not surprising given that $M_1$ and $\beta$ are unconstrained, as we can see in Fig.~\ref{fig:MCMC_flamingo}. 
Fixing $\eta$ or $\theta_\mathrm{inn}$ also results in small biases for the LS8 cosmology. 
Next, we test fixing all BCM parameters except $M_\mathrm{c}$ and $\eta$. We get unbiased results for all FLAMINGO flavours, except the Jet flavour shows $1\sigma$ shift for the $\Omega_\mathrm{m}$ parameter. When fixing all the BCM except $M_\mathrm{c}$, the effect on the Jet feedback description is enhanced. We note that varying only one or two BCM parameters results in much better-constrained cosmological parameters. If the baryonic feedback inferred from future observations will be consistent with the FLAMINGO descriptions that were tested here, it is probably sufficient for current Stage-III and first data releases of Stage-IV surveys to vary only $M_\mathrm{c}$ and $\eta$, or even only $M_\mathrm{c}$. Lastly, we notice that varying parameters such as $M_\mathrm{1}$ might not influence the estimation of cosmological parameters, but can play a vital role in constraining other BCM, such as $M_\mathrm{c}$.

Focusing now on the difference between $\xi_\kappa + \langle \Map^2 \rangle + \langle \Map^3 \rangle$ and $\xi_\kappa + \langle \Map^2 \rangle$, which we show in Fig.~\ref{fig:Flamingo_variation_cosmology_2nd}, we see a huge difference in constraining power, particularly for $\Omega_\mathrm{m}$ and some of the BCM parameters such as $M_\mathrm{c}$. We note here that for $\xi_\kappa + \langle \Map^2 \rangle$ we use the best-fitting parameters obtained from $\xi_\kappa + \langle \Map^2 \rangle$ itself, which are also listed in Table \ref{tab:baryon_best}. Furthermore, we observe an enhancement in the scatter of the best-fitting parameters for $\xi_\kappa + \langle \Map^2 \rangle$, which indicates stronger projection effects. The reason is that the $\langle \Map^3 \rangle$ part of the data vector helps to constrain the BCM parameters, which in turn reduces the projection effects. Lastly, we notice the gain in fixing all BCM parameters except $M_\mathrm{c}$ is greater if we include $\langle \Map^3 \rangle$ to the data vector. The reason is that $\langle \Map^3 \rangle$ significantly helps to constrain the $M_\mathrm{c}$ parameter, which in turn helps to constrain cosmological parameters.

\begin{table}[h]
    \renewcommand{\arraystretch}{1.2}
    \centering
    \caption{Weighted average of the BCM best-fitting parameters over all FLAMINGO simulations considered in this work.}
    \begin{tabular}{l|c|c}
        & 2+3pt Analysis & 2pt Analysis \\
        \hline
        $\log_{10} (M_\mathrm{c} / h^{-1} \mathrm{M}_\odot)$       & 14.09 & 14.57 \\
        $\log_{10} (M_1 / h^{-1} \mathrm{M}_\odot)$               & 12.26 & 12.08 \\
        $\log_{10} (\beta)$                         & 0.35  & $-0.43$ \\
        $\log_{10} (\eta)$                           & $-0.32$ & $-0.42$ \\
        $\log_{10} (\theta_\mathrm{inn})$          & $-0.54$ & $-0.34$ \\
    \end{tabular}
    \label{tab:baryon_best}
\end{table}

\section{Conclusions}
\label{sec:conclusions}

This paper presents novel emulators for modelling the baryonic effects on the matter bispectrum and power spectrum, developed using high-resolution cosmological simulations. These emulators are designed to accurately predict the impact of baryonic processes such as gas cooling, star formation, and feedback from star formation and AGN on the matter power spectrum and bispectrum across a wide range of scales, redshifts, and cosmological scenarios. The ability to model these effects is crucial for large-scale structure surveys, such as \Euclid, which aims to extract cosmological information from the nonlinear regime of structure formation.

To construct the emulators, we exploit the high-resolution \texttt{BACCO} $N$-body simulations, which were post-processed to explore different cosmological scenarios using a cosmology rescaling algorithm. Since direct hydrodynamical simulations are computationally expensive, we incorporated baryonic effects such as gas cooling, star formation, and AGN feedback using the baryonification framework. This approach modifies the matter distribution in gravity-only simulations, effectively mimicking the impact of baryonic processes while maintaining computational efficiency.

Using a GPU-accelerated estimator, we then measured the matter power spectrum $P(k)$ and bispectrum $B(k_1, k_2, k_3)$ across a wide range of scales down to $k_\mathrm{max} \lesssim 20\, \ihMpc$, with shot noise contribution carefully subtracted. The resulting measurements formed the training set of our emulators.

We trained deep neural networks to emulate the baryon correction model of the power spectrum,
${S(k) = P_{\mathrm{Hydro}}(k)/P_{\mathrm{GrO}}(k)}$, and bispectrum $R(k_1, k_2, k_3) = B_{\mathrm{Hydro}}(k_1, k_2, k_3)/B_{\text{GrO}}(k_1, k_2, k_3)$, where `Hydro' refers to hydrodynamical simulations and `GrO' to Gravity-only simulations.
The training dataset consisted of 1200 simulations, sampled using a Latin Hypercube across five baryonic and three cosmological parameters. 

Our emulator is flexible enough to fit the state-of-the-art hydrodynamic FLAMINGO simulations and get unbiased cosmological results. It successfully captures the baryonic suppression of second- and third-order weak lensing statistics across various configurations and scales. This validation underscores the reliability of the emulator as a computationally efficient alternative to running full hydrodynamical simulations, which are prohibitively expensive for exploring the extensive parameter space required by surveys such as \Euclid. We note that the expected accuracy of cosmology scaling+baryonification for the power spectrum is approximately $1\%$ and $3\%$ for the bispectrum, such that the emulator serves as an alternative to hydrodynamical simulations as long as the statistical uncertainty of the data dominates over the modelling error.

Using the emulator, we provided forecasts for the impact of baryonic effects on the matter power and bispectrum measured by \Euclid. Our results highlight the importance of incorporating baryonic corrections when analysing future LSS data. Neglecting these effects could either lead to biased constraints on cosmological parameters of the order of $3\sigma$ in $S_8$ or to a significant reduction of the constraining power by up to $80\%$ for $\Omega_\mathrm{m}$, if the scales affected by baryons are excluded. Both cases would be fatal for the high-precision measurements \Euclid is aiming to deliver. By modelling these effects accurately, our emulator allows for more robust cosmological analyses and ensures that \Euclid can achieve its scientific objectives.

We plan to assess the emulators' accuracy against currently extrapolated scales in future work. This requires larger simulations and more resources to measure spectra from large grids. We also plan to employ our baryonification emulators to extend the forecast presented in this work to a mock \Euclid tomographic analysis. Finally, it would be interesting to validate with hydrodynamical simulations not only the constraints on cosmological parameters, but also the inferred baryonic parameters.

\begin{acknowledgements}
\AckEC  

This work used the DiRAC@Durham facility managed by the Institute for Computational Cosmology on behalf of the STFC DiRAC HPC Facility (\url{www.dirac.ac.uk}). The equipment was funded by BEIS capital funding via STFC capital grants ST/K00042X/1, ST/P002293/1, ST/R002371/1 and ST/S002502/1, Durham University and STFC operations grant ST/R000832/1. DiRAC is part of the National e-Infrastructure. 

This research was enabled in part by support provided by Calcul Quebec\footnote{\url{https://docs.alliancecan.ca/wiki/Narval/en}} and Compute Ontario\footnote{\url{https://docs.alliancecan.ca/wiki/Graham}} and the Digital Research Alliance of Canada\footnote{\url{https://www.alliancecan.ca/en}}. This research used resources from the National Energy Research Scientific Computing Center, supported by the Office of Science of the U.S. Department of Energy under Contract No. DE-AC02-05CH11231.

PB acknowledge financial support from the Canadian Space Agency (Grant 23EXPROSS1), the Waterloo Centre for Astrophysics and the National Science and Engineering Research Council Discovery Grants program. LL is supported by the Austrian Science Fund (FWF) [ESP 357-N]. REA acknowledges sup-
port from project PID2021-128338NB-I00 from the Spanish Ministry of Science
and support from the European Research Executive Agency HORIZON-MSCA-
2021-SE-01 Research and Innovation programme under the Marie Skłodowska-
Curie grant agreement number 101086388 (LACEGAL) MZ is supported by STFC. LP acknowledges support from the DLR grant 50QE2302.

We acknowledge the use of the following software: {\code NumPy} \citep{2020Natur.585..357H}, {\code Numba} \citep{2015llvm.confE...1L}, {\code SciPy} \citep{2020NatMe..17..261V}, {\code JAX} \citep{2021ascl.soft11002B}, {\code TreeCorr} \citep{Jarvis:2004}, {\code CosmoPower} \citep{COSMOPOWER2022}, {\code HEALPix} \citep{Gorskietal2005}. 
\end{acknowledgements}

\section*{Data availability}
The emulators developed in this work are publicly available at \url{https://baccoemu.readthedocs.io/en/latest/}.

% The best way to enter references is to use BibTeX:
\bibliographystyle{aa}
\bibliography{bibliography, Euclid} % if your bibtex file is called example.bib

@ARTICLE{EuclidSkyOverview,
author = {{Euclid Collaboration: Mellier}, Y. and {Abdurro'uf} and {Acevedo~Barroso}, J.A. and others},
	title = {Euclid - I. Overview of the Euclid mission},
	DOI= "10.1051/0004-6361/202450810",
	url= "https://doi.org/10.1051/0004-6361/202450810",
	journal = {A\&A},
	year = 2025,
	volume = 697,
	pages = "A1",
}

@ARTICLE{Ajani-EP29,
       author = {{Euclid Collaboration: Ajani}, V. and {Baldi}, M. and {Barthelemy}, A. and others},
        title = "{\Euclid preparation. XXVIII. Forecasts for ten different higher-order weak lensing statistics}",
      journal = {\aap},
         year = 2023,
        month = jul,
       volume = {675},
          eid = {A120},
        pages = {A120},
          doi = {10.1051/0004-6361/202346017},
archivePrefix = {arXiv},
       eprint = {2301.12890},
 primaryClass = {astro-ph.CO},
       adsurl = {https://ui.adsabs.harvard.edu/abs/2023A&A...675A.120E}
}

@ARTICLE{Blanchard-EP7,
       author = {{Euclid Collaboration: Blanchard}, A. and {Camera}, S. and {Carbone}, C. and others},
        title = "{Euclid preparation. VII. Forecast validation for Euclid cosmological probes}",
      journal = {\aap},
         year = 2020,
        month = oct,
       volume = {642},
          eid = {A191},
        pages = {A191},
          doi = {10.1051/0004-6361/202038071},
archivePrefix = {arXiv},
       eprint = {1910.09273},
 primaryClass = {astro-ph.CO},
       adsurl = {https://ui.adsabs.harvard.edu/abs/2020A&A...642A.191E}
}

@string{june = {June}}

@ARTICLE{2003A&A...397..405B,
       author = {{Bernardeau}, F. and {van Waerbeke}, L. and {Mellier}, Y.},
        title = "{Patterns in the weak shear 3-point correlation function}",
      journal = {\aap},
         year = 2003,
        month = jan,
       volume = {397},
        pages = {405-414},
          doi = {10.1051/0004-6361:20021567},
archivePrefix = {arXiv},
       eprint = {astro-ph/0201029},
 primaryClass = {astro-ph},
       adsurl = {https://ui.adsabs.harvard.edu/abs/2003A&A...397..405B}
}

@ARTICLE{2003MNRAS.344..857T,
       author = {{Takada}, Masahiro and {Jain}, Bhuvnesh},
        title = "{Three-point correlations in weak lensing surveys: model predictions and applications}",
      journal = {\mnras},
         year = 2003,
        month = sep,
       volume = {344},
       number = {3},
        pages = {857-886},
          doi = {10.1046/j.1365-8711.2003.06868.x},
archivePrefix = {arXiv},
       eprint = {astro-ph/0304034},
 primaryClass = {astro-ph},
       adsurl = {https://ui.adsabs.harvard.edu/abs/2003MNRAS.344..857T}
}

@ARTICLE{HSC2023,
       author = {{Dalal}, Roohi and {Li}, Xiangchong and {Nicola}, Andrina and {Zuntz}, Joe and {Strauss}, Michael A. and {Sugiyama}, Sunao and {Zhang}, Tianqing and {Rau}, Markus M. and {Mandelbaum}, Rachel and {Takada}, Masahiro and {More}, Surhud and {Miyatake}, Hironao and {Kannawadi}, Arun and {Shirasaki}, Masato and {Taniguchi}, Takanori and {Takahashi}, Ryuichi and {Osato}, Ken and {Hamana}, Takashi and {Oguri}, Masamune and {Nishizawa}, Atsushi J. and {Malag{\'o}n}, Andr{\'e}s A. Plazas and {Sunayama}, Tomomi and {Alonso}, David and {Slosar}, An{\v{z}}e and {Luo}, Wentao and {Armstrong}, Robert and {Bosch}, James and {Hsieh}, Bau-Ching and {Komiyama}, Yutaka and {Lupton}, Robert H. and {Lust}, Nate B. and {MacArthur}, Lauren A. and {Miyazaki}, Satoshi and {Murayama}, Hitoshi and {Nishimichi}, Takahiro and {Okura}, Yuki and {Price}, Paul A. and {Tait}, Philip J. and {Tanaka}, Masayuki and {Wang}, Shiang-Yu},
        title = "{Hyper Suprime-Cam Year 3 results: Cosmology from cosmic shear power spectra}",
      journal = {\prd},
         year = 2023,
        month = dec,
       volume = {108},
       number = {12},
          eid = {123519},
        pages = {123519},
          doi = {10.1103/PhysRevD.108.123519},
archivePrefix = {arXiv},
       eprint = {2304.00701},
 primaryClass = {astro-ph.CO},
       adsurl = {https://ui.adsabs.harvard.edu/abs/2023PhRvD.108l3519D}
}

@ARTICLE{2023PhRvD.108l3517M,
       author = {{Miyatake}, Hironao and {Sugiyama}, Sunao and {Takada}, Masahiro and {Nishimichi}, Takahiro and {Li}, Xiangchong and {Shirasaki}, Masato and {More}, Surhud and {Kobayashi}, Yosuke and {Nishizawa}, Atsushi J. and {Rau}, Markus M. and {Zhang}, Tianqing and {Takahashi}, Ryuichi and {Dalal}, Roohi and {Mandelbaum}, Rachel and {Strauss}, Michael A. and {Hamana}, Takashi and {Oguri}, Masamune and {Osato}, Ken and {Luo}, Wentao and {Kannawadi}, Arun and {Hsieh}, Bau-Ching and {Armstrong}, Robert and {Bosch}, James and {Komiyama}, Yutaka and {Lupton}, Robert H. and {Lust}, Nate B. and {MacArthur}, Lauren A. and {Miyazaki}, Satoshi and {Murayama}, Hitoshi and {Okura}, Yuki and {Price}, Paul A. and {Sunayama}, Tomomi and {Tait}, Philip J. and {Tanaka}, Masayuki and {Wang}, Shiang-Yu},
        title = "{Hyper Suprime-Cam Year 3 results: Cosmology from galaxy clustering and weak lensing with HSC and SDSS using the emulator based halo model}",
      journal = {\prd},
         year = 2023,
        month = dec,
       volume = {108},
       number = {12},
          eid = {123517},
        pages = {123517},
          doi = {10.1103/PhysRevD.108.123517},
archivePrefix = {arXiv},
       eprint = {2304.00704},
 primaryClass = {astro-ph.CO},
       adsurl = {https://ui.adsabs.harvard.edu/abs/2023PhRvD.108l3517M}
}

@ARTICLE{2022PhRvD.105b3514A,
       author = {{Amon}, A. and {Gruen}, D. and {Troxel}, M.~A. and {MacCrann}, N. and {Dodelson}, S. and {Choi}, A. and {Doux}, C. and {Secco}, L.~F. and {Samuroff}, S. and {Krause}, E. and {Cordero}, J. and {Myles}, J. and {DeRose}, J. and {Wechsler}, R.~H. and {Gatti}, M. and {Navarro-Alsina}, A. and {Bernstein}, G.~M. and {Jain}, B. and {Blazek}, J. and {Alarcon}, A. and {Fert{\'e}}, A. and {Lemos}, P. and {Raveri}, M. and {Campos}, A. and {Prat}, J. and {S{\'a}nchez}, C. and {Jarvis}, M. and {Alves}, O. and {Andrade-Oliveira}, F. and {Baxter}, E. and {Bechtol}, K. and {Becker}, M.~R. and {Bridle}, S.~L. and {Camacho}, H. and {Carnero Rosell}, A. and {Carrasco Kind}, M. and {Cawthon}, R. and {Chang}, C. and {Chen}, R. and {Chintalapati}, P. and {Crocce}, M. and {Davis}, C. and {Diehl}, H.~T. and {Drlica-Wagner}, A. and {Eckert}, K. and {Eifler}, T.~F. and {Elvin-Poole}, J. and {Everett}, S. and {Fang}, X. and {Fosalba}, P. and {Friedrich}, O. and {Gaztanaga}, E. and {Giannini}, G. and {Gruendl}, R.~A. and {Harrison}, I. and {Hartley}, W.~G. and {Herner}, K. and {Huang}, H. and {Huff}, E.~M. and {Huterer}, D. and {Kuropatkin}, N. and {Leget}, P. and {Liddle}, A.~R. and {McCullough}, J. and {Muir}, J. and {Pandey}, S. and {Park}, Y. and {Porredon}, A. and {Refregier}, A. and {Rollins}, R.~P. and {Roodman}, A. and {Rosenfeld}, R. and {Ross}, A.~J. and {Rykoff}, E.~S. and {Sanchez}, J. and {Sevilla-Noarbe}, I. and {Sheldon}, E. and {Shin}, T. and {Troja}, A. and {Tutusaus}, I. and {Tutusaus}, I. and {Varga}, T.~N. and {Weaverdyck}, N. and {Yanny}, B. and {Yin}, B. and {Zhang}, Y. and {Zuntz}, J. and {Aguena}, M. and {Allam}, S. and {Annis}, J. and {Bacon}, D. and {Bertin}, E. and {Bhargava}, S. and {Brooks}, D. and {Buckley-Geer}, E. and {Burke}, D.~L. and {Carretero}, J. and {Costanzi}, M. and {da Costa}, L.~N. and {Pereira}, M.~E.~S. and {De Vicente}, J. and {Desai}, S. and {Dietrich}, J.~P. and {Doel}, P. and {Ferrero}, I. and {Flaugher}, B. and {Frieman}, J. and {Garc{\'\i}a-Bellido}, J. and {Gaztanaga}, E. and {Gerdes}, D.~W. and {Giannantonio}, T. and {Gschwend}, J. and {Gutierrez}, G. and {Hinton}, S.~R. and {Hollowood}, D.~L. and {Honscheid}, K. and {Hoyle}, B. and {James}, D.~J. and {Kron}, R. and {Kuehn}, K. and {Lahav}, O. and {Lima}, M. and {Lin}, H. and {Maia}, M.~A.~G. and {Marshall}, J.~L. and {Martini}, P. and {Melchior}, P. and {Menanteau}, F. and {Miquel}, R. and {Mohr}, J.~J. and {Morgan}, R. and {Ogando}, R.~L.~C. and {Palmese}, A. and {Paz-Chinch{\'o}n}, F. and {Petravick}, D. and {Pieres}, A. and {Romer}, A.~K. and {Sanchez}, E. and {Scarpine}, V. and {Schubnell}, M. and {Serrano}, S. and {Smith}, M. and {Soares-Santos}, M. and {Tarle}, G. and {Thomas}, D. and {To}, C. and {Weller}, J. and {DES Collaboration}},
        title = "{Dark Energy Survey Year 3 results: Cosmology from cosmic shear and robustness to data calibration}",
      journal = {\prd},
         year = 2022,
        month = jan,
       volume = {105},
       number = {2},
          eid = {023514},
        pages = {023514},
          doi = {10.1103/PhysRevD.105.023514},
archivePrefix = {arXiv},
       eprint = {2105.13543},
 primaryClass = {astro-ph.CO},
       adsurl = {https://ui.adsabs.harvard.edu/abs/2022PhRvD.105b3514A}
}

@ARTICLE{2011MNRAS.417.2020S,
       author = {{Semboloni}, Elisabetta and {Hoekstra}, Henk and {Schaye}, Joop and {van Daalen}, Marcel P. and {McCarthy}, Ian G.},
        title = "{Quantifying the effect of baryon physics on weak lensing tomography}",
      journal = {\mnras},
         year = 2011,
        month = nov,
       volume = {417},
       number = {3},
        pages = {2020-2035},
          doi = {10.1111/j.1365-2966.2011.19385.x},
archivePrefix = {arXiv},
       eprint = {1105.1075},
 primaryClass = {astro-ph.CO},
       adsurl = {https://ui.adsabs.harvard.edu/abs/2011MNRAS.417.2020S}
}

@ARTICLE{LoVerde2008,
       author = {{LoVerde}, Marilena and {Afshordi}, Niayesh},
        title = "{Extended Limber approximation}",
      journal = {\prd},
         year = 2008,
        month = dec,
       volume = {78},
       number = {12},
          eid = {123506},
        pages = {123506},
          doi = {10.1103/PhysRevD.78.123506},
archivePrefix = {arXiv},
       eprint = {0809.5112},
 primaryClass = {astro-ph},
       adsurl = {https://ui.adsabs.harvard.edu/abs/2008PhRvD..78l3506L}
}

@article{Asgari:2020,
 adsurl = {https://ui.adsabs.harvard.edu/abs/2020A&A...634A.127A},
 archiveprefix = {arXiv},
 author = {{Asgari}, Marika and {Tr{\"o}ster}, Tilman and {Heymans}, Catherine and {Hildebrandt}, Hendrik and {van den Busch}, Jan Luca and {Wright}, Angus H. and {Choi}, Ami and {Erben}, Thomas and {Joachimi}, Benjamin and {Joudaki}, Shahab and {Kannawadi}, Arun and {Kuijken}, Konrad and {Lin}, Chieh-An and {Schneider}, Peter and {Zuntz}, Joe},
 doi = {10.1051/0004-6361/201936512},
 eid = {A127},
 eprint = {1910.05336},
 journal = {\aap},
 month = {February},
 pages = {A127},
 primaryclass = {astro-ph.CO},
 title = {{KiDS+VIKING-450 and DES-Y1 combined: Mitigating baryon feedback uncertainty with COSEBIs}},
 volume = {634},
 year = {2020}
}

@ARTICLE{Gorskietal2005,
   author = {{G{\'o}rski}, K.~M. and {Hivon}, E. and {Banday}, A.~J. and 
	{Wandelt}, B.~D. and {Hansen}, F.~K. and {Reinecke}, M. and 
	{Bartelmann}, M.},
    title = "{HEALPix: A Framework for High-Resolution Discretization and Fast Analysis of Data Distributed on the Sphere}",
  journal = {\apj},
   eprint = {arXiv:astro-ph/0409513},
     year = 2005,
    month = apr,
   volume = 622,
    pages = {759-771},
      doi = {10.1086/427976},
   adsurl = {http://adsabs.harvard.edu/abs/2005ApJ...622..759G}
}

@article{Bartelmann:2001,
 adsurl = {https://ui.adsabs.harvard.edu/abs/2001PhR...340..291B},
 archiveprefix = {arXiv},
 author = {{Bartelmann}, M. and {Schneider}, P.},
 doi = {10.1016/S0370-1573(00)00082-X},
 eprint = {astro-ph/9912508},
 journal = {\physrep},
 month = {January},
 number = {4-5},
 pages = {291-472},
 primaryclass = {astro-ph},
 title = {{Weak gravitational lensing}},
 volume = {340},
 year = {2001}
}

@article{Bartelmann:2010,
 adsurl = {https://ui.adsabs.harvard.edu/abs/2010CQGra..27w3001B},
 archiveprefix = {arXiv},
 author = {{Bartelmann}, Matthias},
 doi = {10.1088/0264-9381/27/23/233001},
 eid = {233001},
 eprint = {1010.3829},
 journal = {Classical and Quantum Gravity},
 month = {December},
 number = {23},
 pages = {233001},
 primaryclass = {astro-ph.CO},
 title = {{TOPICAL REVIEW Gravitational lensing}},
 volume = {27},
 year = {2010}
}

@article{Bernardeau:1997,
 adsurl = {https://ui.adsabs.harvard.edu/abs/1997A&A...322....1B},
 archiveprefix = {arXiv},
 author = {{Bernardeau}, F. and {van Waerbeke}, L. and {Mellier}, Y.},
 eprint = {astro-ph/9609122},
 journal = {\aap},
 month = {June},
 pages = {1-18},
 primaryclass = {astro-ph},
 title = {{Weak lensing statistics as a probe of \{OMEGA\} and power spectrum.}},
 volume = {322},
 year = {1997}
}

@ARTICLE{Burger2023,
       author = {{Burger}, Pierre A. and {Friedrich}, Oliver and {Harnois-D{\'e}raps}, Joachim and {Schneider}, Peter and {Asgari}, Marika and {Bilicki}, Maciej and {Hildebrandt}, Hendrik and {Wright}, Angus H. and {Castro}, Tiago and {Dolag}, Klaus and {Heymans}, Catherine and {Joachimi}, Benjamin and {Kuijken}, Konrad and {Martinet}, Nicolas and {Shan}, HuanYuan and {Tr{\"o}ster}, Tilman},
        title = "{KiDS-1000 cosmology: Constraints from density split statistics}",
      journal = {\aap},
         year = 2023,
        month = jan,
       volume = {669},
          eid = {A69},
        pages = {A69},
          doi = {10.1051/0004-6361/202244673},
archivePrefix = {arXiv},
       eprint = {2208.02171},
 primaryClass = {astro-ph.CO},
       adsurl = {https://ui.adsabs.harvard.edu/abs/2023A&A...669A..69B}
}

@article{COSMOPOWER2022,
 adsurl = {https://ui.adsabs.harvard.edu/abs/2022MNRAS.511.1771S},
 archiveprefix = {arXiv},
 author = {{Spurio Mancini}, Alessio and {Piras}, Davide and {Alsing}, Justin and {Joachimi}, Benjamin and {Hobson}, Michael P.},
 doi = {10.1093/mnras/stac064},
 eprint = {2106.03846},
 journal = {\mnras},
 month = {April},
 number = {2},
 pages = {1771-1788},
 primaryclass = {astro-ph.CO},
 title = {{COSMOPOWER: emulating cosmological power spectra for accelerated Bayesian inference from next-generation surveys}},
 volume = {511},
 year = {2022}
}

@article{Crittenden:2002,
 adsurl = {https://ui.adsabs.harvard.edu/abs/2002ApJ...568...20C},
 archiveprefix = {arXiv},
 author = {{Crittenden}, Robert G. and {Natarajan}, Priyamvada and {Pen}, Ue-Li and {Theuns}, Tom},
 doi = {10.1086/338838},
 eprint = {astro-ph/0012336},
 journal = {\apj},
 month = {March},
 number = {1},
 pages = {20-27},
 primaryclass = {astro-ph},
 title = {{Discriminating Weak Lensing from Intrinsic Spin Correlations Using the Curl-Gradient Decomposition}},
 volume = {568},
 year = {2002}
}

@ARTICLE{Schaller2024,
       author = {{Schaller}, Matthieu and {Borrow}, Josh and {Draper}, Peter W. and {Ivkovic}, Mladen and {McAlpine}, Stuart and {Vandenbroucke}, Bert and {Bah{\'e}}, Yannick and {Chaikin}, Evgenii and {Chalk}, Aidan B.~G. and {Chan}, Tsang Keung and {Correa}, Camila and {van Daalen}, Marcel and {Elbers}, Willem and {Gonnet}, Pedro and {Hausammann}, Lo{\"\i}c and {Helly}, John and {Hu{\v{s}}ko}, Filip and {Kegerreis}, Jacob A. and {Nobels}, Folkert S.~J. and {Ploeckinger}, Sylvia and {Revaz}, Yves and {Roper}, William J. and {Ruiz-Bonilla}, Sergio and {Sandnes}, Thomas D. and {Uyttenhove}, Yolan and {Willis}, James S. and {Xiang}, Zhen},
        title = "{SWIFT: A modern highly-parallel gravity and smoothed particle hydrodynamics solver for astrophysical and cosmological applications}",
      journal = {\mnras},
         year = 2024,
        month = may,
       volume = {530},
       number = {2},
        pages = {2378-2419},
          doi = {10.1093/mnras/stae922},
archivePrefix = {arXiv},
       eprint = {2305.13380},
 primaryClass = {astro-ph.IM},
       adsurl = {https://ui.adsabs.harvard.edu/abs/2024MNRAS.530.2378S}
}

@ARTICLE{Borrow2022,
       author = {{Borrow}, Josh and {Schaller}, Matthieu and {Bower}, Richard G. and {Schaye}, Joop},
        title = "{SPHENIX: smoothed particle hydrodynamics for the next generation of galaxy formation simulations}",
      journal = {\mnras},
         year = 2022,
        month = apr,
       volume = {511},
       number = {2},
        pages = {2367-2389},
          doi = {10.1093/mnras/stab3166},
archivePrefix = {arXiv},
       eprint = {2012.03974},
 primaryClass = {astro-ph.GA},
       adsurl = {https://ui.adsabs.harvard.edu/abs/2022MNRAS.511.2367B}
}

@ARTICLE{Husko2022,
       author = {{Hu{\v{s}}ko}, Filip and {Lacey}, Cedric G. and {Schaye}, Joop and {Schaller}, Matthieu and {Nobels}, Folkert S.~J.},
        title = "{Spin-driven jet feedback in idealized simulations of galaxy groups and clusters}",
      journal = {\mnras},
         year = 2022,
        month = nov,
       volume = {516},
       number = {3},
        pages = {3750-3772},
          doi = {10.1093/mnras/stac2278},
archivePrefix = {arXiv},
       eprint = {2206.06402},
 primaryClass = {astro-ph.GA},
       adsurl = {https://ui.adsabs.harvard.edu/abs/2022MNRAS.516.3750H}
}

@ARTICLE{Chaikin2023,
       author = {{Chaikin}, Evgenii and {Schaye}, Joop and {Schaller}, Matthieu and {Ben{\'\i}tez-Llambay}, Alejandro and {Nobels}, Folkert S.~J. and {Ploeckinger}, Sylvia},
        title = "{A thermal-kinetic subgrid model for supernova feedback in simulations of galaxy formation}",
      journal = {\mnras},
         year = 2023,
        month = aug,
       volume = {523},
       number = {3},
        pages = {3709-3731},
          doi = {10.1093/mnras/stad1626},
archivePrefix = {arXiv},
       eprint = {2211.04619},
 primaryClass = {astro-ph.GA},
       adsurl = {https://ui.adsabs.harvard.edu/abs/2023MNRAS.523.3709C}
}

@ARTICLE{Wiersma2009,
       author = {{Wiersma}, Robert P.~C. and {Schaye}, Joop and {Theuns}, Tom and {Dalla Vecchia}, Claudio and {Tornatore}, Luca},
        title = "{Chemical enrichment in cosmological, smoothed particle hydrodynamics simulations}",
      journal = {\mnras},
         year = 2009,
        month = oct,
       volume = {399},
       number = {2},
        pages = {574-600},
          doi = {10.1111/j.1365-2966.2009.15331.x},
archivePrefix = {arXiv},
       eprint = {0902.1535},
 primaryClass = {astro-ph.CO},
       adsurl = {https://ui.adsabs.harvard.edu/abs/2009MNRAS.399..574W}
}

@ARTICLE{Ploeckinger2020,
       author = {{Ploeckinger}, Sylvia and {Schaye}, Joop},
        title = "{Radiative cooling rates, ion fractions, molecule abundances, and line emissivities including self-shielding and both local and metagalactic radiation fields}",
      journal = {\mnras},
         year = 2020,
        month = oct,
       volume = {497},
       number = {4},
        pages = {4857-4883},
          doi = {10.1093/mnras/staa2172},
archivePrefix = {arXiv},
       eprint = {2006.14322},
 primaryClass = {astro-ph.GA},
       adsurl = {https://ui.adsabs.harvard.edu/abs/2020MNRAS.497.4857P}
}

@ARTICLE{Lange2023,
       author = {{Lange}, Johannes U.},
        title = "{NAUTILUS: boosting Bayesian importance nested sampling with deep learning}",
      journal = {\mnras},
         year = 2023,
        month = oct,
       volume = {525},
       number = {2},
        pages = {3181-3194},
          doi = {10.1093/mnras/stad2441},
archivePrefix = {arXiv},
       eprint = {2306.16923},
 primaryClass = {astro-ph.IM},
       adsurl = {https://ui.adsabs.harvard.edu/abs/2023MNRAS.525.3181L}
}

@ARTICLE{2015A&C....13...12N,
       author = {{Nelson}, D. and {Pillepich}, A. and {Genel}, S. and {Vogelsberger}, M. and {Springel}, V. and {Torrey}, P. and {Rodriguez-Gomez}, V. and {Sijacki}, D. and {Snyder}, G.~F. and {Griffen}, B. and {Marinacci}, F. and {Blecha}, L. and {Sales}, L. and {Xu}, D. and {Hernquist}, L.},
        title = "{The illustris simulation: Public data release}",
      journal = {Astronomy and Computing},
         year = 2015,
        month = nov,
       volume = {13},
        pages = {12-37},
          doi = {10.1016/j.ascom.2015.09.003},
archivePrefix = {arXiv},
       eprint = {1504.00362},
 primaryClass = {astro-ph.CO},
       adsurl = {https://ui.adsabs.harvard.edu/abs/2015A&C....13...12N}
}

@ARTICLE{2014MNRAS.441.1270L,
       author = {{Le Brun}, Amandine M.~C. and {McCarthy}, Ian G. and {Schaye}, Joop and {Ponman}, Trevor J.},
        title = "{Towards a realistic population of simulated galaxy groups and clusters}",
      journal = {\mnras},
         year = 2014,
        month = jun,
       volume = {441},
       number = {2},
        pages = {1270-1290},
          doi = {10.1093/mnras/stu608},
archivePrefix = {arXiv},
       eprint = {1312.5462},
 primaryClass = {astro-ph.CO},
       adsurl = {https://ui.adsabs.harvard.edu/abs/2014MNRAS.441.1270L}
}

@ARTICLE{2018MNRAS.476.2999M,
       author = {{McCarthy}, Ian G. and {Bird}, Simeon and {Schaye}, Joop and {Harnois-Deraps}, Joachim and {Font}, Andreea S. and {van Waerbeke}, Ludovic},
        title = "{The BAHAMAS project: the CMB-large-scale structure tension and the roles of massive neutrinos and galaxy formation}",
      journal = {\mnras},
         year = 2018,
        month = may,
       volume = {476},
       number = {3},
        pages = {2999-3030},
          doi = {10.1093/mnras/sty377},
archivePrefix = {arXiv},
       eprint = {1712.02411},
 primaryClass = {astro-ph.CO},
       adsurl = {https://ui.adsabs.harvard.edu/abs/2018MNRAS.476.2999M}
}

@ARTICLE{2024MNRAS.528.2308P,
       author = {{Pakmor}, R{\"u}diger and {Bieri}, Rebekka and {van de Voort}, Freeke and {Werhahn}, Maria and {Fattahi}, Azadeh and {Guillet}, Thomas and {Pfrommer}, Christoph and {Springel}, Volker and {Talbot}, Rosie Y.},
        title = "{Magnetic field amplification in cosmological zoom simulations from dwarf galaxies to galaxy groups}",
      journal = {\mnras},
         year = 2024,
        month = feb,
       volume = {528},
       number = {2},
        pages = {2308-2325},
          doi = {10.1093/mnras/stae112},
archivePrefix = {arXiv},
       eprint = {2309.13104},
 primaryClass = {astro-ph.GA},
       adsurl = {https://ui.adsabs.harvard.edu/abs/2024MNRAS.528.2308P}
}

@ARTICLE{2015MNRAS.446..521S,
       author = {{Schaye}, Joop and {Crain}, Robert A. and {Bower}, Richard G. and {Furlong}, Michelle and {Schaller}, Matthieu and {Theuns}, Tom and {Dalla Vecchia}, Claudio and {Frenk}, Carlos S. and {McCarthy}, I.~G. and {Helly}, John C. and {Jenkins}, Adrian and {Rosas-Guevara}, Y.~M. and {White}, Simon D.~M. and {Baes}, Maarten and {Booth}, C.~M. and {Camps}, Peter and {Navarro}, Julio F. and {Qu}, Yan and {Rahmati}, Alireza and {Sawala}, Till and {Thomas}, Peter A. and {Trayford}, James},
        title = "{The EAGLE project: simulating the evolution and assembly of galaxies and their environments}",
      journal = {\mnras},
         year = 2015,
        month = jan,
       volume = {446},
       number = {1},
        pages = {521-554},
          doi = {10.1093/mnras/stu2058},
archivePrefix = {arXiv},
       eprint = {1407.7040},
 primaryClass = {astro-ph.GA},
       adsurl = {https://ui.adsabs.harvard.edu/abs/2015MNRAS.446..521S}
}

@ARTICLE{2020MNRAS.498.2887F,
       author = {{Foreman}, Simon and {Coulton}, William and {Villaescusa-Navarro}, Francisco and {Barreira}, Alexandre},
        title = "{Baryonic effects on the matter bispectrum}",
      journal = {\mnras},
         year = 2020,
        month = oct,
       volume = {498},
       number = {2},
        pages = {2887-2911},
          doi = {10.1093/mnras/staa2523},
archivePrefix = {arXiv},
       eprint = {1910.03597},
 primaryClass = {astro-ph.CO},
       adsurl = {https://ui.adsabs.harvard.edu/abs/2020MNRAS.498.2887F}
}

@ARTICLE{2019JCAP...03..020S,
       author = {{Schneider}, Aurel and {Teyssier}, Romain and {Stadel}, Joachim and {Chisari}, Nora Elisa and {Le Brun}, Amandine M.~C. and {Amara}, Adam and {Refregier}, Alexandre},
        title = "{Quantifying baryon effects on the matter power spectrum and the weak lensing shear correlation}",
      journal = {JCAP},
         year = 2019,
        month = mar,
       volume = {03},
       number = {3},
          eid = {020},
        pages = {020},
          doi = {10.1088/1475-7516/2019/03/020},
archivePrefix = {arXiv},
       eprint = {1810.08629},
 primaryClass = {astro-ph.CO},
       adsurl = {https://ui.adsabs.harvard.edu/abs/2019JCAP...03..020S}
}

@ARTICLE{2021MNRAS.503.3596A,
       author = {{Aric{\`o}}, Giovanni and {Angulo}, Raul E. and {Hern{\'a}ndez-Monteagudo}, Carlos and {Contreras}, Sergio and {Zennaro}, Matteo},
        title = "{Simultaneous modelling of matter power spectrum and bispectrum in the presence of baryons}",
      journal = {\mnras},
         year = 2021,
        month = may,
       volume = {503},
       number = {3},
        pages = {3596-3609},
          doi = {10.1093/mnras/stab699},
archivePrefix = {arXiv},
       eprint = {2009.14225},
 primaryClass = {astro-ph.CO},
       adsurl = {https://ui.adsabs.harvard.edu/abs/2021MNRAS.503.3596A}
}

@ARTICLE{2013ApJ...770...57B,
       author = {{Behroozi}, Peter S. and {Wechsler}, Risa H. and {Conroy}, Charlie},
        title = "{The Average Star Formation Histories of Galaxies in Dark Matter Halos from z = 0-8}",
      journal = {\apj},
         year = 2013,
        month = jun,
       volume = {770},
       number = {1},
          eid = {57},
        pages = {57},
          doi = {10.1088/0004-637X/770/1/57},
archivePrefix = {arXiv},
       eprint = {1207.6105},
 primaryClass = {astro-ph.CO},
       adsurl = {https://ui.adsabs.harvard.edu/abs/2013ApJ...770...57B}
}

@ARTICLE{2020MNRAS.495.4800A,
       author = {{Aric{\`o}}, Giovanni and {Angulo}, Raul E. and {Hern{\'a}ndez-Monteagudo}, Carlos and {Contreras}, Sergio and {Zennaro}, Matteo and {Pellejero-Iba{\~n}ez}, Marcos and {Rosas-Guevara}, Yetli},
        title = "{Modelling the large-scale mass density field of the universe as a function of cosmology and baryonic physics}",
      journal = {\mnras},
         year = 2020,
        month = jul,
       volume = {495},
       number = {4},
        pages = {4800-4819},
          doi = {10.1093/mnras/staa1478},
archivePrefix = {arXiv},
       eprint = {1911.08471},
 primaryClass = {astro-ph.CO},
       adsurl = {https://ui.adsabs.harvard.edu/abs/2020MNRAS.495.4800A}
}

@ARTICLE{2020Natur.585..357H,
       author = {{Harris}, Charles R. and {Millman}, K. Jarrod and {van der Walt}, St{\'e}fan J. and {Gommers}, Ralf and {Virtanen}, Pauli and {Cournapeau}, David and {Wieser}, Eric and {Taylor}, Julian and {Berg}, Sebastian and {Smith}, Nathaniel J. and {Kern}, Robert and {Picus}, Matti and {Hoyer}, Stephan and {van Kerkwijk}, Marten H. and {Brett}, Matthew and {Haldane}, Allan and {del R{\'\i}o}, Jaime Fern{\'a}ndez and {Wiebe}, Mark and {Peterson}, Pearu and {G{\'e}rard-Marchant}, Pierre and {Sheppard}, Kevin and {Reddy}, Tyler and {Weckesser}, Warren and {Abbasi}, Hameer and {Gohlke}, Christoph and {Oliphant}, Travis E.},
        title = "{Array programming with NumPy}",
      journal = {\nat},
         year = 2020,
        month = sep,
       volume = {585},
       number = {7825},
        pages = {357-362},
          doi = {10.1038/s41586-020-2649-2},
archivePrefix = {arXiv},
       eprint = {2006.10256},
 primaryClass = {cs.MS},
       adsurl = {https://ui.adsabs.harvard.edu/abs/2020Natur.585..357H}
}

@ARTICLE{2020NatMe..17..261V,
       author = {{Virtanen}, Pauli and {Gommers}, Ralf and {Oliphant}, Travis E. and {Haberland}, Matt and {Reddy}, Tyler and {Cournapeau}, David and {Burovski}, Evgeni and {Peterson}, Pearu and {Weckesser}, Warren and {Bright}, Jonathan and {van der Walt}, St{\'e}fan J. and {Brett}, Matthew and {Wilson}, Joshua and {Millman}, K. Jarrod and {Mayorov}, Nikolay and {Nelson}, Andrew R.~J. and {Jones}, Eric and {Kern}, Robert and {Larson}, Eric and {Carey}, C.~J. and {Polat}, {\.I}lhan and {Feng}, Yu and {Moore}, Eric W. and {VanderPlas}, Jake and {Laxalde}, Denis and {Perktold}, Josef and {Cimrman}, Robert and {Henriksen}, Ian and {Quintero}, E.~A. and {Harris}, Charles R. and {Archibald}, Anne M. and {Ribeiro}, Ant{\^o}nio H. and {Pedregosa}, Fabian and {van Mulbregt}, Paul and {SciPy 1. 0 Contributors}},
        title = "{SciPy 1.0: fundamental algorithms for scientific computing in Python}",
      journal = {Nature Methods},
         year = 2020,
        month = feb,
       volume = {17},
        pages = {261-272},
          doi = {10.1038/s41592-019-0686-2},
archivePrefix = {arXiv},
       eprint = {1907.10121},
 primaryClass = {cs.MS},
       adsurl = {https://ui.adsabs.harvard.edu/abs/2020NatMe..17..261V}
}

@INPROCEEDINGS{2015llvm.confE...1L,
       author = {{Lam}, Siu Kwan and {Pitrou}, Antoine and {Seibert}, Stanley},
        title = "{Numba: A LLVM-based Python JIT Compiler}",
    booktitle = {Proc. Second Workshop on the LLVM Compiler Infrastructure in HPC},
         year = 2015,
        month = nov,
        pages = {1-6},
          doi = {10.1145/2833157.2833162},
       adsurl = {https://ui.adsabs.harvard.edu/abs/2015llvm.confE...1L}
}

@ARTICLE{2021JCAP...12..046G,
       author = {{Giri}, Sambit K. and {Schneider}, Aurel},
        title = "{Emulation of baryonic effects on the matter power spectrum and constraints from galaxy cluster data}",
      journal = {JCAP},
         year = 2021,
        month = dec,
       volume = {12},
       number = {12},
          eid = {046},
        pages = {046},
          doi = {10.1088/1475-7516/2021/12/046},
archivePrefix = {arXiv},
       eprint = {2108.08863},
 primaryClass = {astro-ph.CO},
       adsurl = {https://ui.adsabs.harvard.edu/abs/2021JCAP...12..046G}
}

@ARTICLE{2021MNRAS.506.4070A,
       author = {{Aric{\`o}}, Giovanni and {Angulo}, Raul E. and {Contreras}, Sergio and {Ondaro-Mallea}, Lurdes and {Pellejero-Iba{\~n}ez}, Marcos and {Zennaro}, Matteo},
        title = "{The BACCO simulation project: a baryonification emulator with neural networks}",
      journal = {\mnras},
         year = 2021,
        month = sep,
       volume = {506},
       number = {3},
        pages = {4070-4082},
          doi = {10.1093/mnras/stab1911},
archivePrefix = {arXiv},
       eprint = {2011.15018},
 primaryClass = {astro-ph.CO},
       adsurl = {https://ui.adsabs.harvard.edu/abs/2021MNRAS.506.4070A}
}

@ARTICLE{2015JCAP...12..049S,
       author = {{Schneider}, Aurel and {Teyssier}, Romain},
        title = "{A new method to quantify the effects of baryons on the matter power spectrum}",
      journal = {JCAP},
         year = 2015,
        month = dec,
       volume = {12},
       number = {12},
        pages = {049-049},
          doi = {10.1088/1475-7516/2015/12/049},
archivePrefix = {arXiv},
       eprint = {1510.06034},
 primaryClass = {astro-ph.CO},
       adsurl = {https://ui.adsabs.harvard.edu/abs/2015JCAP...12..049S}
}

@ARTICLE{2016JCAP...04..047S,
       author = {{Schneider}, Aurel and {Teyssier}, Romain and {Potter}, Doug and {Stadel}, Joachim and {Onions}, Julian and {Reed}, Darren S. and {Smith}, Robert E. and {Springel}, Volker and {Pearce}, Frazer R. and {Scoccimarro}, Roman},
        title = "{Matter power spectrum and the challenge of percent accuracy}",
      journal = {JCAP},
         year = 2016,
        month = apr,
       volume = {04},
       number = {4},
          eid = {047},
        pages = {047},
          doi = {10.1088/1475-7516/2016/04/047},
archivePrefix = {arXiv},
       eprint = {1503.05920},
 primaryClass = {astro-ph.CO},
       adsurl = {https://ui.adsabs.harvard.edu/abs/2016JCAP...04..047S}
}

@ARTICLE{2018MNRAS.480.3962C,
       author = {{Chisari}, N.~E. and {Richardson}, M.~L.~A. and {Devriendt}, J. and {Dubois}, Y. and {Schneider}, A. and {Le Brun}, A.~M.~C. and {Beckmann}, R.~S. and {Peirani}, S. and {Slyz}, A. and {Pichon}, C.},
        title = "{The impact of baryons on the matter power spectrum from the Horizon-AGN cosmological hydrodynamical simulation}",
      journal = {\mnras},
         year = 2018,
        month = nov,
       volume = {480},
       number = {3},
        pages = {3962-3977},
          doi = {10.1093/mnras/sty2093},
archivePrefix = {arXiv},
       eprint = {1801.08559},
 primaryClass = {astro-ph.CO},
       adsurl = {https://ui.adsabs.harvard.edu/abs/2018MNRAS.480.3962C}
}

@ARTICLE{2021MNRAS.507.5869A,
       author = {{Angulo}, Raul E. and {Zennaro}, Matteo and {Contreras}, Sergio and {Aric{\`o}}, Giovanni and {Pellejero-Iba{\~n}ez}, Marcos and {St{\"u}cker}, Jens},
        title = "{The BACCO simulation project: exploiting the full power of large-scale structure for cosmology}",
      journal = {\mnras},
         year = 2021,
        month = nov,
       volume = {507},
       number = {4},
        pages = {5869-5881},
          doi = {10.1093/mnras/stab2018},
archivePrefix = {arXiv},
       eprint = {2004.06245},
 primaryClass = {astro-ph.CO},
       adsurl = {https://ui.adsabs.harvard.edu/abs/2021MNRAS.507.5869A}
}

@ARTICLE{2012ApJ...761..152T,
       author = {{Takahashi}, Ryuichi and {Sato}, Masanori and {Nishimichi}, Takahiro and {Taruya}, Atsushi and {Oguri}, Masamune},
        title = "{Revising the Halofit Model for the Nonlinear Matter Power Spectrum}",
      journal = {\apj},
         year = 2012,
        month = dec,
       volume = {761},
       number = {2},
          eid = {152},
        pages = {152},
          doi = {10.1088/0004-637X/761/2/152},
archivePrefix = {arXiv},
       eprint = {1208.2701},
 primaryClass = {astro-ph.CO},
       adsurl = {https://ui.adsabs.harvard.edu/abs/2012ApJ...761..152T}
}

@ARTICLE{2010MNRAS.405..143A,
       author = {{Angulo}, R.~E. and {White}, S.~D.~M.},
        title = "{One simulation to fit them all - changing the background parameters of a cosmological N-body simulation}",
      journal = {\mnras},
         year = 2010,
        month = jun,
       volume = {405},
       number = {1},
        pages = {143-154},
          doi = {10.1111/j.1365-2966.2010.16459.x},
archivePrefix = {arXiv},
       eprint = {0912.4277},
 primaryClass = {astro-ph.CO},
       adsurl = {https://ui.adsabs.harvard.edu/abs/2010MNRAS.405..143A}
}

@ARTICLE{2012MNRAS.426.2046A,
       author = {{Angulo}, R.~E. and {Springel}, V. and {White}, S.~D.~M. and {Jenkins}, A. and {Baugh}, C.~M. and {Frenk}, C.~S.},
        title = "{Scaling relations for galaxy clusters in the Millennium-XXL simulation}",
      journal = {\mnras},
         year = 2012,
        month = nov,
       volume = {426},
       number = {3},
        pages = {2046-2062},
          doi = {10.1111/j.1365-2966.2012.21830.x},
archivePrefix = {arXiv},
       eprint = {1203.3216},
 primaryClass = {astro-ph.CO},
       adsurl = {https://ui.adsabs.harvard.edu/abs/2012MNRAS.426.2046A}
}

@ARTICLE{2016MNRAS.462L...1A,
       author = {{Angulo}, Raul E. and {Pontzen}, Andrew},
        title = "{Cosmological N-body simulations with suppressed variance}",
      journal = {\mnras},
         year = 2016,
        month = oct,
       volume = {462},
       number = {1},
        pages = {L1-L5},
          doi = {10.1093/mnrasl/slw098},
archivePrefix = {arXiv},
       eprint = {1603.05253},
 primaryClass = {astro-ph.CO},
       adsurl = {https://ui.adsabs.harvard.edu/abs/2016MNRAS.462L...1A}
}

@ARTICLE{Schaye2008,
       author = {{Schaye}, Joop and {Dalla Vecchia}, Claudio},
        title = "{On the relation between the Schmidt and Kennicutt-Schmidt star formation laws and its implications for numerical simulations}",
      journal = {\mnras},
         year = 2008,
        month = jan,
       volume = {383},
       number = {3},
        pages = {1210-1222},
          doi = {10.1111/j.1365-2966.2007.12639.x},
archivePrefix = {arXiv},
       eprint = {0709.0292},
 primaryClass = {astro-ph},
       adsurl = {https://ui.adsabs.harvard.edu/abs/2008MNRAS.383.1210S}
}

@ARTICLE{Booth2009,
       author = {{Booth}, C.~M. and {Schaye}, Joop},
        title = "{Cosmological simulations of the growth of supermassive black holes and feedback from active galactic nuclei: method and tests}",
      journal = {\mnras},
         year = 2009,
        month = sep,
       volume = {398},
       number = {1},
        pages = {53-74},
          doi = {10.1111/j.1365-2966.2009.15043.x},
archivePrefix = {arXiv},
       eprint = {0904.2572},
 primaryClass = {astro-ph.CO},
       adsurl = {https://ui.adsabs.harvard.edu/abs/2009MNRAS.398...53B}
}

@ARTICLE{Elbers2021,
       author = {{Elbers}, Willem and {Frenk}, Carlos S. and {Jenkins}, Adrian and {Li}, Baojiu and {Pascoli}, Silvia},
        title = "{An optimal non-linear method for simulating relic neutrinos}",
      journal = {\mnras},
         year = 2021,
        month = oct,
       volume = {507},
       number = {2},
        pages = {2614-2631},
          doi = {10.1093/mnras/stab2260},
archivePrefix = {arXiv},
       eprint = {2010.07321},
 primaryClass = {astro-ph.CO},
       adsurl = {https://ui.adsabs.harvard.edu/abs/2021MNRAS.507.2614E}
}

@article{DES2021,
 adsurl = {https://ui.adsabs.harvard.edu/abs/2022PhRvD.105b3520A},
 archiveprefix = {arXiv},
 author = {{DES Collaboration} and {Abbott}, T.~M.~C. and {Aguena}, M. and {Alarcon}, A. and {Allam}, S. and {Alves}, O. and {Amon}, A. and {Andrade-Oliveira}, F. and {Annis}, J. and {Avila}, S. and {Bacon}, D. and {Baxter}, E. and {Bechtol}, K. and {Becker}, M.~R. and {Bernstein}, G.~M. and {Bhargava}, S. and {Birrer}, S. and {Blazek}, J. and {Brandao-Souza}, A. and {Bridle}, S.~L. and {Brooks}, D. and {Buckley-Geer}, E. and {Burke}, D.~L. and {Camacho}, H. and {Campos}, A. and {Carnero Rosell}, A. and {Carrasco Kind}, M. and {Carretero}, J. and {Castander}, F.~J. and {Cawthon}, R. and {Chang}, C. and {Chen}, A. and {Chen}, R. and {Choi}, A. and {Conselice}, C. and {Cordero}, J. and {Costanzi}, M. and {Crocce}, M. and {da Costa}, L.~N. and {da Silva Pereira}, M.~E. and {Davis}, C. and {Davis}, T.~M. and {De Vicente}, J. and {DeRose}, J. and {Desai}, S. and {Di Valentino}, E. and {Diehl}, H.~T. and {Dietrich}, J.~P. and {Dodelson}, S. and {Doel}, P. and {Doux}, C. and {Drlica-Wagner}, A. and {Eckert}, K. and {Eifler}, T.~F. and {Elsner}, F. and {Elvin-Poole}, J. and {Everett}, S. and {Evrard}, A.~E. and {Fang}, X. and {Farahi}, A. and {Fernandez}, E. and {Ferrero}, I. and {Fert{\'e}}, A. and {Fosalba}, P. and {Friedrich}, O. and {Frieman}, J. and {Garc{\'\i}a-Bellido}, J. and {Gatti}, M. and {Gaztanaga}, E. and {Gerdes}, D.~W. and {Giannantonio}, T. and {Giannini}, G. and {Gruen}, D. and {Gruendl}, R.~A. and {Gschwend}, J. and {Gutierrez}, G. and {Harrison}, I. and {Hartley}, W.~G. and {Herner}, K. and {Hinton}, S.~R. and {Hollowood}, D.~L. and {Honscheid}, K. and {Hoyle}, B. and {Huff}, E.~M. and {Huterer}, D. and {Jain}, B. and {James}, D.~J. and {Jarvis}, M. and {Jeffrey}, N. and {Jeltema}, T. and {Kovacs}, A. and {Krause}, E. and {Kron}, R. and {Kuehn}, K. and {Kuropatkin}, N. and {Lahav}, O. and {Leget}, P. -F. and {Lemos}, P. and {Liddle}, A.~R. and {Lidman}, C. and {Lima}, M. and {Lin}, H. and {MacCrann}, N. and {Maia}, M.~A.~G. and {Marshall}, J.~L. and {Martini}, P. and {McCullough}, J. and {Melchior}, P. and {Mena-Fern{\'a}ndez}, J. and {Menanteau}, F. and {Miquel}, R. and {Mohr}, J.~J. and {Morgan}, R. and {Muir}, J. and {Myles}, J. and {Nadathur}, S. and {Navarro-Alsina}, A. and {Nichol}, R.~C. and {Ogando}, R.~L.~C. and {Omori}, Y. and {Palmese}, A. and {Pandey}, S. and {Park}, Y. and {Paz-Chinch{\'o}n}, F. and {Petravick}, D. and {Pieres}, A. and {Plazas Malag{\'o}n}, A.~A. and {Porredon}, A. and {Prat}, J. and {Raveri}, M. and {Rodriguez-Monroy}, M. and {Rollins}, R.~P. and {Romer}, A.~K. and {Roodman}, A. and {Rosenfeld}, R. and {Ross}, A.~J. and {Rykoff}, E.~S. and {Samuroff}, S. and {S{\'a}nchez}, C. and {Sanchez}, E. and {Sanchez}, J. and {Sanchez Cid}, D. and {Scarpine}, V. and {Schubnell}, M. and {Scolnic}, D. and {Secco}, L.~F. and {Serrano}, S. and {Sevilla-Noarbe}, I. and {Sheldon}, E. and {Shin}, T. and {Smith}, M. and {Soares-Santos}, M. and {Suchyta}, E. and {Swanson}, M.~E.~C. and {Tabbutt}, M. and {Tarle}, G. and {Thomas}, D. and {To}, C. and {Troja}, A. and {Troxel}, M.~A. and {Tucker}, D.~L. and {Tutusaus}, I. and {Varga}, T.~N. and {Walker}, A.~R. and {Weaverdyck}, N. and {Wechsler}, R. and {Weller}, J. and {Yanny}, B. and {Yin}, B. and {Zhang}, Y. and {Zuntz}, J. and {DES Collaboration}},
 doi = {10.1103/PhysRevD.105.023520},
 eid = {023520},
 eprint = {2105.13549},
 journal = {\prd},
 month = {January},
 number = {2},
 pages = {023520},
 primaryclass = {astro-ph.CO},
 title = {{Dark Energy Survey Year 3 results: Cosmological constraints from galaxy clustering and weak lensing}},
 volume = {105},
 year = {2022}
}

@article{healpy,
  doi = {10.21105/joss.01298},
  url = {https://doi.org/10.21105/joss.01298},
  year = {2019},
  month = mar,
  publisher = {The Open Journal},
  volume = {4},
  number = {35},
  pages = {1298},
  author = {Andrea Zonca and Leo Singer and Daniel Lenz and Martin Reinecke and Cyrille Rosset and Eric Hivon and Krzysztof Gorski},
  title = {healpy: equal area pixelization and spherical harmonics transforms for data on the sphere in Python},
  journal = {Journal of Open Source Software}
}

@ARTICLE{Kilbinger2015,
       author = {{Kilbinger}, Martin},
        title = "{Cosmology with cosmic shear observations: a review}",
      journal = {Reports on Progress in Physics},
         year = 2015,
        month = jul,
       volume = {78},
       number = {8},
          eid = {086901},
        pages = {086901},
          doi = {10.1088/0034-4885/78/8/086901},
archivePrefix = {arXiv},
       eprint = {1411.0115},
 primaryClass = {astro-ph.CO},
       adsurl = {https://ui.adsabs.harvard.edu/abs/2015RPPh...78h6901K}
}

@ARTICLE{Jenkins1998,
       author = {{Jenkins}, A. and {Frenk}, C.~S. and {Pearce}, F.~R. and
         {Thomas}, P.~A. and {Colberg}, J.~M. and {White}, S.~D.~M. and
         {Couchman}, H.~M.~P. and {Peacock}, J.~A. and {Efstathiou}, G. and
         {Nelson}, A.~H.},
        title = "{Evolution of Structure in Cold Dark Matter Universes}",
      journal = {\apj},
         year = 1998,
        month = may,
       volume = {499},
       number = {1},
        pages = {20-40},
          doi = {10.1086/305615},
archivePrefix = {arXiv},
       eprint = {astro-ph/9709010},
 primaryClass = {astro-ph},
       adsurl = {https://ui.adsabs.harvard.edu/abs/1998ApJ...499...20J}
}

@ARTICLE{Burger2024a,
       author = {{Burger}, Pierre A. and {Porth}, Lucas and {Heydenreich}, Sven and {Linke}, Laila and {Wielders}, Niek and {Schneider}, Peter and {Asgari}, Marika and {Castro}, Tiago and {Dolag}, Klaus and {Harnois-D{\'e}raps}, Joachim and {Hildebrandt}, Hendrik and {Kuijken}, Konrad and {Martinet}, Nicolas},
        title = "{KiDS-1000 cosmology: Combined second- and third-order shear statistics}",
      journal = {\aap},
         year = 2024,
        month = mar,
       volume = {683},
          eid = {A103},
        pages = {A103},
          doi = {10.1051/0004-6361/202347986},
archivePrefix = {arXiv},
       eprint = {2309.08602},
 primaryClass = {astro-ph.CO},
       adsurl = {https://ui.adsabs.harvard.edu/abs/2024A&A...683A.103B}
}

@article{Halder:2021,
 adsurl = {https://ui.adsabs.harvard.edu/abs/2021MNRAS.506.2780H},
 archiveprefix = {arXiv},
 author = {{Halder}, Anik and {Friedrich}, Oliver and {Seitz}, Stella and {Varga}, Tamas N.},
 doi = {10.1093/mnras/stab1801},
 eprint = {2102.10177},
 journal = {\mnras},
 month = {September},
 number = {2},
 pages = {2780-2803},
 primaryclass = {astro-ph.CO},
 title = {{The integrated three-point correlation function of cosmic shear}},
 volume = {506},
 year = {2021}
}

@article{Halder:2022,
 adsurl = {https://ui.adsabs.harvard.edu/abs/2022MNRAS.515.4639H},
 archiveprefix = {arXiv},
 author = {{Halder}, Anik and {Barreira}, Alexandre},
 doi = {10.1093/mnras/stac2046},
 eprint = {2201.05607},
 journal = {\mnras},
 month = {September},
 number = {3},
 pages = {4639-4654},
 primaryclass = {astro-ph.CO},
 title = {{Response approach to the integrated shear 3-point correlation function: the impact of baryonic effects on small scales}},
 volume = {515},
 year = {2022}
}

@article{heymans2021,
 adsurl = {https://ui.adsabs.harvard.edu/abs/2021A&A...646A.140H},
 archiveprefix = {arXiv},
 author = {{Heymans}, Catherine and {Tr{\"o}ster}, Tilman and {Asgari}, Marika and {Blake}, Chris and {Hildebrandt}, Hendrik and {Joachimi}, Benjamin and {Kuijken}, Konrad and {Lin}, Chieh-An and {S{\'a}nchez}, Ariel G. and {van den Busch}, Jan Luca and {Wright}, Angus H. and {Amon}, Alexandra and {Bilicki}, Maciej and {de Jong}, Jelte and {Crocce}, Martin and {Dvornik}, Andrej and {Erben}, Thomas and {Fortuna}, Maria Cristina and {Getman}, Fedor and {Giblin}, Benjamin and {Glazebrook}, Karl and {Hoekstra}, Henk and {Joudaki}, Shahab and {Kannawadi}, Arun and {K{\"o}hlinger}, Fabian and {Lidman}, Chris and {Miller}, Lance and {Napolitano}, Nicola R. and {Parkinson}, David and {Schneider}, Peter and {Shan}, HuanYuan and {Valentijn}, Edwin A. and {Verdoes Kleijn}, Gijs and {Wolf}, Christian},
 doi = {10.1051/0004-6361/202039063},
 eid = {A140},
 eprint = {2007.15632},
 journal = {\aap},
 month = {February},
 pages = {A140},
 primaryclass = {astro-ph.CO},
 title = {{KiDS-1000 Cosmology: Multi-probe weak gravitational lensing and spectroscopic galaxy clustering constraints}},
 volume = {646},
 year = {2021}
}

@article{Hoekstra:2008,
 adsurl = {https://ui.adsabs.harvard.edu/abs/2008ARNPS..58...99H},
 archiveprefix = {arXiv},
 author = {{Hoekstra}, Henk and {Jain}, Bhuvnesh},
 doi = {10.1146/annurev.nucl.58.110707.171151},
 eprint = {0805.0139},
 journal = {Annual Review of Nuclear and Particle Science},
 month = {November},
 number = {1},
 pages = {99-123},
 primaryclass = {astro-ph},
 title = {{Weak Gravitational Lensing and Its Cosmological Applications}},
 volume = {58},
 year = {2008}
}

@article{Jarvis:2004,
 adsurl = {https://ui.adsabs.harvard.edu/abs/2004MNRAS.352..338J},
 archiveprefix = {arXiv},
 author = {{Jarvis}, M. and {Bernstein}, G. and {Jain}, B.},
 doi = {10.1111/j.1365-2966.2004.07926.x},
 eprint = {astro-ph/0307393},
 journal = {\mnras},
 month = {July},
 number = {1},
 pages = {338-352},
 primaryclass = {astro-ph},
 title = {{The skewness of the aperture mass statistic}},
 volume = {352},
 year = {2004}
}

@article{Kaiser:1997,
 adsurl = {https://ui.adsabs.harvard.edu/abs/1997ApJ...484..545K},
 archiveprefix = {arXiv},
 author = {{Kaiser}, Nick and {Jaffe}, Andrew},
 doi = {10.1086/304357},
 eprint = {astro-ph/9609043},
 journal = {\apj},
 month = {July},
 number = {2},
 pages = {545-554},
 primaryclass = {astro-ph},
 title = {{Bending of Light by Gravity Waves}},
 volume = {484},
 year = {1997}
}

@ARTICLE{2025arXiv250303964G,
       author = {{Gomes}, R.~C.~H. and {Sugiyama}, S. and {Jain}, B. and {Jarvis}, M. and {Anbajagane}, D. and {Gatti}, M. and {Gebauer}, D. and {Gong}, Z. and {Halder}, A. and {Marques}, G.~A. and {Pandey}, S. and {Marshall}, J.~L. and {Allam}, S. and {Alves}, O. and {Andrade-Oliveira}, F. and {Bacon}, D. and {Blazek}, J. and {Bocquet}, S. and {Brooks}, D. and {Carnero Rosell}, A. and {Carretero}, J. and {da Costa}, L.~N. and {Doel}, P. and {Doux}, C. and {Everett}, S. and {Flaugher}, B. and {Frieman}, J. and {Garc{\'\i}a-Bellido}, J. and {Gaztanaga}, E. and {Gruen}, D. and {Gruendl}, R.~A. and {Gutierrez}, G. and {Herner}, K. and {Hinton}, S.~R. and {Hollowood}, D.~L. and {Honscheid}, K. and {Huterer}, D. and {James}, D.~J. and {Jeffrey}, N. and {Mena-Fern{\'a}ndez}, J. and {Miquel}, R. and {Muir}, J. and {Ogando}, R.~L.~C. and {Pereira}, M.~E.~S. and {Pieres}, A. and {Plazas Malag{\'o}n}, A.~A. and {Samuroff}, S. and {Sanchez}, E. and {Sanchez Cid}, D. and {Santiago}, B. and {Sevilla-Noarbe}, I. and {Smith}, M. and {Suchyta}, E. and {Swanson}, M.~E.~C. and {Tarle}, G. and {To}, C. and {Vikram}, V. and {Weaverdyck}, N. and {Weller}, J.},
        title = "{Cosmology with second and third-order shear statistics for the Dark Energy Survey: Methods and simulated analysis}",
      journal = {},
         year = 2025,
        month = mar,
          eid = {arXiv:2503.03964},
        pages = {arXiv:2503.03964},
archivePrefix = {arXiv},
       eprint = {2503.03964},
 primaryClass = {astro-ph.CO},
       adsurl = {https://ui.adsabs.harvard.edu/abs/2025arXiv250303964G}
}

@ARTICLE{Anbajagane2024,
       author = {{Anbajagane}, Dhayaa and {Pandey}, Shivam and {Chang}, Chihway},
        title = "{Map-level baryonification: Efficient modelling of higher-order correlations in the weak lensing and thermal Sunyaev-Zeldovich fields}",
      journal = {The Open Journal of Astrophysics},
         year = 2024,
        month = dec,
       volume = {7},
          eid = {108},
        pages = {108},
          doi = {10.33232/001c.126788},
archivePrefix = {arXiv},
       eprint = {2409.03822},
 primaryClass = {astro-ph.CO},
       adsurl = {https://ui.adsabs.harvard.edu/abs/2024OJAp....7E.108A}
}

@ARTICLE{Zennaro2019,
       author = {{Zennaro}, Matteo and {Angulo}, Ra{\'u}l E. and {Aric{\`o}}, Giovanni and {Contreras}, Sergio and {Pellejero-Ib{\'a}{\~n}ez}, Marcos},
        title = "{How to add massive neutrinos to your {\ensuremath{\Lambda}}CDM simulation - extending cosmology rescaling algorithms}",
      journal = {\mnras},
         year = 2019,
        month = nov,
       volume = {489},
       number = {4},
        pages = {5938-5951},
          doi = {10.1093/mnras/stz2612},
archivePrefix = {arXiv},
       eprint = {1905.08696},
 primaryClass = {astro-ph.CO},
       adsurl = {https://ui.adsabs.harvard.edu/abs/2019MNRAS.489.5938Z}
}

@ARTICLE{Contreras2020,
       author = {{Contreras}, S. and {Angulo}, R.~E. and {Zennaro}, M. and {Aric{\`o}}, G. and {Pellejero-Iba{\~n}ez}, M.},
        title = "{3 per cent-accurate predictions for the clustering of dark matter, haloes, and subhaloes, over a wide range of cosmologies and scales}",
      journal = {\mnras},
         year = 2020,
        month = dec,
       volume = {499},
       number = {4},
        pages = {4905-4917},
          doi = {10.1093/mnras/staa3117},
archivePrefix = {arXiv},
       eprint = {2001.03176},
 primaryClass = {astro-ph.CO},
       adsurl = {https://ui.adsabs.harvard.edu/abs/2020MNRAS.499.4905C}
}

@ARTICLE{Hinshaw2013,
       author = {{Hinshaw}, G. and {Larson}, D. and {Komatsu}, E. and {Spergel}, D.~N. and {Bennett}, C.~L. and {Dunkley}, J. and {Nolta}, M.~R. and {Halpern}, M. and {Hill}, R.~S. and {Odegard}, N. and {Page}, L. and {Smith}, K.~M. and {Weiland}, J.~L. and {Gold}, B. and {Jarosik}, N. and {Kogut}, A. and {Limon}, M. and {Meyer}, S.~S. and {Tucker}, G.~S. and {Wollack}, E. and {Wright}, E.~L.},
        title = "{Nine-year Wilkinson Microwave Anisotropy Probe (WMAP) Observations: Cosmological Parameter Results}",
      journal = {\apjs},
         year = 2013,
        month = oct,
       volume = {208},
       number = {2},
          eid = {19},
        pages = {19},
          doi = {10.1088/0067-0049/208/2/19},
archivePrefix = {arXiv},
       eprint = {1212.5226},
 primaryClass = {astro-ph.CO},
       adsurl = {https://ui.adsabs.harvard.edu/abs/2013ApJS..208...19H}
}

@ARTICLE{2013MNRAS.434..148S,
       author = {{Semboloni}, Elisabetta and {Hoekstra}, Henk and {Schaye}, Joop},
        title = "{Effect of baryonic feedback on two- and three-point shear statistics: prospects for detection and improved modelling}",
      journal = {\mnras},
         year = 2013,
        month = sep,
       volume = {434},
       number = {1},
        pages = {148-162},
          doi = {10.1093/mnras/stt1013},
archivePrefix = {arXiv},
       eprint = {1210.7303},
 primaryClass = {astro-ph.CO},
       adsurl = {https://ui.adsabs.harvard.edu/abs/2013MNRAS.434..148S}
}

@article{Limber:1954,
 adsurl = {https://ui.adsabs.harvard.edu/abs/1954ApJ...119..655L},
 author = {{Limber}, D. Nelson},
 doi = {10.1086/145870},
 journal = {\apj},
 month = {May},
 pages = {655},
 title = {{The Analysis of Counts of the Extragalactic Nebulae in Terms of a Fluctuating Density Field. II.}},
 volume = {119},
 year = {1954}
}

@article{Munshi:2008,
 adsurl = {https://ui.adsabs.harvard.edu/abs/2008PhR...462...67M},
 archiveprefix = {arXiv},
 author = {{Munshi}, Dipak and {Valageas}, Patrick and {van Waerbeke}, Ludovic and {Heavens}, Alan},
 doi = {10.1016/j.physrep.2008.02.003},
 eprint = {astro-ph/0612667},
 journal = {\physrep},
 month = {June},
 number = {3},
 pages = {67-121},
 primaryclass = {astro-ph},
 title = {{Cosmology with weak lensing surveys}},
 volume = {462},
 year = {2008}
}

@article{Percival2021,
 adsurl = {https://ui.adsabs.harvard.edu/abs/2022MNRAS.510.3207P},
 archiveprefix = {arXiv},
 author = {{Percival}, Will J. and {Friedrich}, Oliver and {Sellentin}, Elena and {Heavens}, Alan},
 doi = {10.1093/mnras/stab3540},
 eprint = {2108.10402},
 journal = {\mnras},
 month = {March},
 number = {3},
 pages = {3207-3221},
 primaryclass = {astro-ph.IM},
 title = {{Matching Bayesian and frequentist coverage probabilities when using an approximate data covariance matrix}},
 volume = {510},
 year = {2022}
}

@ARTICLE{Schaller2024_emu,
       author = {{Schaller}, Matthieu and {Schaye}, Joop and {Kugel}, Roi and {Broxterman}, Jeger C. and {van Daalen}, Marcel P.},
        title = "{The FLAMINGO project: baryon effects on the matter power spectrum}",
      journal = {\mnras},
         year = 2025,
        month = may,
       volume = {539},
       number = {2},
        pages = {1337-1351},
          doi = {10.1093/mnras/staf569},
archivePrefix = {arXiv},
       eprint = {2410.17109},
 primaryClass = {astro-ph.CO},
       adsurl = {https://ui.adsabs.harvard.edu/abs/2025MNRAS.539.1337S}
}

@article{planck2020,
 adsurl = {https://ui.adsabs.harvard.edu/abs/2020A&A...641A...6P},
 archiveprefix = {arXiv},
 author = {{Planck Collaboration} and {Aghanim}, N. and {Akrami}, Y. and {Ashdown}, M. and {Aumont}, J. and {Baccigalupi}, C. and {Ballardini}, M. and {Banday}, A.~J. and {Barreiro}, R.~B. and {Bartolo}, N. and {Basak}, S. and {Battye}, R. and {Benabed}, K. and {Bernard}, J. -P. and {Bersanelli}, M. and {Bielewicz}, P. and {Bock}, J.~J. and {Bond}, J.~R. and {Borrill}, J. and {Bouchet}, F.~R. and {Boulanger}, F. and {Bucher}, M. and {Burigana}, C. and {Butler}, R.~C. and {Calabrese}, E. and {Cardoso}, J. -F. and {Carron}, J. and {Challinor}, A. and {Chiang}, H.~C. and {Chluba}, J. and {Colombo}, L.~P.~L. and {Combet}, C. and {Contreras}, D. and {Crill}, B.~P. and {Cuttaia}, F. and {de Bernardis}, P. and {de Zotti}, G. and {Delabrouille}, J. and {Delouis}, J. -M. and {Di Valentino}, E. and {Diego}, J.~M. and {Dor{\'e}}, O. and {Douspis}, M. and {Ducout}, A. and {Dupac}, X. and {Dusini}, S. and {Efstathiou}, G. and {Elsner}, F. and {En{\ss}lin}, T.~A. and {Eriksen}, H.~K. and {Fantaye}, Y. and {Farhang}, M. and {Fergusson}, J. and {Fernandez-Cobos}, R. and {Finelli}, F. and {Forastieri}, F. and {Frailis}, M. and {Fraisse}, A.~A. and {Franceschi}, E. and {Frolov}, A. and {Galeotta}, S. and {Galli}, S. and {Ganga}, K. and {G{\'e}nova-Santos}, R.~T. and {Gerbino}, M. and {Ghosh}, T. and {Gonz{\'a}lez-Nuevo}, J. and {G{\'o}rski}, K.~M. and {Gratton}, S. and {Gruppuso}, A. and {Gudmundsson}, J.~E. and {Hamann}, J. and {Handley}, W. and {Hansen}, F.~K. and {Herranz}, D. and {Hildebrandt}, S.~R. and {Hivon}, E. and {Huang}, Z. and {Jaffe}, A.~H. and {Jones}, W.~C. and {Karakci}, A. and {Keih{\"a}nen}, E. and {Keskitalo}, R. and {Kiiveri}, K. and {Kim}, J. and {Kisner}, T.~S. and {Knox}, L. and {Krachmalnicoff}, N. and {Kunz}, M. and {Kurki-Suonio}, H. and {Lagache}, G. and {Lamarre}, J. -M. and {Lasenby}, A. and {Lattanzi}, M. and {Lawrence}, C.~R. and {Le Jeune}, M. and {Lemos}, P. and {Lesgourgues}, J. and {Levrier}, F. and {Lewis}, A. and {Liguori}, M. and {Lilje}, P.~B. and {Lilley}, M. and {Lindholm}, V. and {L{\'o}pez-Caniego}, M. and {Lubin}, P.~M. and {Ma}, Y. -Z. and {Mac{\'\i}as-P{\'e}rez}, J.~F. and {Maggio}, G. and {Maino}, D. and {Mandolesi}, N. and {Mangilli}, A. and {Marcos-Caballero}, A. and {Maris}, M. and {Martin}, P.~G. and {Martinelli}, M. and {Mart{\'\i}nez-Gonz{\'a}lez}, E. and {Matarrese}, S. and {Mauri}, N. and {McEwen}, J.~D. and {Meinhold}, P.~R. and {Melchiorri}, A. and {Mennella}, A. and {Migliaccio}, M. and {Millea}, M. and {Mitra}, S. and {Miville-Desch{\^e}nes}, M. -A. and {Molinari}, D. and {Montier}, L. and {Morgante}, G. and {Moss}, A. and {Natoli}, P. and {N{\o}rgaard-Nielsen}, H.~U. and {Pagano}, L. and {Paoletti}, D. and {Partridge}, B. and {Patanchon}, G. and {Peiris}, H.~V. and {Perrotta}, F. and {Pettorino}, V. and {Piacentini}, F. and {Polastri}, L. and {Polenta}, G. and {Puget}, J. -L. and {Rachen}, J.~P. and {Reinecke}, M. and {Remazeilles}, M. and {Renzi}, A. and {Rocha}, G. and {Rosset}, C. and {Roudier}, G. and {Rubi{\~n}o-Mart{\'\i}n}, J.~A. and {Ruiz-Granados}, B. and {Salvati}, L. and {Sandri}, M. and {Savelainen}, M. and {Scott}, D. and {Shellard}, E.~P.~S. and {Sirignano}, C. and {Sirri}, G. and {Spencer}, L.~D. and {Sunyaev}, R. and {Suur-Uski}, A. -S. and {Tauber}, J.~A. and {Tavagnacco}, D. and {Tenti}, M. and {Toffolatti}, L. and {Tomasi}, M. and {Trombetti}, T. and {Valenziano}, L. and {Valiviita}, J. and {Van Tent}, B. and {Vibert}, L. and {Vielva}, P. and {Villa}, F. and {Vittorio}, N. and {Wandelt}, B.~D. and {Wehus}, I.~K. and {White}, M. and {White}, S.~D.~M. and {Zacchei}, A. and {Zonca}, A.},
 doi = {10.1051/0004-6361/201833910},
 eid = {A6},
 eprint = {1807.06209},
 journal = {\aap},
 month = {September},
 pages = {A6},
 primaryclass = {astro-ph.CO},
 title = {{Planck 2018 results. VI. Cosmological parameters}},
 volume = {641},
 year = {2020}
}

@article{Schneider:1998,
 adsurl = {https://ui.adsabs.harvard.edu/abs/1998MNRAS.296..873S},
 archiveprefix = {arXiv},
 author = {{Schneider}, Peter and {van Waerbeke}, Ludovic and {Jain}, Bhuvnesh and {Kruse}, Guido},
 doi = {10.1046/j.1365-8711.1998.01422.x},
 eprint = {astro-ph/9708143},
 journal = {\mnras},
 month = {June},
 number = {4},
 pages = {873-892},
 primaryclass = {astro-ph},
 title = {{A new measure for cosmic shear}},
 volume = {296},
 year = {1998}
}

@article{Schneider:2005,
 adsurl = {https://ui.adsabs.harvard.edu/abs/2005A&A...431....9S},
 archiveprefix = {arXiv},
 author = {{Schneider}, P. and {Kilbinger}, M. and {Lombardi}, M.},
 doi = {10.1051/0004-6361:20034217},
 eprint = {astro-ph/0308328},
 journal = {\aap},
 month = {February},
 pages = {9-25},
 primaryclass = {astro-ph},
 title = {{The three-point correlation function of cosmic shear. II. Relation to the bispectrum of the projected mass density and generalized third-order aperture measures}},
 volume = {431},
 year = {2005}
}

@article{Takahashi2017,
 adsurl = {https://ui.adsabs.harvard.edu/abs/2017ApJ...850...24T},
 archiveprefix = {arXiv},
 author = {{Takahashi}, Ryuichi and {Hamana}, Takashi and {Shirasaki}, Masato and {Namikawa}, Toshiya and {Nishimichi}, Takahiro and {Osato}, Ken and {Shiroyama}, Kosei},
 doi = {10.3847/1538-4357/aa943d},
 eid = {24},
 eprint = {1706.01472},
 journal = {\apj},
 month = {November},
 number = {1},
 pages = {24},
 primaryclass = {astro-ph.CO},
 title = {{Full-sky Gravitational Lensing Simulation for Large-area Galaxy Surveys and Cosmic Microwave Background Experiments}},
 volume = {850},
 year = {2017}
}

@article{Takahashi:2020,
 adsurl = {https://ui.adsabs.harvard.edu/abs/2020ApJ...895..113T},
 archiveprefix = {arXiv},
 author = {{Takahashi}, Ryuichi and {Nishimichi}, Takahiro and {Namikawa}, Toshiya and {Taruya}, Atsushi and {Kayo}, Issha and {Osato}, Ken and {Kobayashi}, Yosuke and {Shirasaki}, Masato},
 doi = {10.3847/1538-4357/ab908d},
 eid = {113},
 eprint = {1911.07886},
 journal = {\apj},
 month = {June},
 number = {2},
 pages = {113},
 primaryclass = {astro-ph.CO},
 title = {{Fitting the Nonlinear Matter Bispectrum by the Halofit Approach}},
 volume = {895},
 year = {2020}
}

@ARTICLE{vanDaalen2011,
       author = {{van Daalen}, Marcel P. and {Schaye}, Joop and {Booth}, C.~M. and {Dalla Vecchia}, Claudio},
        title = "{The effects of galaxy formation on the matter power spectrum: a challenge for precision cosmology}",
      journal = {\mnras},
         year = 2011,
        month = aug,
       volume = {415},
       number = {4},
        pages = {3649-3665},
          doi = {10.1111/j.1365-2966.2011.18981.x},
archivePrefix = {arXiv},
       eprint = {1104.1174},
 primaryClass = {astro-ph.CO},
       adsurl = {https://ui.adsabs.harvard.edu/abs/2011MNRAS.415.3649V}
}

@ARTICLE{Heydenreich2023,
       author = {{Heydenreich}, Sven and {Linke}, Laila and {Burger}, Pierre and {Schneider}, Peter},
        title = "{A roadmap to cosmological parameter analysis with third-order shear statistics. I. Modelling and validation}",
      journal = {\aap},
         year = 2023,
        month = apr,
       volume = {672},
          eid = {A44},
        pages = {A44},
          doi = {10.1051/0004-6361/202244820},
archivePrefix = {arXiv},
       eprint = {2208.11686},
 primaryClass = {astro-ph.CO},
       adsurl = {https://ui.adsabs.harvard.edu/abs/2023A&A...672A..44H}
}

@BOOK{Peebles1980,
       author = {{Peebles}, P.~J.~E.},
        title = "{The large-scale structure of the universe}",
         year = 1980,
        publisher = {Princeton University Press},
       adsurl = {https://ui.adsabs.harvard.edu/abs/1980lssu.book.....P}
}

@ARTICLE{2010MNRAS.402.1536S,
       author = {{Schaye}, Joop and {Dalla Vecchia}, Claudio and {Booth}, C.~M. and {Wiersma}, Robert P.~C. and {Theuns}, Tom and {Haas}, Marcel R. and {Bertone}, Serena and {Duffy}, Alan R. and {McCarthy}, I.~G. and {van de Voort}, Freeke},
        title = "{The physics driving the cosmic star formation history}",
      journal = {\mnras},
         year = 2010,
        month = mar,
       volume = {402},
       number = {3},
        pages = {1536-1560},
          doi = {10.1111/j.1365-2966.2009.16029.x},
archivePrefix = {arXiv},
       eprint = {0909.5196},
 primaryClass = {astro-ph.CO},
       adsurl = {https://ui.adsabs.harvard.edu/abs/2010MNRAS.402.1536S}
}

@ARTICLE{Zhou2025,
       author = {{Zhou}, Alan Junzhe and {Gatti}, Marco and {Anbajagane}, Dhayaa and {Dodelson}, Scott and {Schaller}, Matthieu and {Schaye}, Joop},
        title = "{Map-level baryonification: unified treatment of weak lensing two-point and higher-order statistics}",
      journal = {JCAP},
         year = 2025,
        month = sep,
       volume = {09},
       number = {9},
          eid = {073},
        pages = {073},
          doi = {10.1088/1475-7516/2025/09/073},
archivePrefix = {arXiv},
       eprint = {2505.07949},
 primaryClass = {astro-ph.CO},
       adsurl = {https://ui.adsabs.harvard.edu/abs/2025JCAP...09..073Z}
}

@article{2021ascl.soft11002B,
       author = {{Bradbury}, James and {Frostig}, Roy and {Hawkins}, Peter and {Johnson}, Matthew James and {Leary}, Chris and {Maclaurin}, Dougal and {Necula}, George and {Paszke}, Adam and {VanderPlas}, Jake and {Wanderman-Milne}, Skye and {Zhang}, Qiao},
        title = "{JAX: Autograd and XLA}",
        journal = {Astrophysics Source Code Library},
 howpublished = {Astrophysics Source Code Library, record ascl:2111.002},
         year = 2021,
        month = nov,
          eprint = {ascl:2111.002},
       adsurl = {https://ui.adsabs.harvard.edu/abs/2021ascl.soft11002B}
}

@ARTICLE{Springel2005,
       author = {{Springel}, Volker},
        title = "{The cosmological simulation code GADGET-2}",
      journal = {\mnras},
         year = 2005,
        month = dec,
       volume = {364},
       number = {4},
        pages = {1105-1134},
          doi = {10.1111/j.1365-2966.2005.09655.x},
archivePrefix = {arXiv},
       eprint = {astro-ph/0505010},
 primaryClass = {astro-ph},
       adsurl = {https://ui.adsabs.harvard.edu/abs/2005MNRAS.364.1105S}
}

@ARTICLE{McCarthy2023,
       author = {{McCarthy}, Ian G. and {Salcido}, Jaime and {Schaye}, Joop and {Kwan}, Juliana and {Elbers}, Willem and {Kugel}, Roi and {Schaller}, Matthieu and {Helly}, John C. and {Braspenning}, Joey and {Frenk}, Carlos S. and {van Daalen}, Marcel P. and {Vandenbroucke}, Bert and {Conley}, Jonah T. and {Font}, Andreea S. and {Upadhye}, Amol},
        title = "{The FLAMINGO project: revisiting the S$_{8}$ tension and the role of baryonic physics}",
      journal = {\mnras},
         year = 2023,
        month = dec,
       volume = {526},
       number = {4},
        pages = {5494-5519},
          doi = {10.1093/mnras/stad3107},
archivePrefix = {arXiv},
       eprint = {2309.07959},
 primaryClass = {astro-ph.CO},
       adsurl = {https://ui.adsabs.harvard.edu/abs/2023MNRAS.526.5494M}
}

@ARTICLE{Kugel2023,
       author = {{Kugel}, Roi and {Schaye}, Joop and {Schaller}, Matthieu and {Helly}, John C. and {Braspenning}, Joey and {Elbers}, Willem and {Frenk}, Carlos S. and {McCarthy}, Ian G. and {Kwan}, Juliana and {Salcido}, Jaime and {van Daalen}, Marcel P. and {Vandenbroucke}, Bert and {Bah{\'e}}, Yannick M. and {Borrow}, Josh and {Chaikin}, Evgenii and {Hu{\v{s}}ko}, Filip and {Jenkins}, Adrian and {Lacey}, Cedric G. and {Nobels}, Folkert S.~J. and {Vernon}, Ian},
        title = "{FLAMINGO: calibrating large cosmological hydrodynamical simulations with machine learning}",
      journal = {\mnras},
         year = 2023,
        month = dec,
       volume = {526},
       number = {4},
        pages = {6103-6127},
          doi = {10.1093/mnras/stad2540},
archivePrefix = {arXiv},
       eprint = {2306.05492},
 primaryClass = {astro-ph.CO},
       adsurl = {https://ui.adsabs.harvard.edu/abs/2023MNRAS.526.6103K}
}

@ARTICLE{Schaye2023,
       author = {{Schaye}, Joop and {Kugel}, Roi and {Schaller}, Matthieu and {Helly}, John C. and {Braspenning}, Joey and {Elbers}, Willem and {McCarthy}, Ian G. and {van Daalen}, Marcel P. and {Vandenbroucke}, Bert and {Frenk}, Carlos S. and {Kwan}, Juliana and {Salcido}, Jaime and {Bah{\'e}}, Yannick M. and {Borrow}, Josh and {Chaikin}, Evgenii and {Hahn}, Oliver and {Hu{\v{s}}ko}, Filip and {Jenkins}, Adrian and {Lacey}, Cedric G. and {Nobels}, Folkert S.~J.},
        title = "{The FLAMINGO project: cosmological hydrodynamical simulations for large-scale structure and galaxy cluster surveys}",
      journal = {\mnras},
         year = 2023,
        month = dec,
       volume = {526},
       number = {4},
        pages = {4978-5020},
          doi = {10.1093/mnras/stad2419},
archivePrefix = {arXiv},
       eprint = {2306.04024},
 primaryClass = {astro-ph.CO},
       adsurl = {https://ui.adsabs.harvard.edu/abs/2023MNRAS.526.4978S}
}

@ARTICLE{Jeger2024,
       author = {{Broxterman}, Jeger C. and {Schaller}, Matthieu and {Schaye}, Joop and {Hoekstra}, Henk and {Kuijken}, Konrad and {Helly}, John C. and {Kugel}, Roi and {Braspenning}, Joey and {Elbers}, Willem and {Frenk}, Carlos S. and {Kwan}, Juliana and {McCarthy}, Ian G. and {Salcido}, Jaime and {van Daalen}, Marcel P. and {Vandenbroucke}, Bert},
        title = "{The FLAMINGO project: baryonic impact on weak gravitational lensing convergence peak counts}",
      journal = {\mnras},
         year = 2024,
        month = apr,
       volume = {529},
       number = {3},
        pages = {2309-2326},
          doi = {10.1093/mnras/stae698},
archivePrefix = {arXiv},
       eprint = {2312.08450},
 primaryClass = {astro-ph.CO},
       adsurl = {https://ui.adsabs.harvard.edu/abs/2024MNRAS.529.2309B}
}

@article{BiG,
  author       = {Laila Linke},
  title        = {llinke1/BiG: First Release},
  month        = nov,
  year         = 2024,
  journal    = {Zenodo},
  version      = {v1.0.0},
  eprint          = {10.5281/zenodo.14181881},
  url          = {https://doi.org/10.5281/zenodo.14181881}
}

@ARTICLE{Scoccimarroe:2000,
       author = {{Scoccimarro}, Rom{\'a}n},
        title = "{The Bispectrum: From Theory to Observations}",
      journal = {\apj},
         year = 2000,
        month = dec,
       volume = {544},
       number = {2},
        pages = {597-615},
          doi = {10.1086/317248},
archivePrefix = {arXiv},
       eprint = {astro-ph/0004086},
 primaryClass = {astro-ph},
       adsurl = {https://ui.adsabs.harvard.edu/abs/2000ApJ...544..597S}
}

@ARTICLE{Oddo:2020,
       author = {{Oddo}, Andrea and {Sefusatti}, Emiliano and {Porciani}, Cristiano and {Monaco}, Pierluigi and {S{\'a}nchez}, Ariel G.},
        title = "{Toward a robust inference method for the galaxy bispectrum: likelihood function and model selection}",
      journal = {JCAP},
         year = 2020,
        month = mar,
       volume = {03},
       number = {3},
          eid = {056},
        pages = {056},
          doi = {10.1088/1475-7516/2020/03/056},
archivePrefix = {arXiv},
       eprint = {1908.01774},
 primaryClass = {astro-ph.CO},
       adsurl = {https://ui.adsabs.harvard.edu/abs/2020JCAP...03..056O}
}

@ARTICLE{Colombi:2009,
       author = {{Colombi}, St{\'e}phane and {Jaffe}, Andrew and {Novikov}, Dmitri and {Pichon}, Christophe},
        title = "{Accurate estimators of power spectra in N-body simulations}",
      journal = {\mnras},
         year = 2009,
        month = feb,
       volume = {393},
       number = {2},
        pages = {511-526},
          doi = {10.1111/j.1365-2966.2008.14176.x},
archivePrefix = {arXiv},
       eprint = {0811.0313},
 primaryClass = {astro-ph},
       adsurl = {https://ui.adsabs.harvard.edu/abs/2009MNRAS.393..511C}
}

@ARTICLE{Hand2018,
       author = {{Hand}, Nick and {Feng}, Yu and {Beutler}, Florian and {Li}, Yin and {Modi}, Chirag and {Seljak}, Uro{\v{s}} and {Slepian}, Zachary},
        title = "{nbodykit: An Open-source, Massively Parallel Toolkit for Large-scale Structure}",
      journal = {\aj},
         year = 2018,
        month = oct,
       volume = {156},
       number = {4},
          eid = {160},
        pages = {160},
          doi = {10.3847/1538-3881/aadae0},
archivePrefix = {arXiv},
       eprint = {1712.05834},
 primaryClass = {astro-ph.IM},
       adsurl = {https://ui.adsabs.harvard.edu/abs/2018AJ....156..160H}
}
%%%%%%%%%%%%%%%%%%%%%%%%%%%%%%%%%%%%%%%%%%%%%%%%%%

%%%%%%%%%%%%%%%%% APPENDICES %%%%%%%%%%%%%%%%%%%%%

%\appendix
\begin{appendix}

\section{Accuracy of bispectrum measurement code}
\label{app:accuracyBiG}

\begin{figure*}
    \centering
    \includegraphics[width=0.49\linewidth]{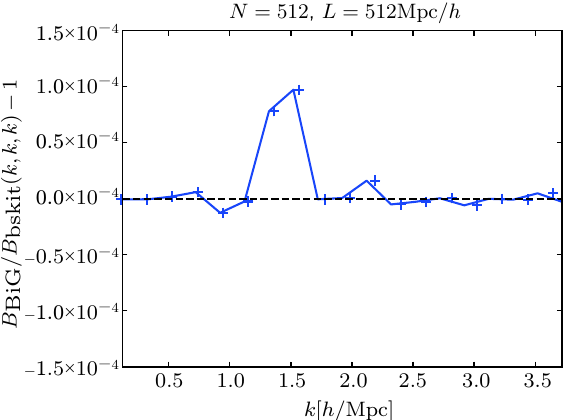}
    \includegraphics[width=0.49\linewidth]{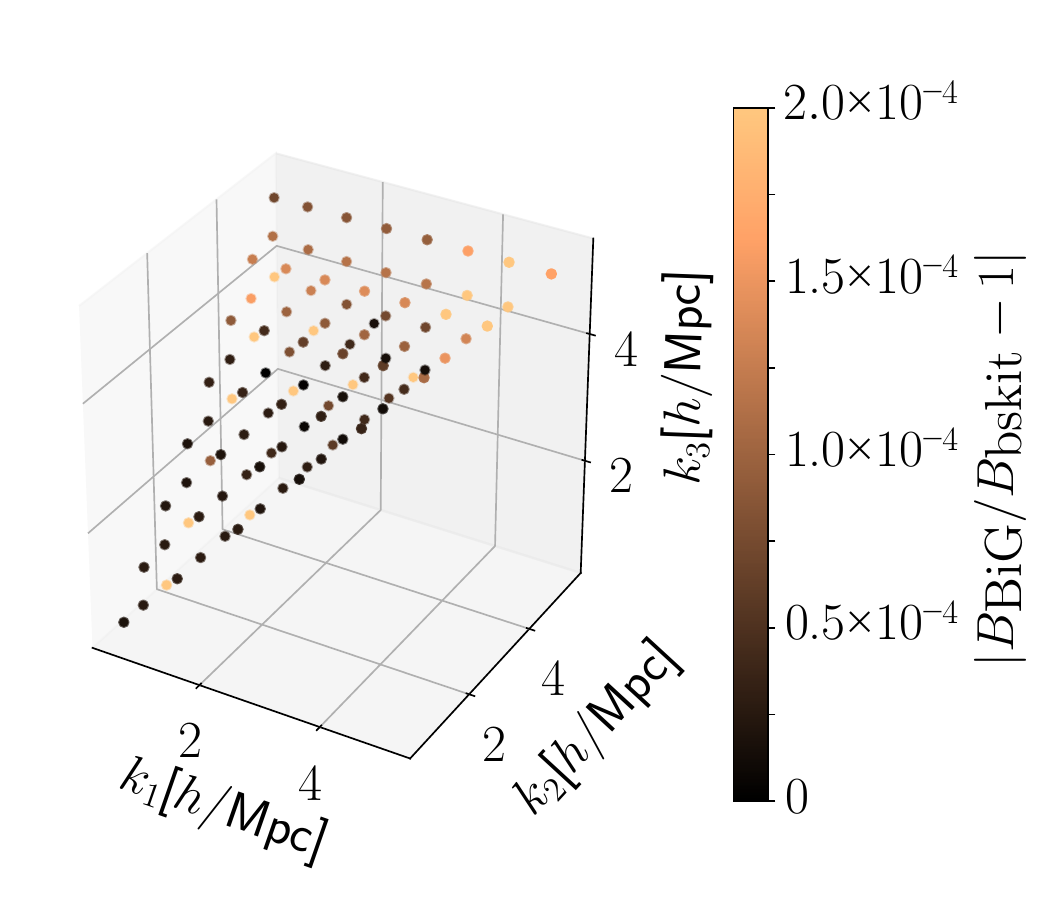}
    \caption{Relative difference of bispectrum estimate by our measurement code BiG and \texttt{bskit} for an example simulation. \emph{Left}: For equilateral triangles, \emph{Right}: for all triangle configurations}
    \label{fig:accuracy}
\end{figure*}

To validate our bispectrum measurement code, we compare it to the established code \texttt{bskit} \citep{Scoccimarroe:2000,2020MNRAS.498.2887F}, which uses \texttt{nbodykit} \citep{Hand2018}. It implements the same algorithm as BiG \citep{BiG} but uses no GPU acceleration. We show in Fig.~\ref{fig:accuracy} the relative difference of the BiG and \texttt{bskit} estimates for the bispectrum of an example simulation with boxsize $L=512 h^{-1}\mathrm{Mpc}$ and grid size $N=512$. The difference is at the numerical precision level, with no deviation being larger than $2\times 10^{-4}$.

\section{Additional accuracy plots of the bispectrum correction model}

\begin{figure*}[ht]
\includegraphics[width=\textwidth]{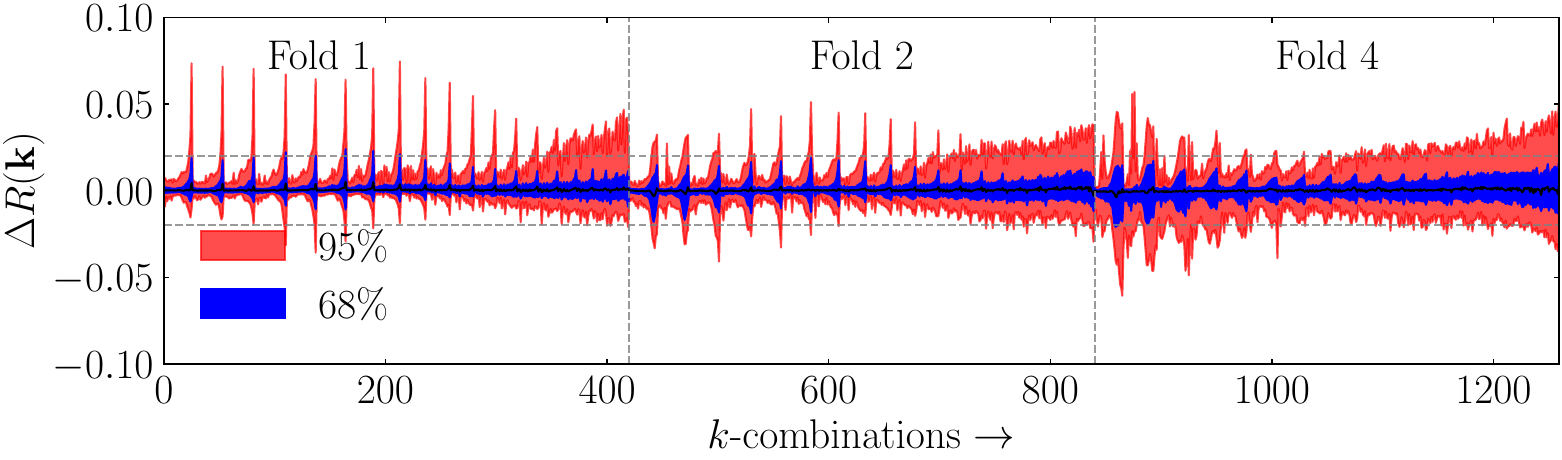}\\
\caption{Accuracy of the emulator for baryonic effects on the matter bispectrum. The $y$-axis is defined as $\Delta R(k) = R^\mathrm{emu}/ R^\mathrm{true}-1$, where $R(\vec{k})=B_{\rm BCM}(\vec{k})/B_{\rm GrO}(\vec{k})$. Here, we flattened and concatenated the three folding. The horizontal dashed lines show the 2$\%$ region. }
\label{fig:emulator_acc_deltaR}
\end{figure*}

\begin{figure*}[ht]
\includegraphics[width=\textwidth]{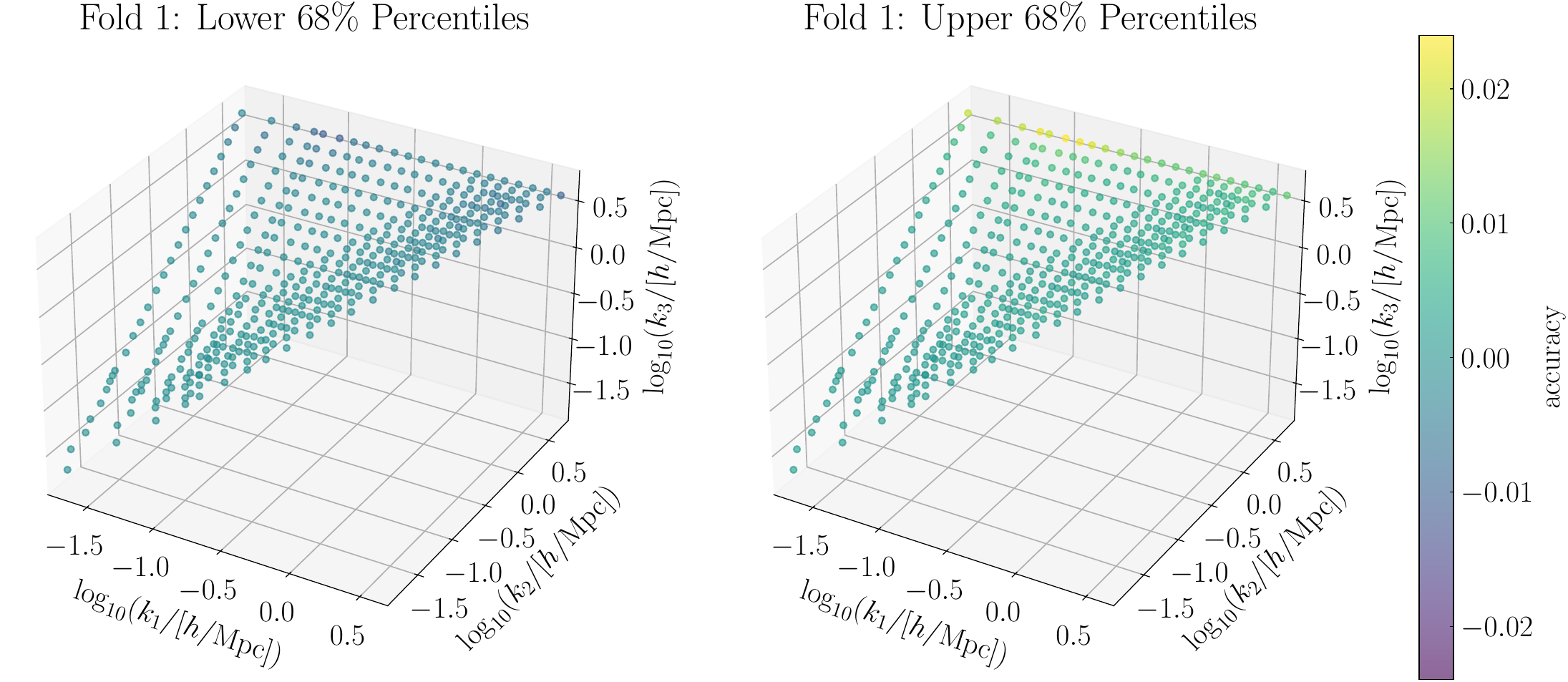}\\
\includegraphics[width=\textwidth]{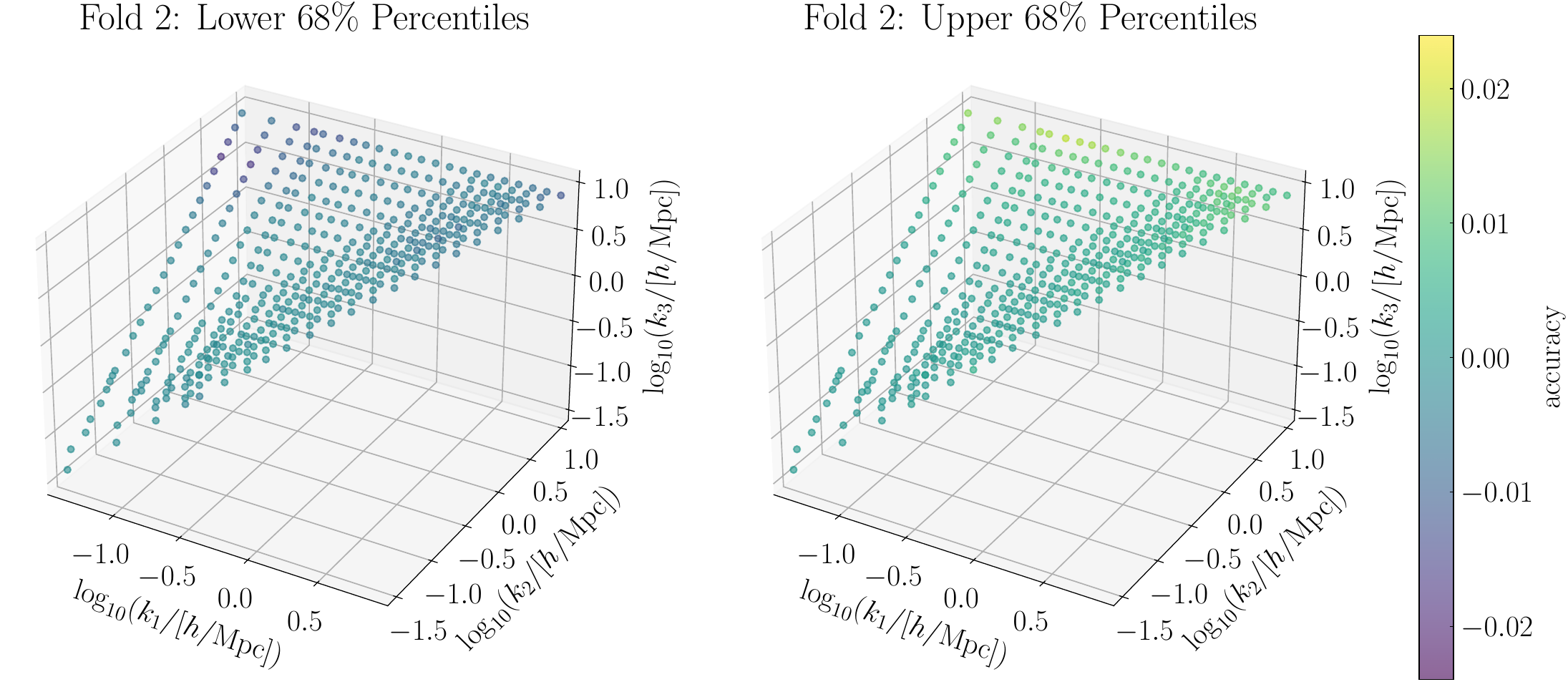}\\
\includegraphics[width=\textwidth]{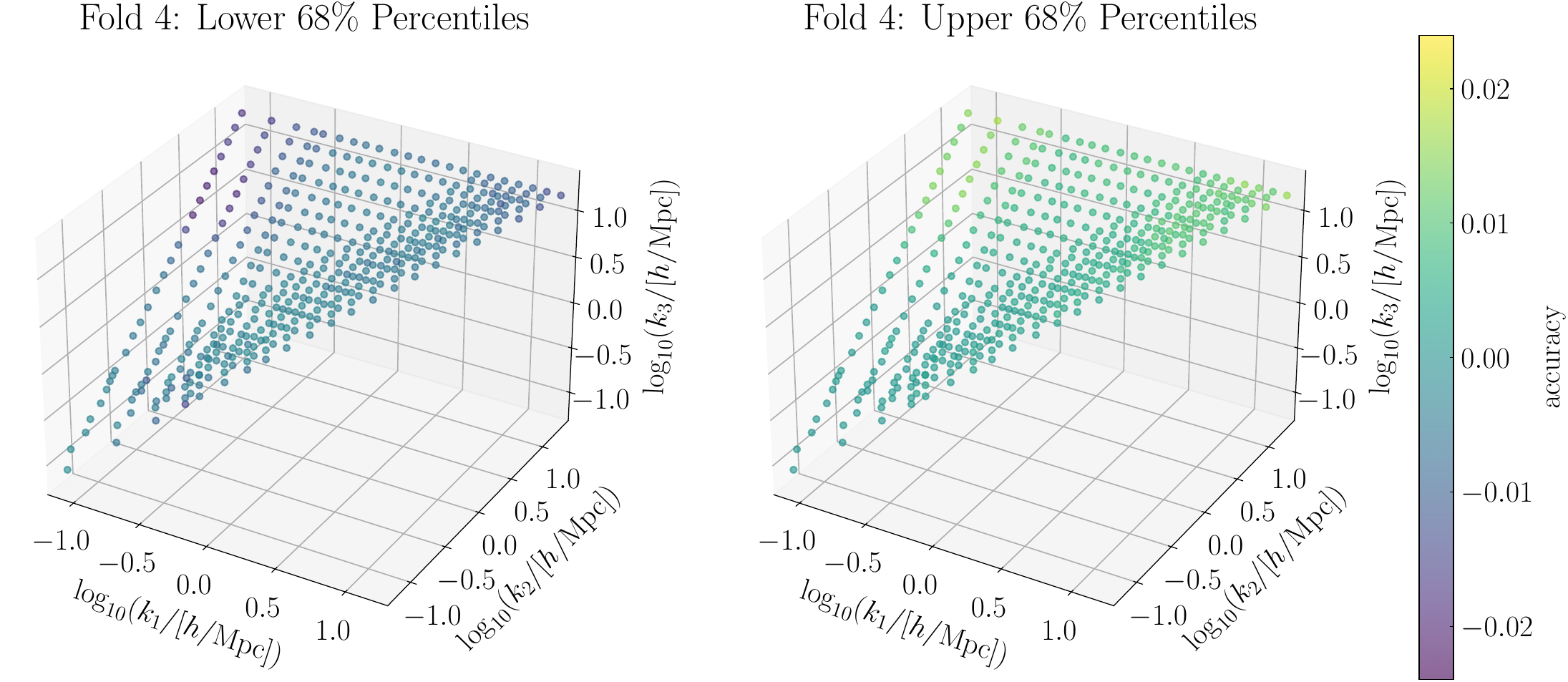}\\
\caption{Similar to Fig.~\ref{fig:emulator_acc_deltaR}, but now illustrated as a three dimensional scatter plot. }
\label{fig:emulator_acc_deltaR_kgrid}
\end{figure*}

This section shows two additional plots that illustrate the accuracy of BCM bispectrum emulator. While in Fig.~\ref{fig:emulator_acc_deltaR} we show the flattened and concatenated measurements from all three folding, we show in Fig.~\ref{fig:emulator_acc_deltaR_kgrid} the same as a three-dimensional scatter plot.

\section{Parameter dependence}
\label{app:BCMdependence}
\begin{figure*}[ht]
\includegraphics[width=\textwidth]{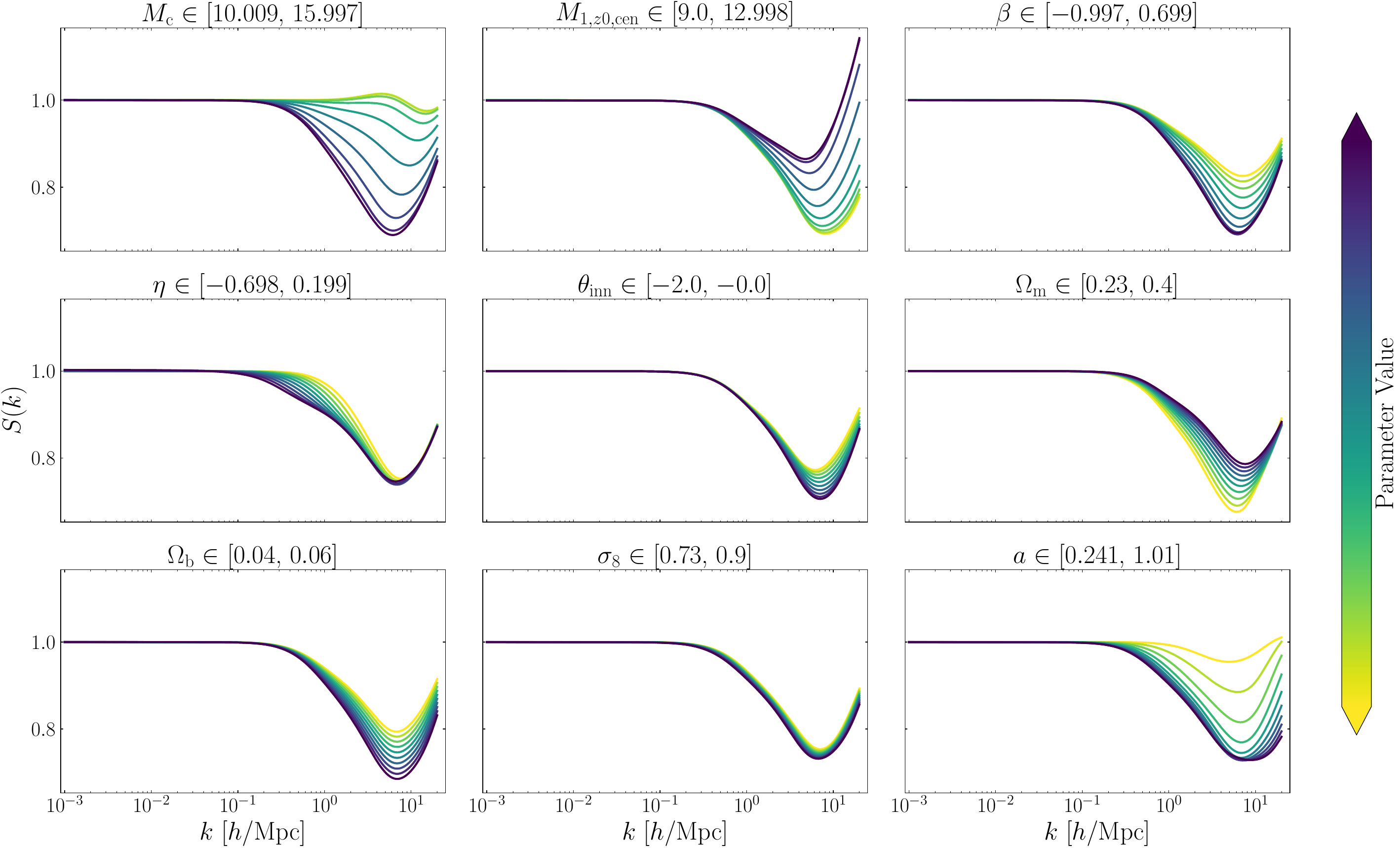}\vspace{0.06\textwidth}
\includegraphics[width=\textwidth]{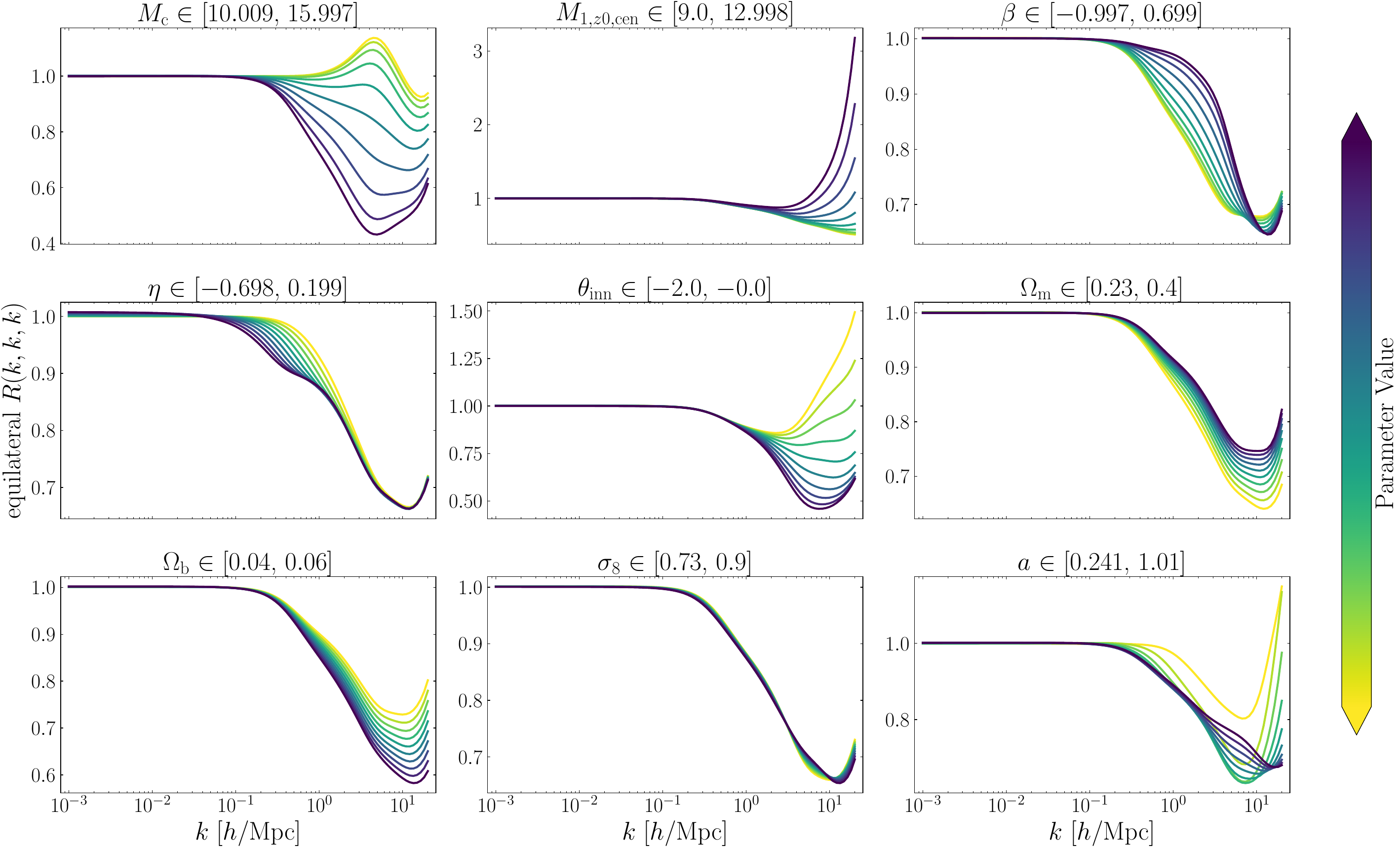}\\
\caption{In this figure, we show the power spectrum's dependence on baryonic feedback effects in the three upper rows. In the bottom three rows, we show the dependence of the bispectrum for $k_1=k_2=k_3$.}
\label{fig:bispec_dependence}
\end{figure*}

\begin{figure*}[ht]
\includegraphics[width=\textwidth]{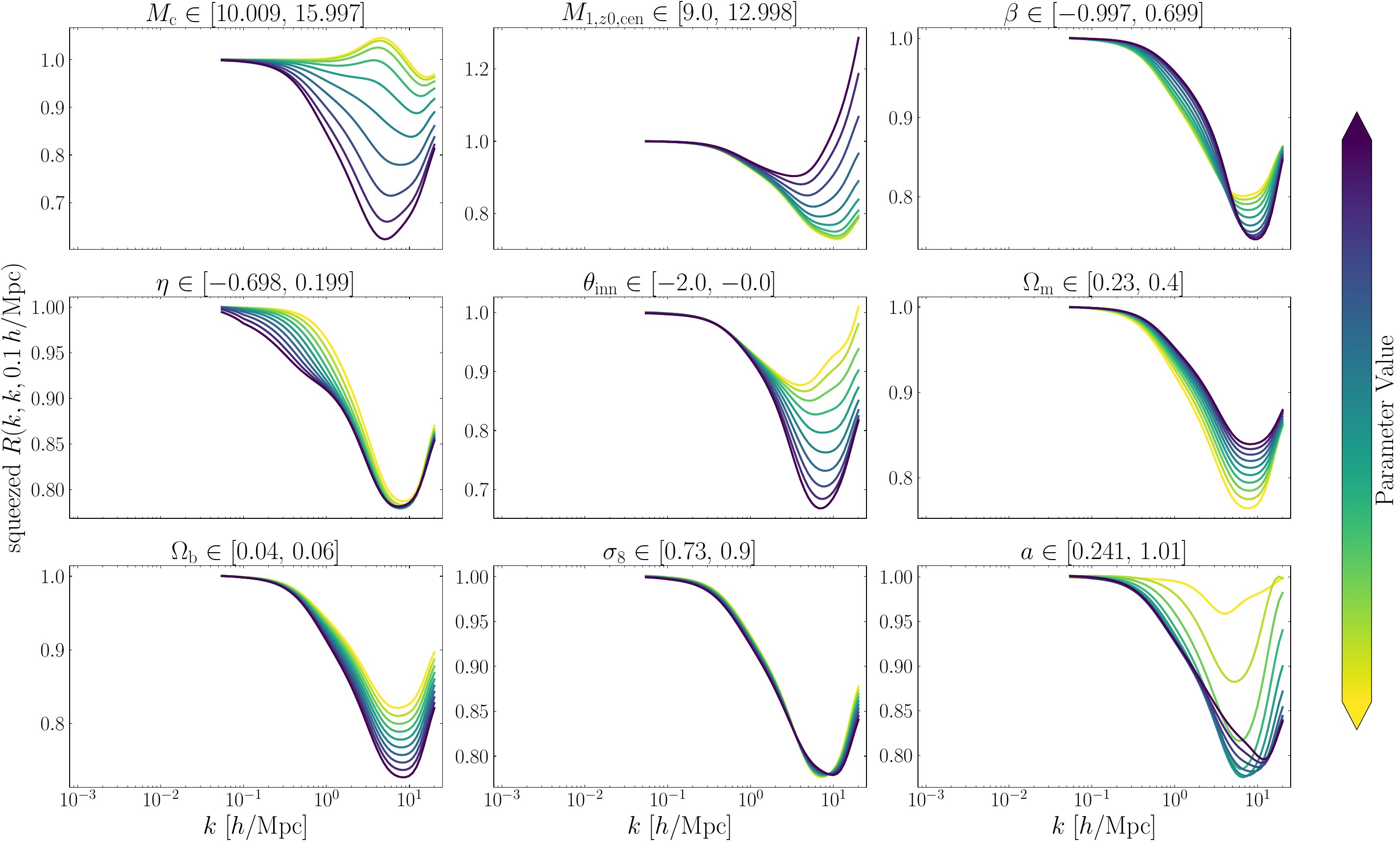}\vspace{0.06\textwidth}
\includegraphics[width=\textwidth]{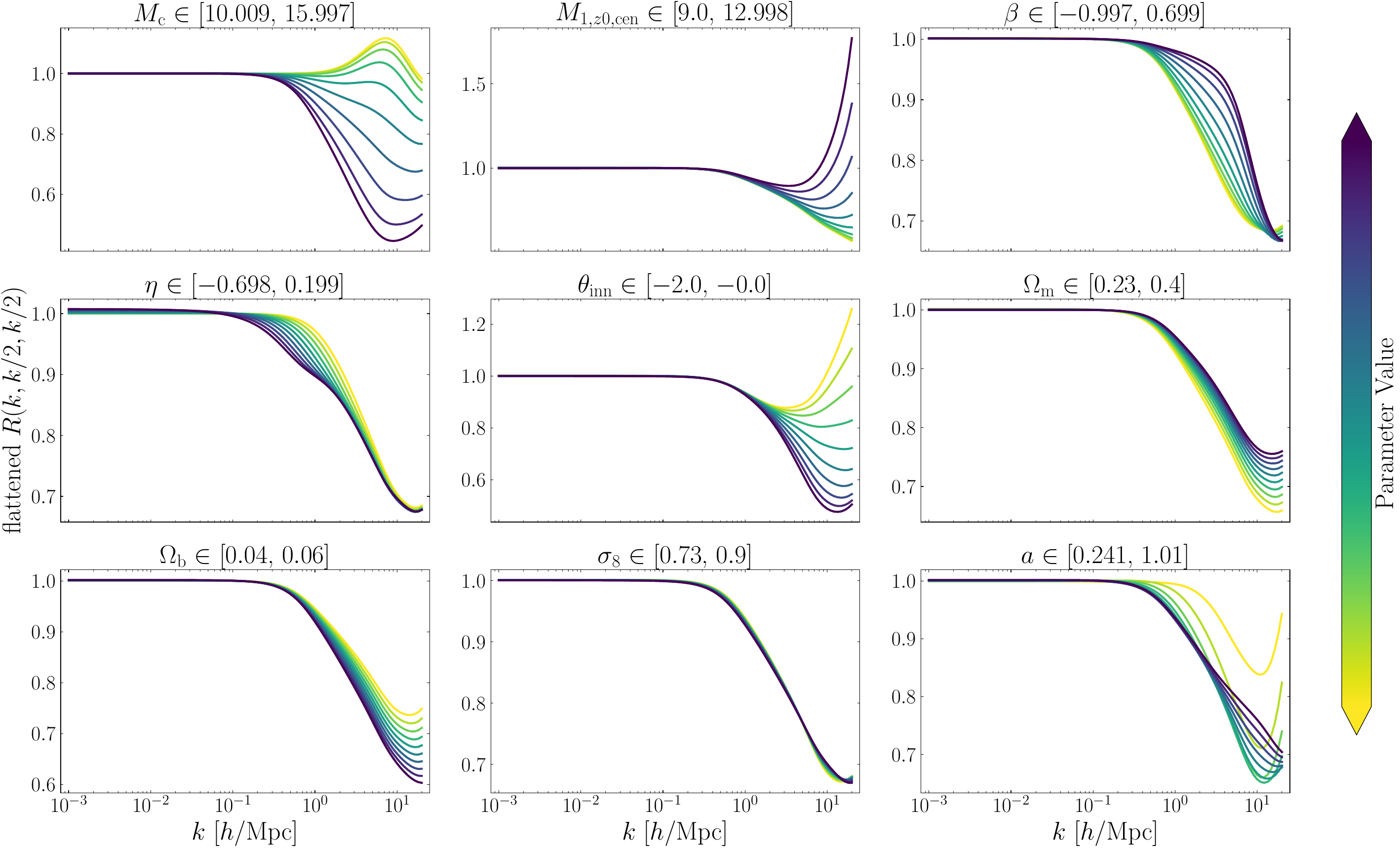}\\
\caption{The same as the bottom three rows of Fig.~\ref{fig:bispec_dependence} but for two different settings of $k_i$ as indicated on the $y$-axis.}
\label{fig:bispec_dependence_squeezed}
\end{figure*}

In this appendix, we explore further the dependences of two- and three-point statistics on both cosmological and baryonic parameters. 
We begin in the upper panel of Fig.~\ref{fig:bispec_dependence}, with the dependence of the power spectrum ratio, $S(k) = P^\mathrm{Hydro}/P^\mathrm{GrO}$, on baryonic feedback effects. 
In the lower panels of Figs.~\ref{fig:bispec_dependence} and \ref{fig:bispec_dependence_squeezed}, we then show the dependence of bispectra ratio, $R(\vec{k}) = B^\mathrm{Hydro}/B^\mathrm{GrO}$ for three different $k$ configurations. Here we especially see that setting $R(k_1,k_2,k_3)=1$ if one $k_i< k_\mathrm{min}$ might result in biased results. 

\begin{figure*}[ht]
\centering
\includegraphics[width=0.45\textwidth]{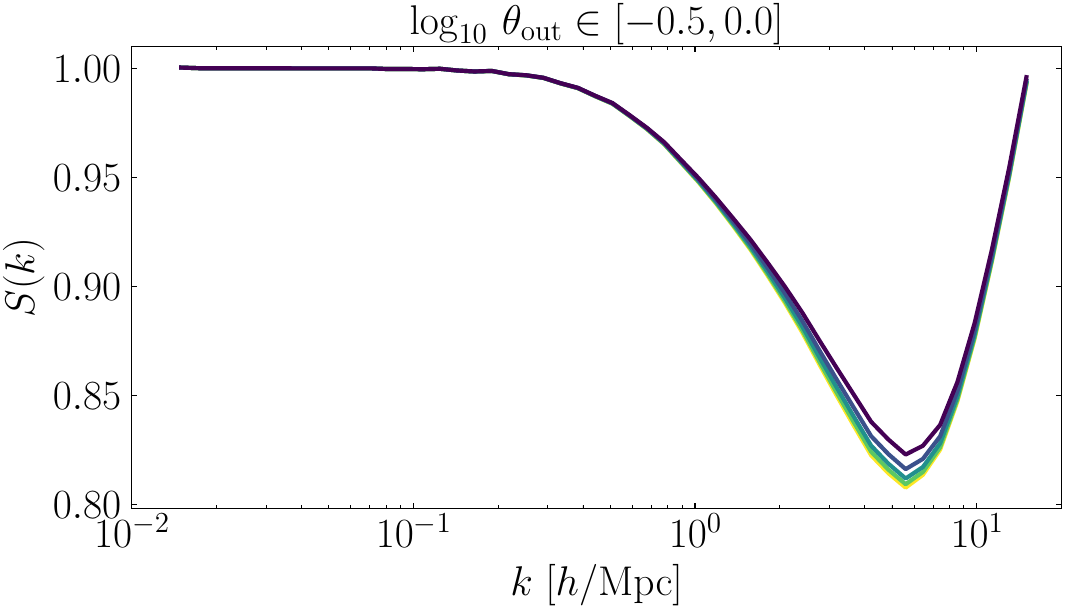}\hspace{0.04\textwidth}
\includegraphics[width=0.45\textwidth]{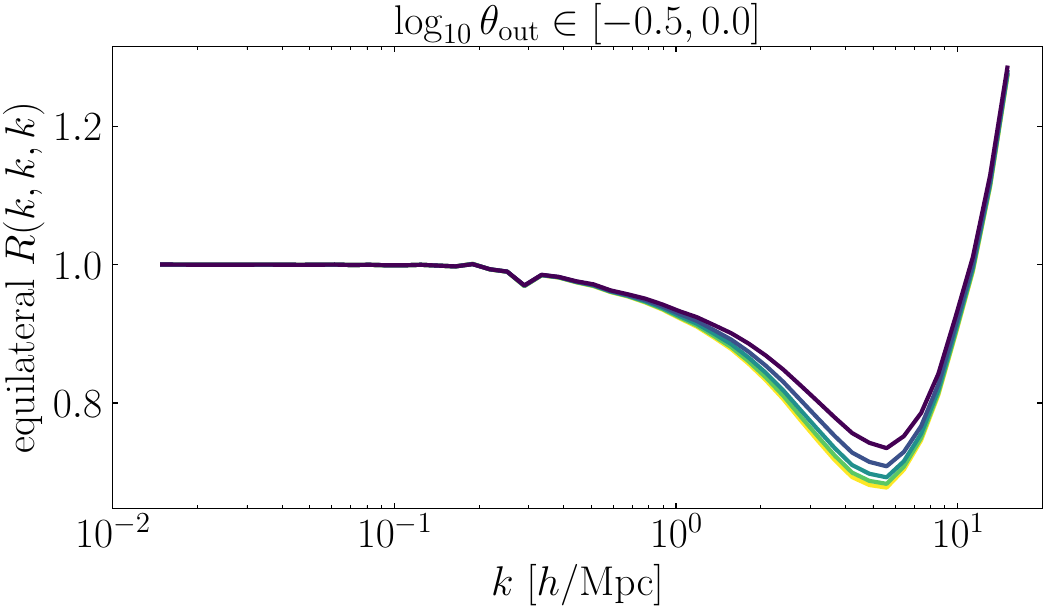}\\[1ex]
\includegraphics[width=0.45\textwidth]{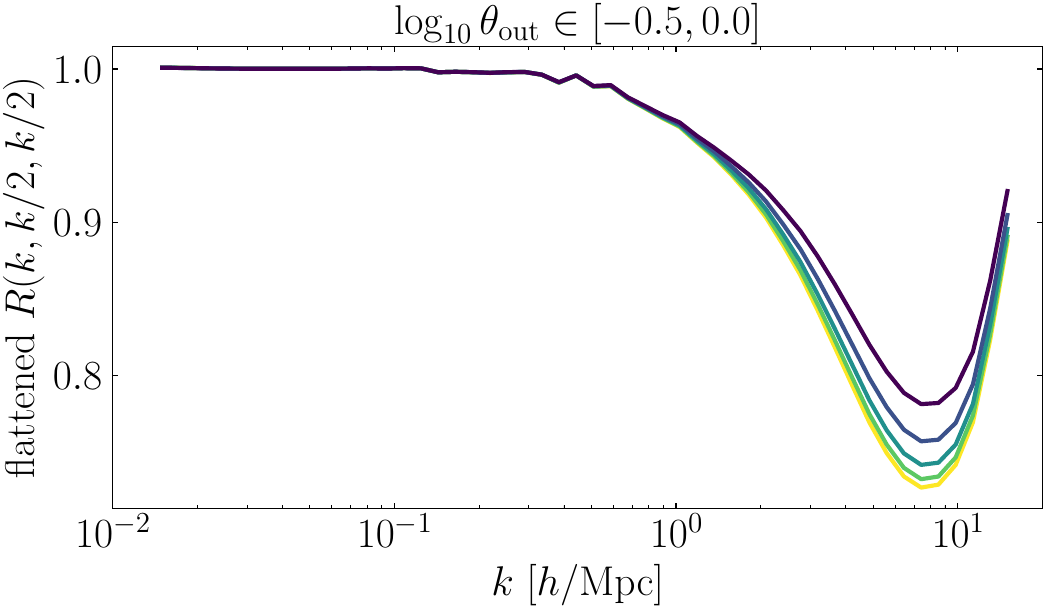}\hspace{0.04\textwidth}
\includegraphics[width=0.45\textwidth]{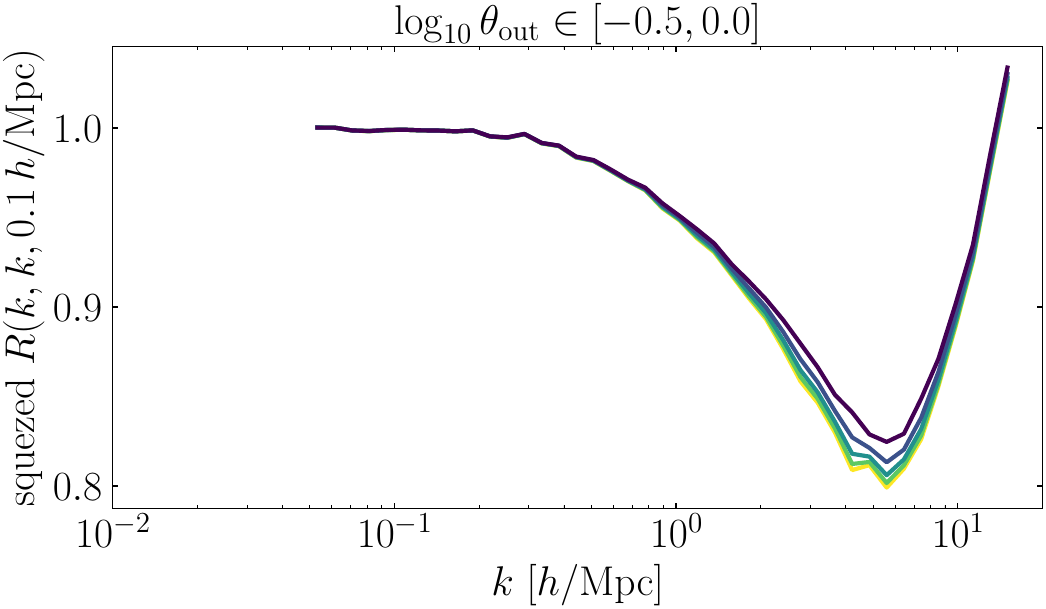}
\caption{Dependence of the $S(k)$ and $R(\boldsymbol{k})$ on the BCM parameter $\theta_\mathrm{out}$. Yellow lines are smaller and blue lines are larger $\theta_\mathrm{out}$.}
\label{fig:spectra_theta_out}
\end{figure*}

\begin{figure*}[ht]
\centering
\includegraphics[width=0.45\textwidth]{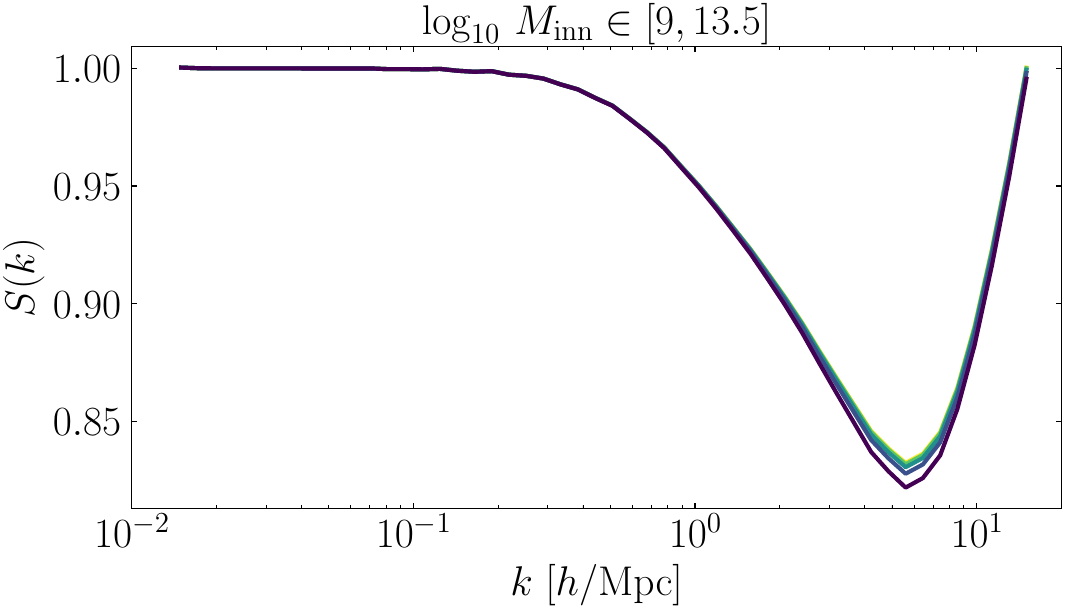}\hspace{0.04\textwidth}
\includegraphics[width=0.45\textwidth]{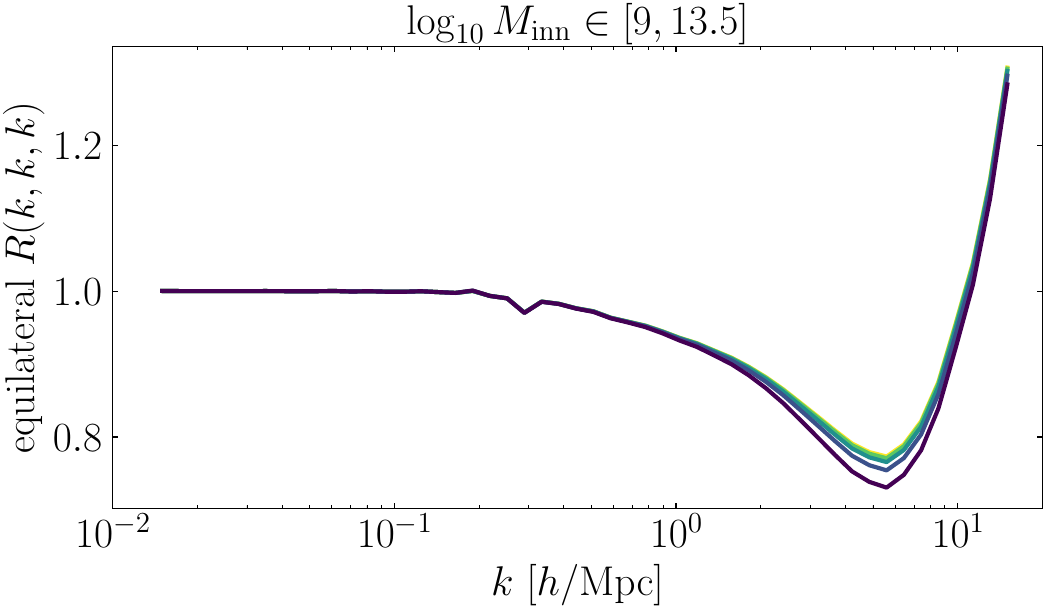}\\[1ex]
\includegraphics[width=0.45\textwidth]{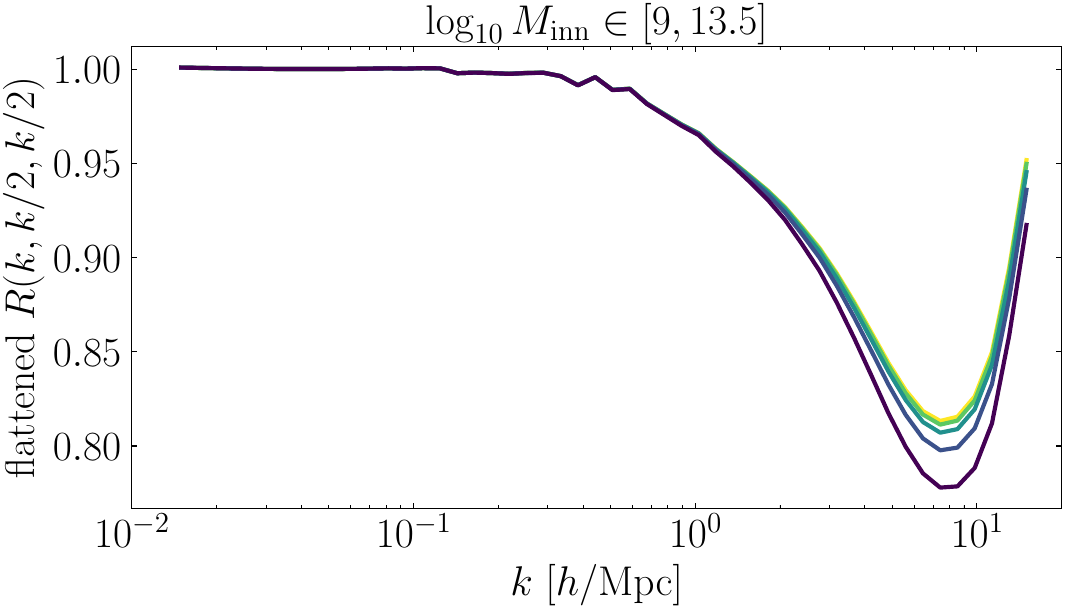}\hspace{0.04\textwidth}
\includegraphics[width=0.45\textwidth]{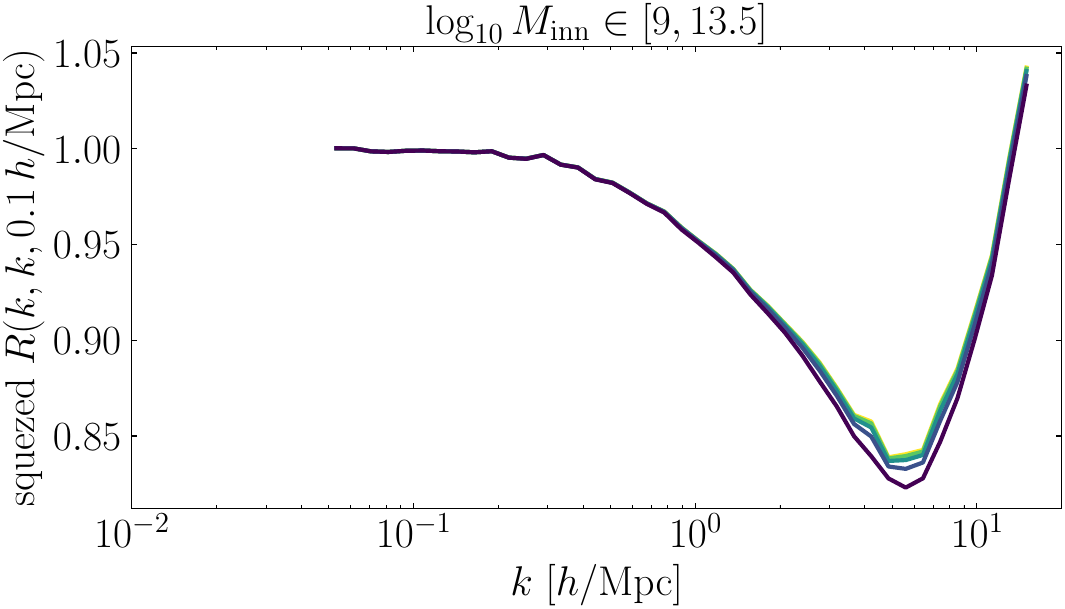}
\caption{Dependence of the $S(k)$ and $R(\vec{k})$ on the BCM parameter $M_\mathrm{inn}$. Yellow lines are smaller and blue lines are larger $M_\mathrm{inn}$.}
\label{fig:spectra_Minn}
\end{figure*}

In Figs.~\ref{fig:spectra_theta_out} and \ref{fig:spectra_Minn}, we display the dependence of power spectrum and bispectrum on the BCM parameter $\theta_\mathrm{out}$ and $M_{\rm inn}$. We apply the same baryonification, compatible with the FLAMINGO suite, to pairs of $N$-body simulations which vary one cosmological parameter at a time, in a wide range displayed in the legend. We note that, for simplicity, we use a simple linear interpolation and do not remove noisy measurements. However, this does not affect our conclusions. The dependence of $\theta_\mathrm{out}$ and $M_{\rm inn}$ is significantly smaller compared to other BCM parameters, such as $M_1$ or $\beta$. Given their subdominant contribution, we decided to fix $\theta_\mathrm{out} =1$ and $M_{\rm inn}=2.3\times 10^{13} \, \Msun$.

\begin{figure*}[ht]
\centering
\includegraphics[width=0.45\textwidth]{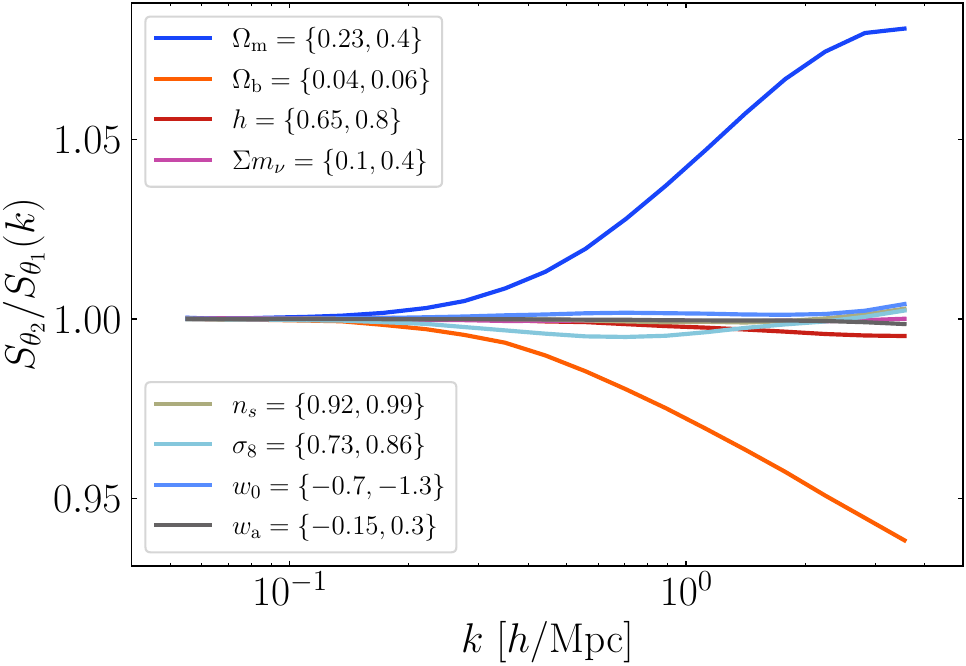}\hspace{0.04\textwidth}
\includegraphics[width=0.45\textwidth]{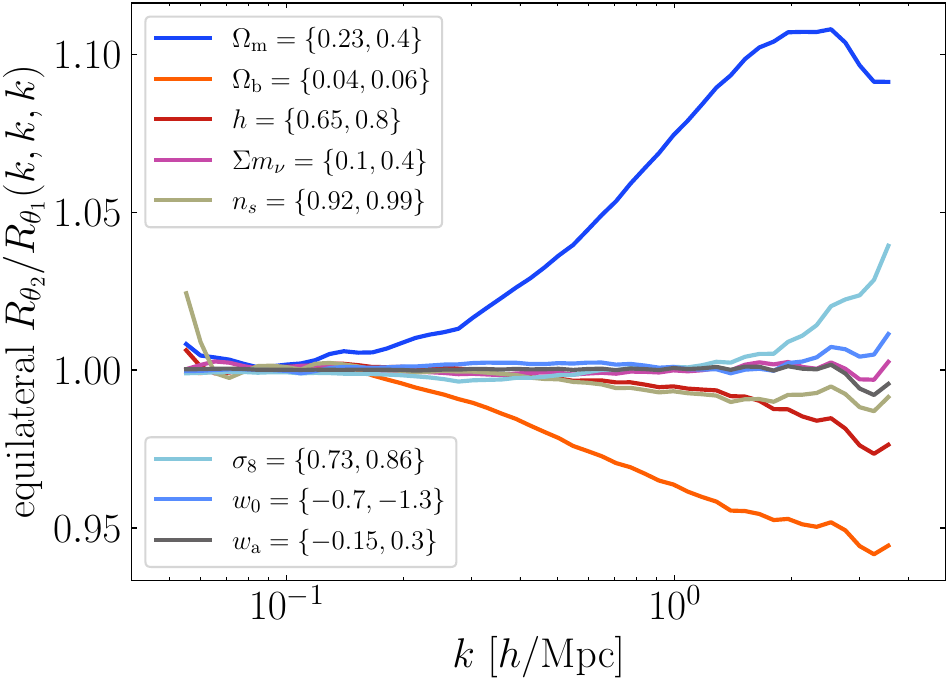}\\[1ex]
\includegraphics[width=0.45\textwidth]{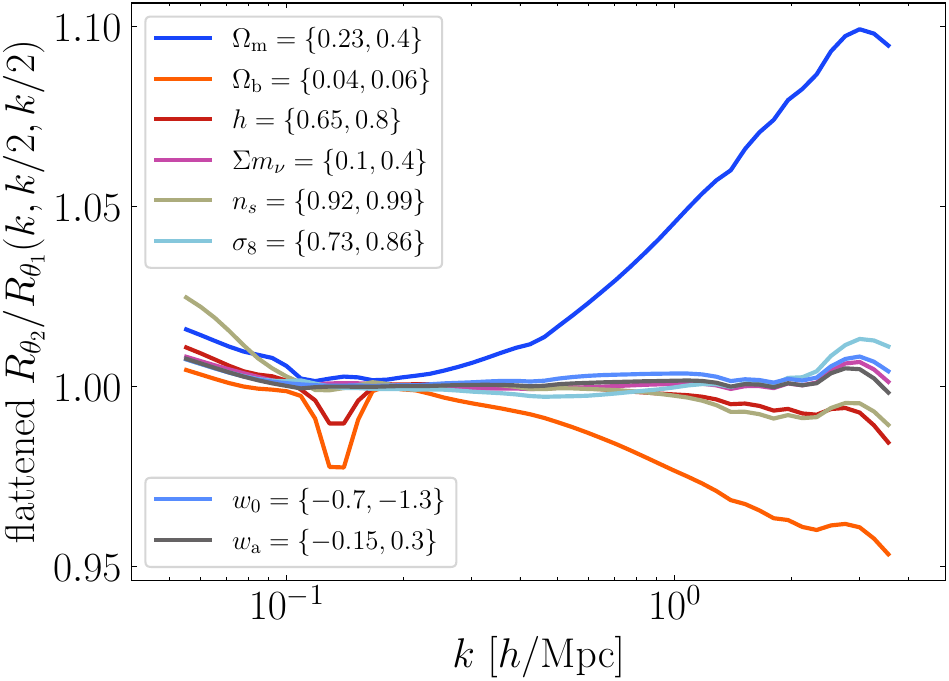}\hspace{0.04\textwidth}
\includegraphics[width=0.45\textwidth]{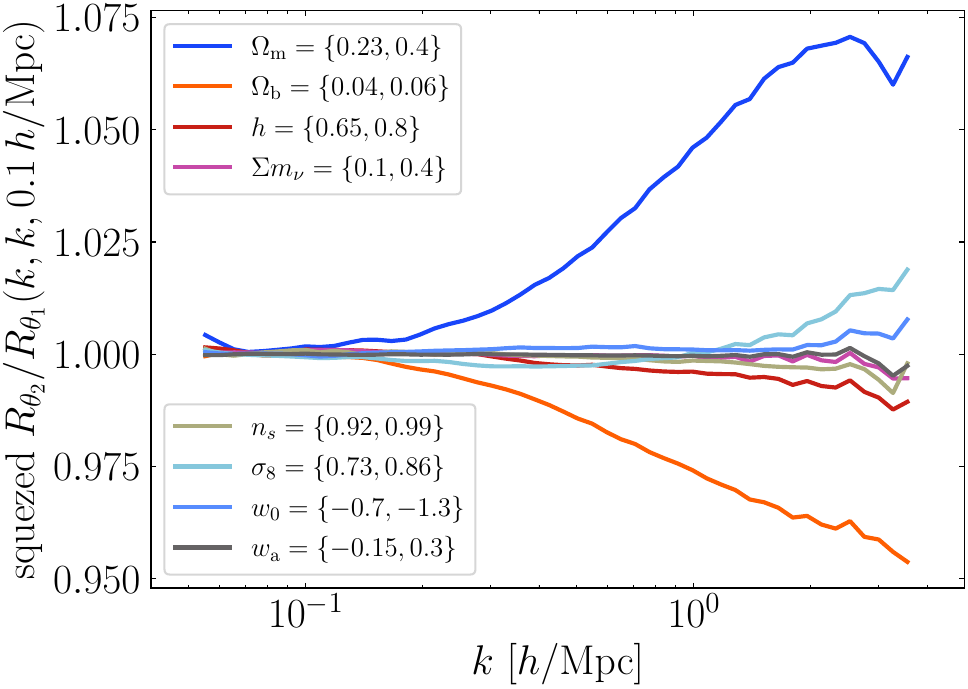}
\caption{Dependence of the $S(k)$ and $R(\boldsymbol{k})$ on cosmological parameters. Here, we plot the ratio between each parameter's lowest and largest values, as indicated in the legend.}
\label{fig:spectra_cosmology}
\end{figure*}

\begin{figure*}[ht]
\includegraphics[width=\textwidth]{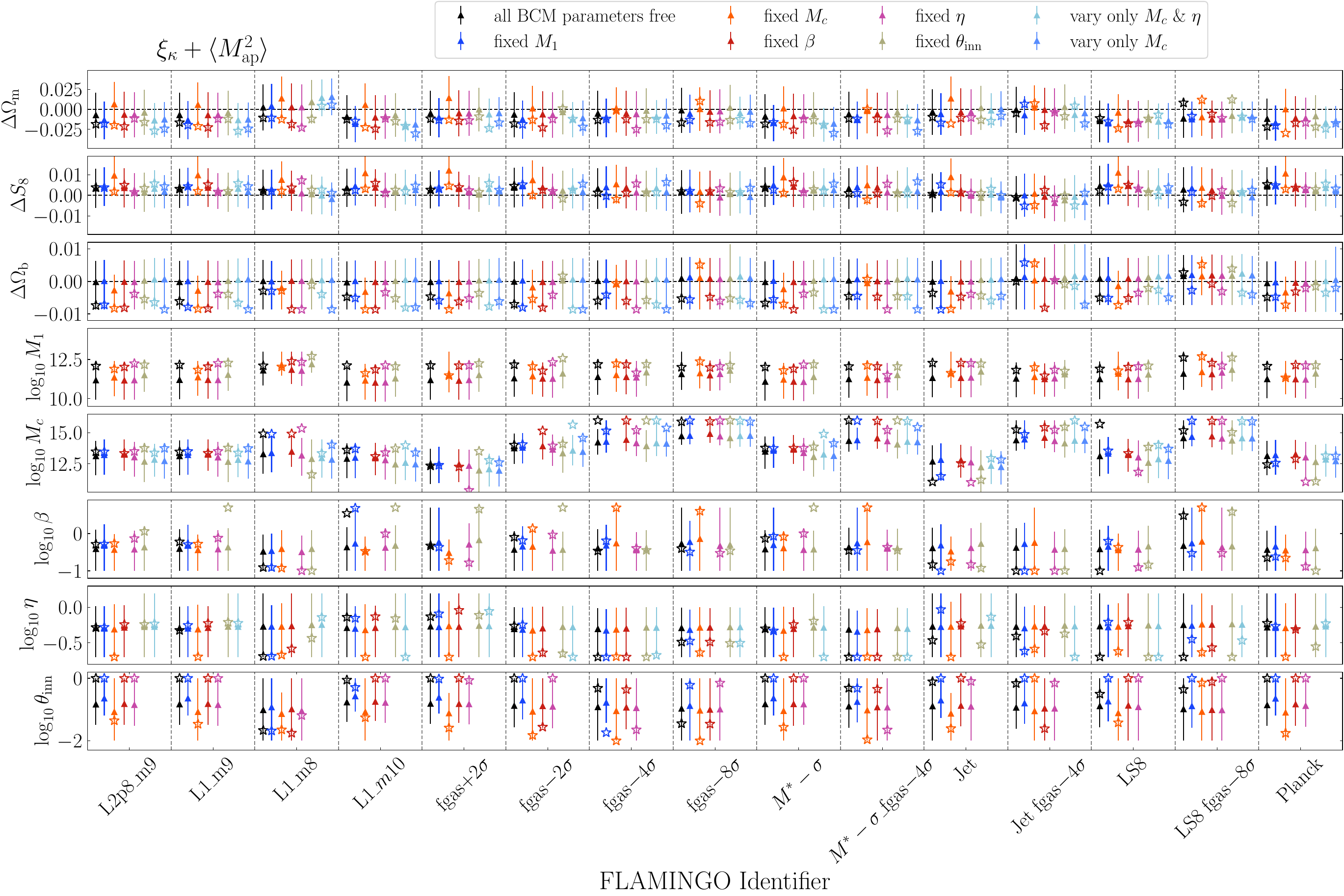}
\caption{Same as Fig.~\ref{fig:Flamingo_variation_cosmology} but second-order statistics alone. The triangles with error bars are estimated from the MCMC chains for all FLAMINGO models described in Sect.~\ref{sec:Flamingo_sim}. Since the LS8 and \Planck cosmology is different we plot $\Delta \Omega_\mathrm{m} = \Omega^\mathrm{best}_\mathrm{m} - \Omega^\mathrm{true}_\mathrm{m}$ and in analogy $\Delta S_8$ and  and $\Delta \Omega_\mathrm{b}$. The different colours show cases where parameters are fixed. The stars indicate the best-fitting parameters resulting from a minimisation process that returns the lowest $\chi^2$. The corresponding figure for $\xi_\kappa + \langle \Map^2 \rangle+ \langle \Map^3 \rangle$ is shown in Fig.~\ref{fig:Flamingo_variation_cosmology}.}
\label{fig:Flamingo_variation_cosmology_2nd}
\end{figure*}

\begin{figure*}[ht]
\includegraphics[width=\textwidth]{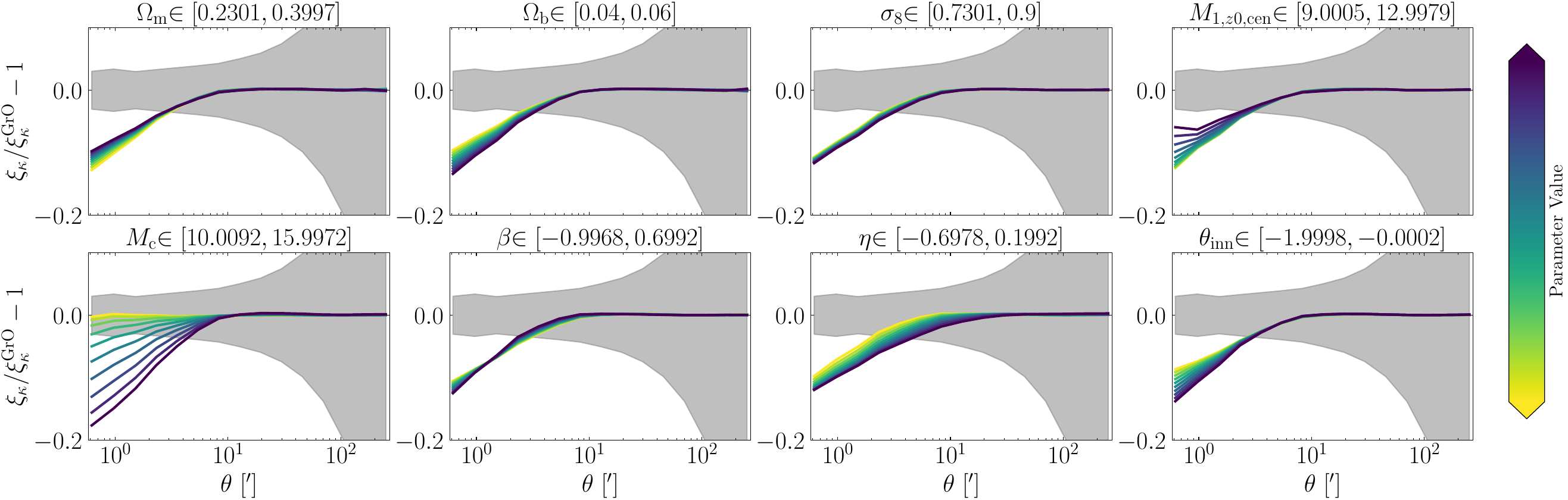}
\includegraphics[width=\textwidth]{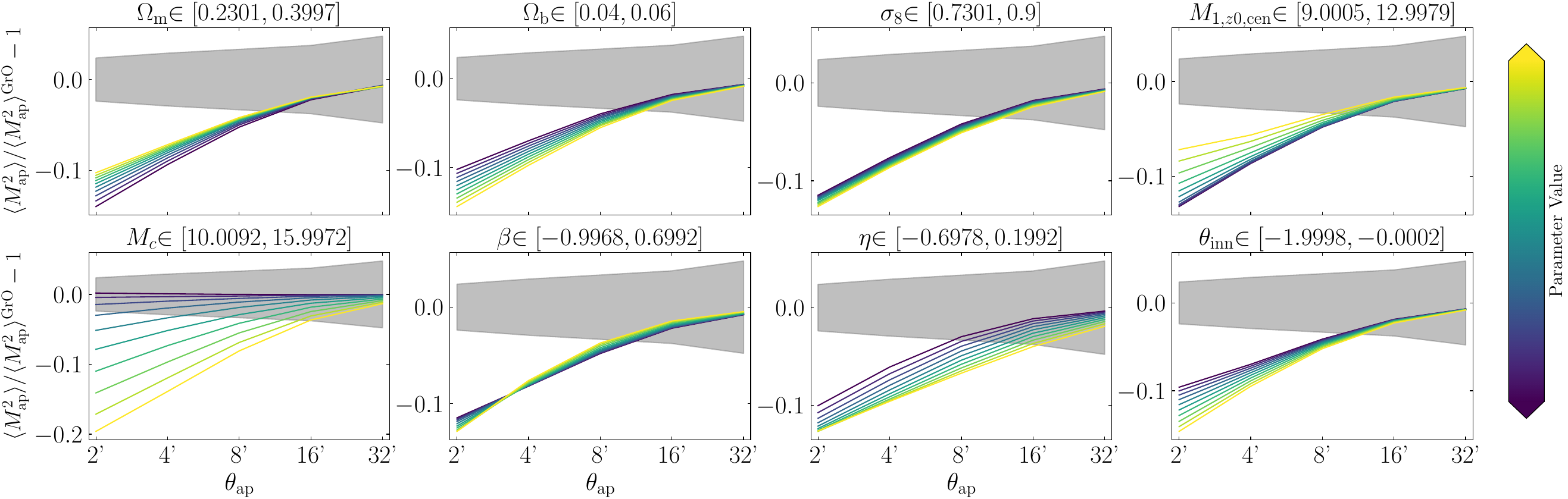}\\
\includegraphics[width=\textwidth]{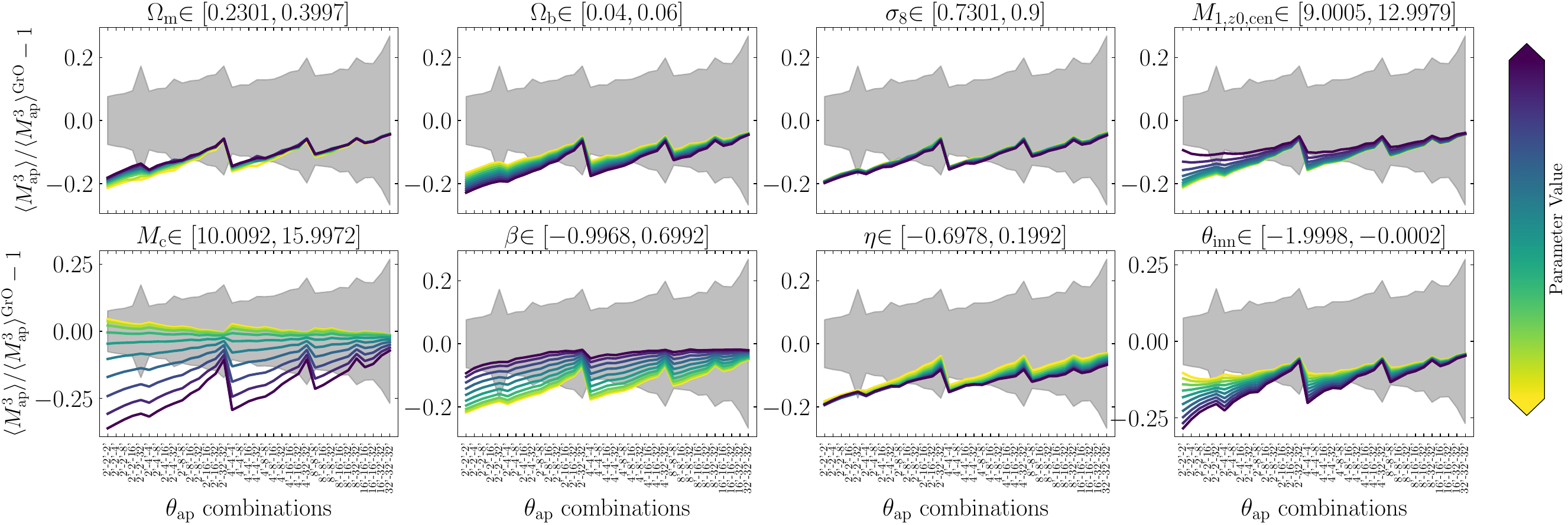}\\
\caption{Dependence of the $\expval{\Map^n}$ and $\xi_\kappa$ to changes of baryonic and cosmological parameters. We scaled the model vectors using the same model just for a GrO case, disentangling the effect of the baryons. The $\theta_\mathrm{ap} = \{\theta_\mathrm{ap,1},\theta_\mathrm{ap,2},\theta_\mathrm{ap,3}\}$ values for the lowest panel are increasing from left to right as in Fig.~\ref{fig:Map3_Flamingo}, increasing first $\theta_\mathrm{ap,3}$, then $\theta_\mathrm{ap,2}$ and lastly $\theta_\mathrm{ap,1}$. \label{LastPage}}
\label{fig:Map23_dependence_GrO}
\end{figure*}

Furthermore, we show in Fig.~\ref{fig:spectra_cosmology} the dependence of $S(k)$ and $R(\boldsymbol{k})$ on cosmological parameters. To do so, we apply the baryonification with a fixed set of baryonic parameters to a set of $N$-body simulations, where one cosmological parameter at a time is varied. We find that, analogously to the results with the power spectrum obtained by \cite{2021MNRAS.506.4070A},  $\Omega_\mathrm{m}$ and $\Omega_\mathrm{b}$ are the most significant, followed by $\sigma_8$. The other parameters, including $h$, $n_{\rm s}$, $\Sigma m_{\nu}$, $w_0$, $w_{\rm a}$, have a smaller impact than the nominal accuracy of the cosmological scaling, and thus we do not model them. Lastly, we show in Fig.~\ref{fig:Flamingo_variation_cosmology_2nd} the equivalent to Fig.~\ref{fig:Flamingo_variation_cosmology} but for second-order statistics alone, and in Fig.~\ref{fig:Map23_dependence_GrO} the dependence of the three measured weak lensing statistics on the effects of baryonic feedback.  

\end{appendix}

\end{document}